\documentclass[11pt]{article}

\usepackage[margin=1in]{geometry} 

\usepackage{algorithmic} 
\usepackage{algorithm} 
\usepackage{setspace} 
\usepackage{rotating} 
\usepackage{amsmath} 
\usepackage{amsfonts, amssymb} 
\usepackage{graphicx} 
\usepackage{epstopdf} 
\usepackage{hyperref} 
\usepackage{bm} 
\usepackage{bbm}
\usepackage{multirow} 
\usepackage{latexsym} 
\usepackage{natbib}
\usepackage{rotating} 
\usepackage{titlesec} 
\usepackage{caption} 
\usepackage{mathtools} 
\usepackage{color} 
\usepackage{enumitem} 
\graphicspath{ {plot/} }
\usepackage{diagbox}
\usepackage{pict2e}
\usepackage{etoolbox}
\usepackage[english]{babel}

\usepackage{cases}
\usepackage{tikz}
\usetikzlibrary{arrows, chains, positioning, quotes, shapes.geometric}

\usepackage{lato}

\usepackage{xr}
\definecolor{mycolor}{RGB}{0,200,200}

\newcommand{\indep}{\perp }

\DeclareMathOperator*{\argmin}{arg\,min}
\DeclareMathOperator*{\median}{median}

\newcommand{\NC}{N}
\newcommand{\NT}{{\overline{M}}}
\newcommand{\MT}{J}
\newcommand{\NI}{n}
\newcommand{\eij}{ {i(-j)} }
\newcommand{\II}{\mathcal{I}}
\newcommand{\PARASUB}{}

\newcommand{\bI}{{\mathbf{I}}}

\newcommand{\bg}{\mathbf{g}}

\newcommand{\bO}{\mathbf{O}}
\newcommand{\bY}{\mathbf{Y}}
\newcommand{\bA}{\mathbf{A}}
\newcommand{\bU}{\mathbf{U}}
\newcommand{\bV}{\mathbf{V}}
\newcommand{\bW}{\mathbf{W}}
\newcommand{\bX}{\mathbf{X}}

\newcommand{\bC}{\mathbf{C}}

\newcommand{\bo}{\mathbf{o}}
\newcommand{\by}{\mathbf{y}}
\newcommand{\ba}{\mathbf{a}}
\newcommand{\bu}{\mathbf{u}}
\newcommand{\bv}{\mathbf{v}}

\newcommand{\bx}{\mathbf{x}}

\newcommand{\bc}{\mathbf{c}}

\newcommand{\paraB}{\mathcal{B}_\gamma}
\newcommand{\Bmat}{B \big( \beta(\bC_i) \big)}

\newcommand{\ind}{\mathbbm{1}}
\newcommand{\T}{^{\intercal}}
\newcommand{\EXP}{\text{E}}
\newcommand{\VAR}{\text{Var}}
\newcommand{\cond}{\, \vert \,}
\newcommand{\con}{\, ; \,}
\newcommand{\pot}[2]{#1^{(#2)}}
\newcommand{\R}{\mathbb{R}}

\newcommand{\iid}{\stackrel{\mathrm{iid}}{\sim}}

\usepackage[normalem]{ulem}
\newcommand\redsout{\bgroup\markoverwith{\textcolor{red}{\rule[0.5ex]{2pt}{0.4pt}}}\ULon}

\hypersetup{
colorlinks=true,
linkcolor=blue,
filecolor=blue,
urlcolor=blue,
citecolor=blue
}

\graphicspath{ {plot/} }

\definecolor{red1}{RGB}{255,64,64}
\definecolor{blue1}{RGB}{128,255,255}
\definecolor{green1}{RGB}{0,205,0}
\definecolor{myblue}{RGB}{34,151,230}

\numberwithin{table}{section}
\numberwithin{figure}{section}

\usepackage{rotating}
\usepackage{caption}

\newtheorem{theorem}{Theorem}[section]
\newtheorem{lemma}[theorem]{Lemma}

\graphicspath{ {plot/} }

\definecolor{red1}{RGB}{255,204,204}
\definecolor{blue1}{RGB}{204,204,255}
\definecolor{light-gray}{gray}{0.7}

\begin{document}

\setlength{\abovedisplayskip}{5pt}
\setlength{\belowdisplayskip}{5pt}
\setlength{\abovedisplayshortskip}{5pt}
\setlength{\belowdisplayshortskip}{5pt}

\title{A More Efficient, Doubly Robust, Nonparametric Estimator of Treatment Effects in Multilevel Studies}
\author{Chan Park and Hyunseung Kang\\Department of Statistics, University of Wisconsin--Madison}
\date{ }
\maketitle
\begin{abstract}
When studying treatment effects in multilevel studies, investigators commonly use (semi-)parametric estimators, which make strong parametric assumptions about the outcome, the treatment, and/or the correlation structure between study units in a cluster. We propose a novel estimator of treatment effects that does not make such assumptions. Specifically, the new estimator is shown to be doubly robust, asymptotically Normal, and often more efficient than existing estimators, all without having to make any parametric modeling assumptions about the outcome, the treatment, and the correlation structure. We achieve this by estimating two non-standard nuisance functions in causal inference, the conditional propensity score and the outcome covariance model, using existing existing machine learning methods designed for independent and identically distributed (i.i.d) data. The new estimator is also demonstrated in simulated and real data where the new estimator is drastically more efficient than existing estimators, especially when studying cluster-specific treatment effects.
\end{abstract}

\section{Introduction} \label{sec:intro}
When studying treatment effects in the social sciences, study units are often clustered together and form dependencies. For example, in educational assessment studies, such as the Trends in International Mathematics and Science Study (TIMSS), the Programme for International Student Assessment (PISA), or the Early Childhood Longitudinal Study (ECLS), students (i.e., the study units) are clustered at the classroom or school level. In the National Study of Learning Mindsets that was part of the workshop at the 2018 Atlantic Causal Inference Conference \citep{Carvalho2019}, students (i.e., the study units) are clustered at the school level. In cross-classified educational studies, students belong to two, non-nested clusters, say school and neighborhood. In all these examples, which are broadly known as multilevel studies, students' data from the same school or the same neighborhood may be correlated due to shared school or neighborhood characteristics and properly accounting for the correlation structure, ideally with minimal assumptions, is paramount to obtain valid and robust inference of treatment effects  \citep{MLbook1}. The main goal of our paper is to study nonparametric, doubly robust estimation of treatment effects in multilevel studies where there are only nonparametric constraints on the observed multilevel data. A useful byproduct of our paper is one formal framework to leverage existing i.i.d.-based machine learning methods to more efficiently estimate treatment effects in clustered/multilevel settings.

There is a long and rich history of studying treatment effects in multilevel studies; see \citet{Review2}, \citet{Review1}, and references therein for the most recent reviews. 
Broadly speaking, to account for correlation in multilevel studies, many of the popular methods use parametric mixed effect models or semiparametric estimating equations
 \citep{GEE,Gelman2006,hongraudenbush06,ArpinoMealli2011,thoemmes2011ps,LZL2013, ArpinoCannas2016}. Some well-known methods include generalized linear mixed effects models (GLMMs) and generalized estimating equations (GEEs). However, a key limitation with these methods is that their respective parametric components are assumed to be correctly specified. 
 A bit more formally,
if $\bO_{ij} = (Y_{ij}, A_{ij}, \bX_{ij})$ denote the outcome, binary treatment, and pre-treatment covariates, respectively, of study unit $j$ in cluster $i$, GLMMs assume that the random effects, which model the correlation between study units in a cluster, are Normally distributed and the outcomes of a pair of units $ij$ and $ij'$, $j\neq j'$, follow a conditional bivariate Normal distribution with a constant off-diagonal covariance matrix.  
GEEs and other semiparametric approaches do not make distributional assumptions, but the outcome model $\EXP ( Y_{ij} \cond A_{ij}, \bX_{ij})$ and/or the propensity score model $\EXP ( A_{ij} \cond \bX_{ij})$ \citep{Rosenbaum1983} are often assumed to be (semi-)parametric \citep{RobinsSNMM1994, Gray1998, Robins1999, Gray2000, RobinsMSM, Goetgeluk2008, Stephens2012, LZL2013, Zetterqvist2016, Liu2018, Review2, Review1, Lee2021}. 
Unfortunately, all the aforementioned methods may lead to inconsistent estimates of treatment effects when the parametric component of their estimators are mis-specified.


Recently, there has been great interest in using supervised machine learning (ML) methods to estimate treatment effects where both of the outcome and the treatment are modeled nonparametrically; see \citet{SL2007}, \citet{Hill2011}, \citet{vvLaan2011}, \citet{AtheyImbens2016}, \citet{Victor2018}, \citet{grf}, \citet{Hahn2020}, \citet{NieWager2020}, and references therein. But, many of these estimators are designed for 
i.i.d. data where $\bO_{ij} \iid P$ for some distribution $P$ and it is not clear how to properly apply these flexible ML methods in multilevel settings where the data within each cluster are correlated, say when $P(\bO_{ij},\bO_{ij'}) \neq P(\bO_{ij})P(\bO_{ij'})$ for $j\neq j'$.



We present a new type of nonparametric, doubly robust and asymptotically Normal estimators of treatment effects in multilevel studies. At a high level, our estimator allows both (i) the outcome regression $\EXP (Y_{ij} \cond A_{ij}, \bX_{ij})$, (ii) the propensity score $\EXP (A_{ij} \cond \bX_{ij})$, and (iii) the correlation structure between study units to be nonparametric.  
In other words, we allow both the ``first-order'' conditional moments (i.e., (i) and (ii)) and the ``second-order'' conditional moments (i.e., (iii)) to be nonparametric. To achieve this, the estimator uses two non-standard nuisance functions in causal inference that nonparametrically capture the correlation structure in the treatment and the outcome: (i) a conditional propensity score, which is the propensity score of a study unit given his/her peers' treatment status and pre-treatment covariates, and (ii) an outcome covariance model, which measures the correlation in the residualized outcomes of study units in the same cluster. Also, the new estimator allows investigators to use existing i.i.d.-based ML methods in causal inference, but in a manner that allows for the estimator to be doubly robust and potentially more efficient in multilevel settings. In real data, we find that our estimator consistently has dramatically smaller variance the variances of competing estimators, especially when studying cluster-specific treatment effects; see Section \ref{sec:application} for details.

\section{Setup}											\label{sec:Setup}
\subsection{Review: Notation}
Let $i=1,\ldots,\NC$ index $\NC$ clusters and let $j=1,\ldots,\NI_{i}$ be the number of individuals (i.e., study units) within cluster $i$. In two-level studies, $i$ would denote clusters and $j$ would denote individuals within a cluster. In three-level observational studies, $i$ would denote clusters jointly defined by second and third levels and $j$ would denote individuals within this cluster; one can also expand the cluster index $i$ using double subscripts, say $st$ where $s$ indexes clusters at the second-level and $t$ indexes clusters at the third-level. 
In cross-classified studies with non-nested clusters, $i$ would denote each cluster defined by interacting non-nested clusters, say school and neighborhood, and $j$ would denote individuals within the interacted cluster. More generally, we can expand the subscript $i$ to better denote two, three, or $k$-level hierarchical structures in a multilevel study. But, for simplicity, we suppress these double or triple subscripts. 

For each study unit $j$ in cluster $i$, let $\bX_{ij}=(\bW_{ij}\T, \bC_i\T )\T$ denote the study unit's pre-treatment covariates where $\bW_{ij}$ denotes unit-level pre-treatment covariates and $\bC_i$ denotes cluster-level pre-treatment covariates. In educational assessment studies like the ECLS, unit-level covariates may include student's gender, age, prior test scores, and socioeconomic status. Cluster-level covariates may include type of school, say private or public schools, location of the school, say rural, suburban, or city, and cluster size $\NI_i$. Often, cluster-level covariates drive the dependence between study units in a cluster. For example, if the outcome is student test scores, school location may be one of the reasons why students from the same school may have similar test scores. In Section \ref{sec:result}, we utilize this observational about multilevel studies to help train the aforementioned outcome covariance model that captures the outcome correlation between study units. 

Let $A_{ij} \in \{0,1\}$ denote treatment assignment where $A_{ij} = 1$ indicates that individual $j$ in cluster $i$ was treated and $A_{ij} = 0$ indicates that individual $j$ in cluster $i$ was untreated. Let $Y_{ij}$ denote the observed outcome of individual $j$ in cluster $i$. We use the following notation for vectors of observed variable: $\bY_i = (Y_{i1}, \ldots, Y_{i\NI_i})\T$, $\bA_i = (A_{i1}, \ldots, A_{i\NI_i})\T$, and $\bX_i = (\bX_{i1}\T, \ldots, \bX_{i\NI_i}\T)\T$. Also, let $\bA_\eij$ be the vector of treatment variables for all study units in cluster $i$ except individual $j$. Overall, for each individual $j$ in cluster $i$, we observe $\bO_{ij} = (Y_{ij}, A_{ij}, \bX_{ij})$ and for each cluster $i$, we observe $\bO_i = (\bO_{i1},\ldots,\bO_{i \NI_i})$. 

We adopt the following notations for convergence, norms, and vectors. For a measurable function $f$, let $\|f\|_{P,2}$ denote the $L_2(P)$-norm of $f$, i.e., $\|f\|_{P,2}= \big\{ \int \| f(\bo) \|_2^2 \, dP(\bo) \big\}^{1/2}$.  
Let $O(\cdot)$ and $o(\cdot)$ be the usual big-o and small-o notation, respectively. We denote $m$-dimensional vectors of ones and zeros as $\bm{1}_m$ and $\bm{0}_m$, respectively, and denote the $m \times m$ identity matrix as $I_m$.
\subsection{Review: Causal Estimands and Causal Identification}
We use the potential outcomes notation \citep{Neyman1923, Rubin1974} 
to define causal effects. Let $\pot{Y_{ij}}{1}$ denote the potential outcome if individual $i$ in cluster $j$ were treated and let $\pot{Y_{ij}}{0}$ denote the potential outcome if the individual were untreated. 
The target estimand of interest is the (weighted) average treatment effect and is denoted as $\tau^*= \EXP  [ w(\bC_i) \{ \pot{Y_{ij}}{1} - \pot{Y_{ij}}{0} \} ]$.  
The weight function $w(\bC_i) \in \R$ is bounded and specified by the investigator based on the scientific question at hand.
For example, if $w(\bC_i) =1$, $\tau^*$ becomes the usual (i.e., unweighted or equally-weighted) average treatment effect. But, if an investigator wants to over-represent certain clusters, say schools in rural areas are weighed more than schools in non-rural areas when taking the average across schools, $w(\bC_i)$ can vary as a function of $\bC_i$. Or, if investigators only want to study certain clusters, we can set $w(\bC_i)=1$ for the clusters of interest, $w(\bC_i) =0$ otherwise, and re-weigh the estimand to study the (conditional) average treatment effects among schools with $w(\bC_i)=1$; see Section \ref{sec:application} for examples. For simplicity, we will refer to both the unweighted/equally-weighted and weighted/unequally-weighted treatment effects as the average treatment effect in the paper.

To identify the average treatment effect $\tau^*$ from observed data, we assume the following.
\begin{itemize}
\item[\hypertarget{(A1)}{(A1)}] (\textit{Stable Unit Treatment Value Assumption,  \citet{Rubin1976, Rubin1978}}) $Y_{ij} = \pot{Y_{ij}}{A_{ij}}$.
\item[\hypertarget{(A2)}{(A2)}] (\textit{Conditional Ignorability}) $\pot{Y_{ij}}{1} , \pot{Y_{ij}}{0} \perp A_{ij} \cond \bX_{ij}$.
\item[\hypertarget{(A3)}{(A3)}] (\textit{Overlap}) There exists a finite constant $\delta > 0$ such that $\delta  \leq P(\bA_{i} = \ba \cond \bX_{i}= \bx) \leq 1- \delta$ for all $(\ba,\bx)$.
\end{itemize}
\noindent Briefly, Assumptions \hyperlink{(A1)}{(A1)} and \hyperlink{(A2)}{(A2)} are well-known identifying assumptions in causal inference. Notably, 
similar to past works in multilevel studies (e.g., \citet{ArpinoMealli2011}, \citet{LZL2013}, \citet{ArpinoCannas2016},  \citet{Lee2021}), we assume that interference is absent. Assumption \hyperlink{(A3)}{(A3)} is a slight generalization of the usual overlap assumption to allow for a vector of treatment indicators in cluster $i$. Assumption \hyperlink{(A3)}{(A3)} implies the usual overlap assumption for the ``individual-level'' propensity score $e^*(a \cond \bx) = P(A_{ij} = a \cond \bX_{ij} = \bx)$, i.e., $c_e \leq P(A_{ij} = a \cond \bX_{ij} = \mathbf{x}) \leq 1-c_e$ for a positive constant $c_e$. In the paper, we refer to $e^*(a \cond \bx)$ as the individual-level propensity score and $P(\bA_{i} = \ba \cond \bX_{i}= \bx)$ as the cluster-level propensity score.
\citet{hongraudenbush06} and \citet{hong2013heterogeneous} discuss the practical merits of Assumptions \hyperlink{(A1)}{(A1)}-\hyperlink{(A3)}{(A3)} when estimating treatment effects in multilevel studies. 

Under Assumptions \hyperlink{(A1)}{(A1)}-\hyperlink{(A3)}{(A3)}, the average treatment effect $\tau^*$ can be written as a function of observed data via
\begin{align} \label{eq-ident}
	& \tau^*  = \EXP \big[ w(\bC_i) \big\{ g^*(1,\bX_{ij}) - g^*(0, \bX_{ij}) \big\} \big]
\end{align}
where $g^*(1,\bx) = \EXP (Y_{ij} \cond A_{ij} = 1, \bX_{ij}=\bx)$ and $g^*(0,\bx) = \EXP (Y_{ij} \cond A_{ij} = 0, \bX_{ij}=\bx)$ are the true outcome regression models under treatment and control, respectively. 
The rest of the paper focuses on estimating $\tau^*$, specifically the functional on the right-hand side of equation \eqref{eq-ident}, when the observed data $\bO_{ij}$ comes from a multilevel study, i.e., when $\bO_{ij} \iid P$ no longer holds.

\subsection{Review: Existing Estimators in Multilevel Studies}								\label{sec:2-3}
To better frame our contribution,  we briefly review existing estimators of $\tau^*$ in multilevel studies. Our review focuses on existing doubly robust estimators of $\tau^*$ as our proposed estimator is also doubly robust; for a more complete review of estimators in multilevel studies, including estimators based on inverse probability weighting, see \cite{MLbook1}, \citet{thoemmes2011ps}, \citet{Review2} and \citet{Review1}.

Broadly speaking, all existing doubly robust estimators of $\tau^*$ in multilevel settings are variations of the augmented inverse probability weighted (AIPW) estimator \citep{Robins1994, Scharfstein1999}, denoted as $\overline{\tau}$ below.
\begin{align}												\label{eq-bartau}
	& \overline{\tau} = \overline{\tau} (e, g)
	= \frac{1}{N} \sum_{i=1}^{\NC} {\varphi}(\bO_i, e,g)  \ , \
		\\ 
		&
		\varphi(\bO_i, e , g) 
		=  \frac{w(\bC_i)}{n_i} \sum_{j =1}^{\NI_i} \bigg[  \frac{A_{ij} \big\{ Y_{ij} - g(1, \bX_{ij}) \big\} }{e(1 \cond \bX_{ij})} + g(1, \bX_{ij}) 
		-   \frac{(1 - A_{ij}) \big\{ Y_{ij} - g(0, \bX_{ij}) \big\}  }{e(0 \cond \bX_{ij})} - g(0, \bX_{ij})   \bigg]. \nonumber 
\end{align}
So long as either the outcome model $g$ or the individual-level propensity score $e$ is correctly specified, but not necessarily both, $\overline{\tau}$ is consistent for $\tau^*$ (i.e., doubly robust). Also, under addition assumptions, $\overline{\tau}$ is asymptotically Normal around $\tau^*$. For completeness, Section \ref{Supp-sec:ExistingMethod} 
of the Supplementary Materials provides one proof of these statistical properties that allow for nonparametrically estimated $g$ and $e$. But, the citations we provide in the next paragraph provide alternative proofs of these statistical properties under different, often more restrictive, assumptions about $g$ and $e$.

Existing doubly robust estimators differ on how $g$ and $e$ in $\overline{\tau}$ are modeled. For example, one popular choice, especially in educational studies like our data example below, is to use generalized linear mixed effect models (GLMMs) \citep{Goldstein2002, hongraudenbush06, ArpinoMealli2011,thoemmes2011ps,ArpinoCannas2016,Shardell2018}, say
\begin{align*}			
	& Y_{ij} = \alpha_0 + A_{ij} \alpha_1 +
	\bX_{ij} \T \bm{\beta} + U_i + \epsilon_{ij} , \quad{}
	&& U_i \iid N(0,\sigma_U^2), \quad{} \epsilon_{ij} \iid N(0, \sigma_\epsilon^2), 
	\\ 
	&
	P \big( A_{ij} = 1 \cond \bX_i, V_i \big)
	=
	\text{expit}
	\big( \bX_{ij} \T \bm{\gamma} + V_i \big), \quad{} 
	&&
	V_i \iid N(0,\sigma_V^2), 
	\nonumber \\
		& g(A_{ij}, \bX_{ij}) = \int P(Y_{ij} \cond A_{ij}, \bX_{ij}, U_i = u) dP(u), \quad{} 
		&& e(1 \cond \bX_{ij}) = \int P \big( A_{ij} = 1 \cond \bX_{ij}, V_i = v \big) dP(v)  \ .
\end{align*}
where $(\bX_{i}, U_i, V_i) \perp \epsilon_{ij}$ and $\bX_{i} \perp (U_i, V_i)$. The terms $U_i$ and $V_i$ are unobserved, Normally distributed random effects that govern the correlation between study units in a cluster.  
Another popular choice is to use generalized estimating equations (GEEs) \citep{GEE, Gray1998, Robins1999, Gray2000, Goetgeluk2008, Stephens2012, Zetterqvist2016}, say $ g(A_{ij},\bX_{ij}) = \alpha_0 + A_{ij} \alpha_1 + \bX_{ij} \T \bm{\beta}$ and $e(1 \cond \bX_{ij}) = \text{expit}
	\big( \bX_{ij} \T \bm{\gamma} \big)$. Unlike GLMMs which parametrize the within-cluster correlations, GEEs leave the correlation structure unspecified. 
%
Unfortunately, both GLMMs and GEEs require parametric models for the outcome regression or the propensity score. If these models are mis-specified, $\overline{\tau}$ may be inconsistent for $\tau^*$.

Finally, a modern, but somewhat perplexing approach would be to use supervised ML models for the outcome regression $g$ and the propensity score $e$. 
We say perplexing because commonly used supervised ML models in causal inference often assume that the data, specifically $\bO_{ij}$, is i.i.d. even though in multilevel settings, it is not. But, as noted by \citet{Carvalho2019} from the aforementioned 2018 Atlantic Causal Inference Conference workshop in Section \ref{sec:intro}, using i.i.d.-based ML methods was common to study treatment effects in multilevel settings, likely due to its simplicity. 
Yet, to the best of our knowledge, there is no theoretical justification as to whether this would be the most appropriate or efficient way of using i.i.d.-based supervised ML models to estimate $\tau^*$ in multilevel studies. Our proposed estimator below can be seen an answer to this question by proposing a different way to use i.i.d.-based modern ML methods to obtain a doubly robust, asymptotically Normal, and a more efficient estimator of $\tau^*$.

\section{Our Approach: A More Efficient, Nonparametric, Doubly Robust Estimator in Multilevel Studies}							\label{sec:result}
\subsection{Key Ideas: Conditional Propensity Score and Outcome Covariance Model} \label{sec:overview_est}
At a high level, our proposed estimator, denoted as $\widehat{\tau}$, retains some of the favorable properties of the existing estimator $\overline{\tau}$ discussed above, notably double robustness. But, in settings where $\overline{\tau}$ is consistent and asymptotically Normal for $\tau^*$ in multilevel studies, $\widehat{\tau}$ can be more efficient than $\overline{\tau}$. For simplicity, this section  introduces $\widehat{\tau}$ when the nuisance functions are known and subsequent sections discusses how to estimate the nuisance functions using i.i.d.-based ML methods.

The new estimator $\widehat{\tau}$ contains two new functions to model the correlation structure between study units: (i) the conditional propensity score, denoted as $\pi^* ( a \cond \ba' , \bx, \bx' ) = P \big( A_{ij} = a \cond \bA_\eij = \ba', \bX_{ij} = \bx, \bX_{\eij} = \bx' \big)$, and (ii) the outcome covariance model $\beta(\bC_i)$. 
These two functions are used inside the proposed estimator $\widehat{\tau}$ as follows:
\begin{align*}
	& \widehat{\tau}(\pi, g, \beta) = \frac{1}{\NC} \sum_{i=1}^\NC \phi(\bO_i,\pi,g,\beta), \\[-0.1cm]
	\nonumber
	&
	\phi(\bO_i, \pi, g, \beta) 
	=  \frac{w(\bC_i)}{\NI_i}  \sum_{j=1}^{\NI_i} \bigg[ \frac{A_{ij} \big[ \{ Y_{ij} - g(1, \bX_{ij}) \} - \beta(\bC_i) \sum_{k\neq j} \{ Y_{ik} - g(A_{ik}, \bX_{ik}) \} \big] }{\pi(1 \cond \bA_{i(-j)}, \bX_{ij}, \bX_{i(-j)})} +  g(1, \bX_{ij}) 
	\nonumber
	\\[-0.1cm]
	&
	\hspace*{2cm}
	- 
	 \frac{(1-A_{ij}) \big[ \{ Y_{ij} - g(0, \bX_{ij})\} - \beta(\bC_i) \sum_{k\neq j} \{Y_{ik} - g(A_{ik}, \bX_{ik})\}\big]}{\pi(0 \cond \bA_{i(-j)}, \bX_{ij}, \bX_{i(-j)})} -  g(0, \bX_{ij}) 
	 \bigg]\ .
	  \nonumber
	  \hspace*{-1cm}
\end{align*}
Compared to the existing estimator $\overline{\tau}$, the new estimator $\widehat{\tau}$ captures the correlation structure in the treatment and the outcome variables. Specifically, for the treatment variable, $\widehat{\tau}$ uses the conditional propensity score $\pi$ to capture the (conditional) correlation between $ij$'s treatment $A_{ij}$ and his/her peers $\bA_\eij$ conditional on $ij$'s and his/her peers' covariates, $\bX_{ij}$ and $\bX_\eij$, respectively. As we show below, the conditional propensity score reduces to the usual, individual-level propensity score in the absence of any correlation between treatment variables. More importantly, as we show in the theorems below, the conditional propensity score is, in some sense, the ``correct'' inverse probability weight to use in order to simultaneously maintain double robustness and achieve efficiency gains in multilevel settings. 
For the outcome variable, $\widehat{\tau}$ uses the outcome covariance model $\beta(\bC_i)$ to capture the relationship between $ij$'s (residual) outcome relative to his/her peers' (residual) outcomes; here, the residual refers to the leftover variation in an individual's outcome after conditioning on his/her own treatment and pre-treatment covariates. Specifically, 
$\beta(\bC_i)$ acts as a (functional) regression coefficient from regressing $ij$'s residual (i.e., $Y_{ij} - g(A_{ij}, \bX_{ij})$) onto the sum of his/her peers' residuals (i.e., $\sum_{k\neq j} \{Y_{ik} - g(A_{ik}, \bX_{ik}) \} $). 

The new estimator $\widehat{\tau}$ collapses to the existing estimator $\overline{\tau}$ under additional assumptions about the conditional propensity score and the correlation structure. For example, if the $ij$'s treatment is independent of his/her peers' treatments and pre-treatment covariates conditional on his/her own pre-treatment covariates, $\pi$ reduces the usual, individual-level propensity score $e$ and the inverse probability weights are identical between $\widehat{\tau}$ and  $\overline{\tau}$. 
Or, if the residuals of everyone are independent of each other, or equivalently, everyone's outcomes are independent of each other conditional on their own treatment and pre-treatment covariates, a hypothetical regression between $ij$'s residuals onto his/her peers residuals would lead to an estimated coefficient of $\beta(\bC_i) = 0$, matching the equivalent expression in equation \eqref{eq-bartau} for $\overline{\tau}$. Finally, if both independence conditions hold, $\widehat{\tau}$ equals $\overline{\tau}$. In short,
$\overline{\tau}$ can be seen as a special case of $\widehat{\tau}$ when there are no correlations between the outcome and the treatment of study units in the same cluster.

Given the conditional propensity score $\pi$ and the outcome regression $g$, let $\beta_{\pi,g}^*$ be the minimizer of the variance of the estimator $\widehat{\tau}(\pi,g,\beta)$ over some model space for the outcome covariance model $\mathcal{B}$ that contains 0, i.e., $\beta_{\pi,g}^* = \argmin_{\beta \in \mathcal{B} } \VAR \big\{ \widehat{\tau}(\pi,g,\beta) \}$ where $0 \in \mathcal{B}$; the next section discusses different model spaces $\mathcal{B}$. Lemma \ref{thm:2} highlights some simple, but useful characteristics of $\widehat{\tau}$ when we assume, for a moment, that the nuisance functions are fixed, non-random functions.
\begin{lemma} \label{thm:2} Suppose Assumptions \hyperlink{(A1)}{(A1)}-\hyperlink{(A3)}{(A3)} and the following assumptions hold.
\begin{itemize}
\item[\hypertarget{(M1)}{(M1)}] (\textit{Nonparametric Model}) For some constants $\NT$, the observed data satisfies the following; see Section \ref{sec:M1} for examples of data generating processes in multilevel studies which satisfy \hyperlink{(M1)}{(M1)}. 
\begin{align*}
 &
P(\bO_1,\ldots,\bO_\NC)  =
\prod_{i=1}^{\NC}   P (\bO_i) \ , \\
&
P (\bO_i) \in \left\{ P \left|
\begin{array}{l}
  \NI_i \leq \NT, \
 \EXP \big( Y_{ij} \cond A_{ij} =a, \bX_{ij} = \bx \big) =  
g(a,\bx) , \\
\EXP  \big( A_{ij} =a \cond  \bA_{i(-j)} = \ba', \bX_{ij} = \bx, \bX_{\eij} = \bx' \big) = \pi( a \cond \ba', \bx, \bx') 
\end{array} \right.
\right\} \ . 
\end{align*}
%
\item[\hypertarget{(M2)}{(M2)}] (\textit{Bounded Moments}) Let $\bg^*(\bA_i, \bX_i) =  (g^*(A_{i1}, \bX_{i1}), \ldots, g^*(A_{i\NI_i}, \bX_{i\NI_i}))\T$. 
For all $(\bA_i,\bX_i)$,  $\EXP \big\{ \big\|  \bY_i - \bg^*(\bA_i, \bX_i) \big\|_2^4 \cond \bA_i, \bX_i \big\}$ is bounded. Also, for any $\bX_i$, there exists $\bA_i$ such that the smallest eigenvalue of 
	$\EXP \big[  \big\{ \bY_i - \bg^*(\bA_i, \bX_i)\big\}^{\otimes 2} \cond \bA_i, \bX_i \big]$ is positive. 
\end{itemize}
Then, the following properties hold for the proposed estimator $\widehat{\tau}$.
\begin{itemize}
	\item[(1)] (\textit{Mean Double Robustness}) For any fixed outcome covariance model $\beta$, we have $\EXP \big\{ \widehat{\tau}(\pi, g,\beta) \big\} = \tau^*$ as long as the conditional propensity score or the outcome model (but not necessarily both) is correct, i.e., $\pi=\pi^*$ or $g=g^*$. 
	\item[(2)] (\textit{Efficiency Gain Under Conditionally Independent Treatment Assignment}) Suppose the cluster-level propensity score is decomposable into the product of individual-level propensity scores, i.e., $P(\bA_i \cond \bX_i ) = \prod_{j = 1}^{\NI_i} P(A_{ij} \cond \bX_{ij} )$. Then, we have $\VAR \big\{ \widehat{\tau}(\pi^*,g,\beta_{\pi^*,g}^* ) \big\} \leq \VAR \big\{  \overline{\tau}(e^*,g) \big\}$  for any outcome model $g$.
\end{itemize}
\end{lemma}
In words, part 1 of Lemma \ref{thm:2} states that the new estimator $\widehat{\tau}$ is doubly robust (in expectation) for any outcome covariance model $\beta$; that is, whatever the investigator specifies for the outcome covariance model $\beta$, $\widehat{\tau}$ will remain unbiased if either the conditional propensity score or the outcome regression, but not necessarily both, is correct. This property is very similar to a well-known property of the GEE estimator where, irrespectively of the specification of the weighting/covariance matrix, the GEE estimator remains consistent. Indeed, as we show in Section \ref{sec:properties}, with a few mild assumptions, notably on how the nuisance functions are nonparametrically estimated, we can also achieve the usual double robustness (in consistency) and critically, asymptotic Normality of $\widehat{\tau}$ for a working, but not necessarily ``correct'' outcome covariance model $\beta$.

Part 2 of Lemma \ref{thm:2} states that if the treatment assignment mechanism is independent conditional on $\bX_i$ and does not depend on others' covariates, the new estimator $\widehat{\tau}$ is at least as efficient than the existing estimator $\overline{\tau}$. The conditional independence assumption is satisfied in most randomized experiments and some observational studies in multilevel settings where each individual is randomly assigned to treatment with equal probability or with probability that is a function of his/her own pre-treatment covariates. However, when the treatment assignment mechanism depends on peers' covariates or treatment, there is no general relationship between $\widehat{\tau}$ and $\overline{\tau}$ in terms of variance unless we make further assumptions about the correlation structure. This is, in part, because the variance term in the asymptotic expansion of $\widehat{\tau}$ may behave like a first-order bias term in clustered settings; see Section \ref{Supp-proof-thm:2} 
of the Supplementary Materials for the exact details. 
Regardless, even if the conditional independence assumption in part 2 of Lemma \ref{thm:2} is not satisfied and the relative efficiency between the two estimators cannot be theoretically characterized, subsequent sections numerically show that $\widehat{\tau}$ almost always have smaller variances than $\overline{\tau}$ or other existing estimators.

\subsection{Interpreting Assumption (M1)}		 \label{sec:M1}
We take a moment to discuss an important condition for our theoretical results, Assumption \hyperlink{(M1)}{(M1)}, and to explain that, despite its initial appearance, the assumption is not strong for estimating the our target estimand $\tau^*$. 
At a high level, Assumption \hyperlink{(M1)}{(M1)} assumes that the cluster-level data $\bO_i$ are i.i.d. from some distribution $P$, which belongs to a family of models $\mathcal{P}$ where the conditional expectation of the individual-level outcome and the individual-level treatment are the same for every unit $ij$. At first, this may seem like a strong assumption. Indeed, if the target quantity of interest is at the \emph{cluster-level}, say the average treatment effect for a cluster-level outcome and the treatment is assigned at the cluster-level, our work based on Assumption \hyperlink{(M1)}{(M1)} is a simple extension of the usual semiparametric theory in causal inference except its applied at the cluster-level; in fact, in this setting, the efficient influence function of the cluster-level estimand can be derived from existing work of \citet{Hahn1998}. However, our quantity of interest $\tau^*$ is at the \emph{individual-level}, specifically the average treatment effect of assigning treatment at the individual-level on an individual-level outcome, and Assumption \hyperlink{(M1)}{(M1)} is likely insufficient to derive the efficient influence function of $\tau^*$ because of the unknown correlation structure of $\bO_{ij}$ within each cluster; see additional details below. To put it differently, the mismatch between our target estimand $\tau^*$, which is defined as a contrast at the \emph{individual-level}, and Assumption  \hyperlink{(M1)}{(M1)} on the observed data, which is assumed to be i.i.d. at the \emph{cluster-level}, makes the problem  more difficult than had we assumed that both the estimand and the data's i.i.d. assumption are at the same level (i.e., both individual-level or cluster-level).

Also, Assumption \hyperlink{(M1)}{(M1)} does not necessarily assume that all functionals of the individual-level observations $\bO_{ij}$ are identical across study units; it only assumes that the individual outcome regression and the conditional propensity score model are the same. For example, it may be possible that the conditional variance of the outcome $\VAR ( Y_{ij} \cond \bA_i, \bX_i )$ may be different across study units where study unit $j=1$ in cluster $i$ may have $\VAR ( Y_{i1} \cond \bA_{i}, \bX_{i} ) = 1$ and study unit $j=2$ in cluster $i$ may have $\VAR ( Y_{i2} \cond \bA_{i}, \bX_{i} ) = A_{i2} + \| \bX_i \|^2 + 1$. Similarly, the functional form of the usual, individual-level propensity score $e(a \cond \bX_{ij})$ may actually be different across study units where $e(1 \cond \bX_{ij}) = 0.5$ for study unit $j=1$ and $e(1 \cond \bX_{ij}) =  \text{expit} \big\{ 1 + 2  \big\| \bX_{ij} \|^2  \big\}$ for study unit $j=2$.

Additionally, Assumption \hyperlink{(M1)}{(M1)} encompasses a large array of existing parametric, semiparametric, or nonparametric models that have been used to understand multilevel observational studies. The list below provides some common examples; all examples assume that the clusters are independent from each other and $\NI_i$ is bounded. Also, Section \ref{Supp-sec:modelM1} 
of the Supplementary Materials provide formal proofs that these models satisfy \hyperlink{(M1)}{(M1)}.
\begin{enumerate}
\item \emph{GLMM and GEE models in Section \ref{sec:2-3}}: Under mild assumptions on the rank of $\bX_{ij}$, 
 these existing parametric or semi-parametric models for multilevel studies are instances of the models allowed under Assumption \hyperlink{(M1)}{(M1)}.
\item \emph{Locally independent, nonparametric latent variable models of \citet{Lord1980} and \citet{Holland1986}}:  Suppose there exists mutually independent latent variables $(\bU_i, \bV_i)$ that are also independent from pre-treatment covariates. These variables factorize the observed outcome and  treatment variables as follows:
\begin{align}				\label{eq-201}
	&
	P ( \bY_i \cond \bA_i, \bX_i, \bU_i , \bV_i )
	=
	\prod_{j=1}^{\NI_i} P(Y_{ij} \cond A_{ij} , \bX_{ij} , \bU_i ),  
	\nonumber
	\\
	&
	P ( \bA_i \cond \bX_i, \bU_i, \bV_i )
	=
	\prod_{j=1}^{\NI_i} P(A_{ij} \cond \bX_{ij} , \bV_i ), \ \bX_i \indep (\bU_i, \bV_i)  \ .
\end{align}

 \begin{figure}[!htb]
	\centering
\begin{tikzpicture}[scale=0.7]
\tikzset{vertex/.style = {draw=none,fill=none,minimum size=1em}}
\tikzset{edge/.style = {->,> = latex'}}
\tikzset{edge2/.style = {-}}
\node[vertex] (v) at (0,0) {$\bV_i$};
\node[vertex] (u) at (0,4) {$\bU_i$};
\node[vertex] (x1) at (0,1) {$\bX_{i1}$};
\node[vertex] (x2) at (0,3) {$\bX_{i2}$};
\node[vertex] (a2) at (5,1) {$A_{i2}$};
\node[vertex] (a1) at (4,0) {$A_{i1}$};
\node[vertex] (y2) at (5,3) {$Y_{i2}$};
\node[vertex] (y1) at (4,4) {$Y_{i1}$};

\draw[edge2] (x1) to (x2);
\draw[edge] (x1) to (a1);
\draw[edge] (x2) to (a2);
\draw[edge] (x1) to (y1);
\draw[edge] (x2) to (y2);
\draw[edge] (a1) to (y1);
\draw[edge] (a2) to (y2);
\draw[edge] (v) to (a1);
\draw[edge] (v) to (a2);
\draw[edge] (u) to (y1);
\draw[edge] (u) to (y2);
\end{tikzpicture}
\caption{A graphical illustration of Assumption \protect\hyperlink{(M1)}{(M1)} under a cluster of size $\NI_i=2$ when there are latent variables $\bU_i$ and $\bV_i$.}
\label{Fig:DAG}
\end{figure}
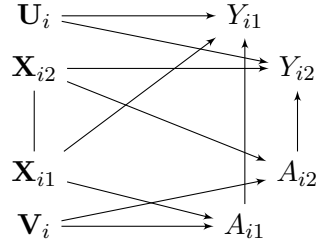

Figure \ref{Fig:DAG} provide a visual illustration of equation \eqref{eq-201} when $\NI_i = 2$. 
The latent variables $\bU_i$ and $\bV_i$ are not unmeasured confounders in that they do not simultaneously affect both the outcome and the treatment. Instead, the latent variables act similarly to random effects in GLMMs in that they are used to model dependence in the observed outcome or the treatment variables among study units in the same cluster. These models are also allowed under Assumption \hyperlink{(M1)}{(M1)}.
\item \emph{Nonparametric, conditionally i.i.d. models}: Conditional on the observed cluster-level variables $\bC_i$, the individual-level observations $\bO_{ij}$ become independent, i.e.,$P (\bO_1,  \ldots , \bO_\NC ) = \prod_{i=1}^{\NC}  P(\bC_i) \prod_{j=1}^{\NI_i} P(\bO_{ij} \cond \bC_i)$. These conditionally independent nonparametric models are a subset of the models allowed under Assumption \hyperlink{(M1)}{(M1)} .
\item \emph{Nonparaemtric mixture models}: Suppose each cluster's data is generated from a mixture model of the form
\begin{align*}
P(\bO_i) = \int P(\bO_i \cond \bm{Z}_i = \bm{z} ) \, dP( \bm{z} ), 
\end{align*}
where $\bm{Z}_i$ represents observed or unobserved mixture labels and the functional forms of $\EXP(Y_{ij} \cond A_{ij} = a, \bX_{ij} = \bx , \bm{Z}_i = \bm{z} )$ and $\EXP  \big( A_{ij} =a \cond  \bA_{i(-j)} = \ba', \bX_{ij} = \bx, \bX_{\eij} = \bx', \bm{Z}_i = \bm{z} \big)$ are the same across all $(i,j)$. 
We remark that \citet{Vaart1996} considered a variant of the above model where the density $P(\bO_i \cond \bm{Z}_i = \bm{z} )$ was parametrized by a finite-dimensional parameter. Under additional assumptions on $\bm{Z}_i$, say the conditional distribution of $\bm{Z}_i \cond (h_1(\bO_{ij}), h_2(\bO_{\eij}) )$ is identical across all $(i,j)$ for any fixed functions $h_1$ and $h_2$, 
the mixture model satisfies \hyperlink{(M1)}{(M1)}; note that these are not the only assumptions on $\bm{Z}_i$ that would satisfy \hyperlink{(M1)}{(M1)} and see Section \ref{Supp-sec:modelM1} 
of the Supplementary Materials for weaker conditions on $\bm{Z}_i$.
\end{enumerate} 
Assumption \hyperlink{(M1)}{(M1)} assumes that the true data generating model can be any one of these models above, plus others that are not listed. Without knowing a priori which model is the correct data generating model or making more assumptions on top of Assumption \hyperlink{(M1)}{(M1)}, deriving the efficient influence function of $\tau^*$ may not be possible and the next best alternative is to develop an estimator that is at least as efficient as the existing ones.

Finally, Assumption \hyperlink{(M1)}{(M1)} bounds the cluster size $\NI_i$ and to the best of our knowledge, there is no established semiparametric efficiency theory that allows the dimension of $\bO_i$ to grow to infinity while the elements of $\bO_i$ remain dependent arbitrarily and asymptotically, i.e., the dependence does not vanish to zero as sample size increases. Instead, by bounding the cluster size, we allow for arbitrary dependence between units in a cluster. Practically, this implies that our estimator's theoretical guarantees in multilevel studies are realized when the cluster size is smaller than the number of clusters, i.e., $\NC \gg \NI_i$. But, in our numerical study in Section \ref{sec:varyingN}, we find that our theoretical guarantees remain plausible even if this is not the case, say if there are $\NC = 25$ clusters with each cluster having $\NI_i = 100$ study units.

\subsection{Nonparametric Estimation of Nuisance Functions $\pi, g$ and $\beta$}  \label{sec:nfunc}
In this section, we discuss nonparametric estimation of the nuisance functions $\pi, g$ and $\beta$ inside the proposed estimator $\widehat{\tau}$ using existing, i.i.d.-based ML methods popular in causal inference. For estimating the conditional propensity score $\pi$ and the outcome regression $g$, we propose to use a slight modification of cross-fitting \citep{Victor2018} at the cluster-level where we randomly split the observed data into non-overlapping folds, say two folds $\II_1$ and $\II_2$, at the cluster level instead of at the individual-level; note that the original version of cross-fitting randomly splits the data at the individual-level and assumes that each study unit's data is i.i.d. One fold, referred to as the auxiliary data and denoted as $\II_k^c$, is used to estimate the nuisance functions and the other fold, referred to as the main data and denoted as $\II_k$, evaluates the estimated nuisance functions obtained from the first fold. We then switch the roll of the two folds to fully use the data for treatment effect estimation. We let $\widehat{\pi}^{(-k)}$ and $\widehat{g}^{(-k)}$ be the estimated conditional propensity score and the outcome regression based on the auxiliary data $\II_k^c$. We also let $\widehat{\pi}^{(-k)}( a \cond \bA_{\eij}, \bX_{ij}, \bX_{\eij} )$ $(a=0,1)$ and $\widehat{g}^{(-k)}  (A_{ij}, \bX_{ij})$ be the evaluation of these functions using cluster $i$ in the main data $\II_k$. 

There are some practical considerations to consider when estimating the conditional propensity score $\pi$ with existing i.i.d-based ML methods. Naively using the raw variables $(\bA_\eij, \bX_{ij}, \bX_\eij)$ as independent variables in existing i.i.d.-based ML methods may not work well because each cluster may have different cluster size, leading to different number of independent variables. For example, without covariates, a cluster of size $2$ will have 1 independent variable while a cluster of size 10 will have 9 independent variables when we directly use $\bA_\eij$ as independent variables in a nonparametric regression estimator. Thankfully, there are many ways to resolve this issue in the literature. \citet{vvLaan2014}, \citet{Sofrygin2016}, and \citet{Ogburn2020arxiv} suggested making structural assumptions on $\pi$ where the investigator uses fixed, dimension-reducing summaries of peers' data $(\bA_\eij, \bX_\eij)$ as independent variables. Some popular examples of these dimension-reducing summaries include the proportion of treated peers, $\overline{A}_\eij = \sum_{\ell \neq j} A_{i\ell}/(\NI_i-1)$, and the average of peers' covariates, $\overline{\bX}_\eij = \sum_{\ell \neq j} \bX_{i\ell}/(\NI_i-1)$. 
With the additional structural assumptions, estimating $\pi$ becomes a classification problem where $A_{ij}$ is the dependent variable and $(\overline{A}_\eij, \bX_{ij}, \overline{\bX}_\eij) $ are the independent variables, respectively. Or, investigators can use additional summaries of peers' data as independent variables or more flexible models of dependence typical in network data \citep{Salathe2010, Keeling2011,Lorna2013, Imai2015}. 


For estimating the outcome covariance model $\beta$, we only use the main sample from the cluster-level cross-fitting procedure above and solve the following optimization problem. 
\begin{equation} \label{eq-genbeta}
 \widehat{\beta}^{(k)} \in \argmin_{\beta \in \mathcal{B} } \frac{2}{\NC} \sum_{i \in \II_k}  
 \left[
 \begin{array}{l}
 	w^2(\bC_i) \bI(\bA_i, \bX_i , \widehat{\pi}^{(-k)} )\T \Bmat 
 	\\
 	\hspace*{1.5cm}
 	\times \widehat{\mathbf{S}}_{i}^{(-k)}  \Bmat \bI(\bA_i, \bX_i , \widehat{\pi}^{(-k)} )\T
 \end{array}
 \right]
\end{equation}
where  $\Bmat = I_{\NI_i} - \beta(\bC_i) \bm{1}_{\NI_i} \bm{1}_{\NI_i} \T , \quad{} 
	\widehat{\mathbf{S}}_{i}^{(-k)} = \big\{ \bY_i - \widehat{\bg}^{(-k)}(\bA_i, \bX_i) \big\}^{\otimes 2}$ and
\begin{align*}
	\bI (\bA_i, \bX_i , \widehat{\pi}^{(-k)}) & =
	\frac{1}{\NI_i}
	\begin{bmatrix}
	\frac{\ind(A_{i1} = 1)}{ \widehat{\pi}^{(-k)}(1 \cond \bA_{i(-1)}, \bX_{i1}, \bX_{i(-1)}) }
	-
	\frac{\ind(A_{i1} = 0)}{ \widehat{\pi}^{(-k)} (0 \cond \bA_{i(-1)}, \bX_{i1}, \bX_{i(-1)}) }
		\\
		\vdots
		\\
		\frac{\ind(A_{i\NI_i} = 1)}{ \widehat{\pi}^{(-k)} (1 \cond \bA_{i(-\NI_i)}, \bX_{i\NI_i}, \bX_{i(-\NI_i)}) }
	-
	\frac{\ind(A_{i\NI_i} = 0)}{ \widehat{\pi}^{(-k)} (0 \cond \bA_{i(-\NI_i)}, \bX_{i\NI_i}, \bX_{i(-\NI_i)}) }
	\end{bmatrix} \ .
\end{align*}
The motivation for equation \eqref{eq-genbeta} comes from the sandwich variance/Huber-White variance estimator \citep{Huber1967, White1980} where $\Bmat$ and $ \widehat{\mathbf{S}}_{i}^{(-k)} $ correspond to the outer and inner part, respectively, of the sandwich estimator. 
Also, from a theoretical perspective, the model space for $\beta$, denoted as $\mathcal{B}$ in \eqref{eq-genbeta}, only needs to be complete and $P$-Donsker with a finite envelope; see Theorem \ref{thm:3} for details. Some examples of complete and $P$-Donsker classes include a collection of bounded functions indexed by a finite-dimensional parameter in a compact set,  a collection of bounded monotone functions, a collection of variationally bounded functions,  $\alpha$-H\"older class with $\alpha> \text{dim}(\bC_i)/2$, and Sobolev class; see Chapter 2 of \citet{VW1996} 
for a textbook discussion about Donsker classes. 

Theoretically speaking, it may be tempting to consider more complex models of $\beta$ beyond $P$-Donsker, say by using flexible ML methods. Indeed, if this is desired, we can use additional steps in the cluster-level cross-fitting procedure described above to estimate $\beta$ and remove the Donsker class assumption. However, in the process of developing the new estimator, we found that this strategy is only useful if the sample size is sufficiently large to estimate the underlying outcome covariance relationships. Also, since $\beta$ do not need to be ``correctly'' estimated to achieve consistency or asymptotic Normality of the proposed estimator (in contrast to the conditional propensity score or the outcome model; see below), 
we find that using the $P$-Donsker class assumption is sensible for $\beta$.

In particular, in our numerical experiments, we find that the following, simple $P$-Donsker model space, denoted as $\paraB$, leads to good performance in our empirical examples:
\begin{align}										\label{eq-ParaB}
	\paraB
	=
	\bigg\{
		\beta(\bC_i) \, \bigg| \,
		&
		\beta(\bC_i)
		=
		\sum_{\ell=1}^\MT \ind\{ L(\bC_i) = \ell \} \gamma_\ell , \ 
		\begin{array}{l}
		\gamma_\ell \in [-B_0,B_0] \text{ for some } B_0>0 \ ,
		\\
		L: \bC_i \rightarrow \{1,\ldots,\MT\}
		\end{array}
		\bigg\} \ .
\end{align}
Functions in $\paraB$ are parametrized by a finite dimensional parameter $\bm{\gamma} = \big( \gamma_1,\ldots,\gamma_\MT \big)\T \in [-B_0,B_0]^{\otimes \MT}$ where $\beta(\bC_i) = \gamma_\ell$ if cluster $i$ belongs to stratum $\ell$ based on a user-specified label function $L$. 
The label function $L$ can be a function about cluster size, say $L(\bC_i) = \ind( \NI_i \leq M_1) + 2 \cdot \ind(M_1 < \NI_i \leq M_2) + 3 \cdot \ind (M_2 < \NI_i)$ with thresholds $M_1$ and $M_2$, or $L$ can be defined over a discrete covariate $C_i^{(d)}$, say $L( \bC_i ) = \sum_{j=1}^\MT j \cdot \ind( C_i^{(d)} = d_j )$. Under $\paraB$, estimation of $\beta(\bC_i)$ is equivalent to solving the following estimating equation for $\bm{\gamma}$ for $\ell=1,\ldots,J$.
\begin{align*}	
	&
	\hspace*{-0.05cm}
	\sum_{i \in \II_k} \hspace*{-0.05cm}  w(\bC_i)^2 \ind \big\{ L(\bC_i) = \ell \big\}
	\Big[ 
	2 \bI(\bA_i, \bX_i , \widehat{\pi}^{(-k)} ) \T
	( I - \bm{1}\bm{1}\T ) \widehat{\mathbf{S}}_i^{(-k)}
		 ( I - \bm{1}\bm{1}\T ) \bI(\bA_i, \bX_i , \widehat{\pi}^{(-k)} ) \widehat{\gamma}_\ell
		 \\
		 \nonumber
		 &
		 \hspace*{0.5cm}
		 -
		 \bI(\bA_i, \bX_i , \widehat{\pi}^{(-k)} ) \T
	\Big\{ 2 \widehat{\mathbf{S}}_i^{(-k)} - \bm{1}\bm{1}\T \widehat{\mathbf{S}}_i^{(-k)} - \widehat{\mathbf{S}}_i^{(-k)} \bm{1}\bm{1}\T \Big\}
		\bI(\bA_i, \bX_i , \widehat{\pi}^{(-k)} )
		 \Big]
		 = 0
		  \ .
\end{align*}
\noindent Once we have estimates of $\pi$, $g$, and $\beta$ from above, we can plug them inside $\widehat{\tau}$ to arrive at the final estimator of $\tau^*$, i.e.
\begin{align}								\label{eq-tauhat}
	&
	\widehat{\tau} 
	= 
	\widehat{\tau} (\widehat{\pi}, \widehat{g}, \widehat{\beta}) =
	\frac{1}{2} \sum_{k=1}^2 	\frac{2}{\NC} \sum_{i \in \II_k}
	\phi(\bO_i, \widehat{\pi}^{(-k)}, \widehat{g}^{(-k)}, \widehat{\beta}^{(k)})
	\ .
\end{align}

Section \ref{Supp-sec:practical} 
of the Supplementary Materials provides additional implementation details and tips, especially for investigators who are more familiar with using parametric or semiparametric methods like GLMM and GEE. Much of these tips, including median adjustment to stabilize cross-fitted estimators, training ML models, and removing outlying clusters, already exist in the literature and the references in the Supplementary Materials contain additional details.

\subsection{Statistical Properties}						\label{sec:properties}
To characterize the statistical properties of $\widehat{\tau}$, we make the following assumptions about the behavior of the estimated nuisance functions. 
\begin{itemize}
	\item[\hypertarget{(EN1)}{(EN1)}] (\textit{Estimated Conditional Propensity Score and Outcome Regression}) 
	For all $k=1,2$, and $(\bA_i,\bX_i)$, we have\\
	$\widehat{\pi}^{(-k)} \big( 1 \cond \bA_\eij, \bX_{ij}, \bX_{\eij}) \in [c_\pi, 1-c_\pi]$, and $\big| \widehat{g}^{(-k)}(A_{ij}, \bX_{ij}) - g^*(A_{ij}, \bX_{ij}) \big| \leq c_g$ for some positive constants $c_\pi$ and $c_g$. 
	\item[\hypertarget{(EN2)}{(EN2)}] (\textit{Convergence Rate of Estimated Conditional Propensity Score and Outcome Regression})
	We consider one of the following three conditions.  
	\begin{itemize}
	\item[\hypertarget{(EN2.PS)}{(EN2.PS)}] We have $\big\| \widehat{\pi}^{(-k)} \big( 1 \cond \bA_\eij, \bX_{ij}, \bX_\eij \big) - \pi^* \big( 1 \cond \bA_\eij,  \bX_{ij}, \bX_\eij \big) \big\|_{P,2} = O_P(r_{\pi,\NC})$ where $r_{\pi,\NC}$ is $o(1)$ and $
	\big\| \widehat{g}^{(-k)}(A_{ij}, \bX_{ij}) - g(A_{ij},\bX_{ij}) \big\|_{P,2} = O_P(r_{g,\NC})$ where $g$ is some function satisfying $\big| g^*(A_{ij},\bX_{ij}) - g(A_{ij},\bX_{ij}) \big| \leq c_g $ for all $(A_{ij},\bX_{ij})$ and $r_{g,\NC}$ is $o(1)$.
	\item[\hypertarget{(EN2.OR)}{(EN2.OR)}] We have $\big\| \widehat{g}^{(-k)}(A_{ij}, \bX_{ij}) - g^*(A_{ij},\bX_{ij}) \big\|_{P,2} = O_P(r_{g,\NC})$ where $r_{g,\NC}$ is $o(1)$ and
	$\big\| \widehat{\pi}^{(-k)} \big( 1 \cond \bA_{\eij}, \bX_{ij}, \bX_\eij \big) - \pi \big( 1 \cond \bA_{\eij}, \bX_{ij}, \bX_\eij \big) \big\|_{P,2} = O_P(r_{\pi,\NC})$ where $\pi$ is some function satisfying $\pi(1 \cond \bA_{\eij}, \bX_{ij}, \bX_\eij) \in [c_\pi,1-c_\pi]$ for all $(\bA_{\eij},\bX_{i})$ and $r_{\pi,\NC}$ is $o(1)$. 
	\item[\hypertarget{(EN2.Both)}{(EN2.Both)}] We have $\big\| \widehat{\pi}^{(-k)} \big( 1 \cond \bA_\eij, \bX_{ij}, \bX_\eij \big) - \pi^* \big( 1 \cond \bA_\eij, \bX_{ij}, \bX_\eij \big) \big\|_{P,2} = O_P(r_{\pi,\NC})$ 	and $
	\big\| \widehat{g}^{(-k)}(A_{ij}, \bX_{ij}) - g^*(A_{ij},\bX_{ij}) \big\|_{P,2} = O_P(r_{g,\NC})$
	where $r_{\pi,\NC}$, $r_{g,\NC}$, and $\NC^{1/2}r_{\pi,\NC} r_{g,\NC}$ are $o(1)$. 
	\end{itemize}
	\item[\hypertarget{(EN3)}{(EN3)}] (\textit{Estimated Outcome Covariance Model}) $\mathcal{B}$ is complete and $P$-Donsker with $\big\| \widehat{\beta}^{(k)}(\bC_i) - \beta^{\dagger,(-k)} (\bC_i) \big\|_{P,2} = O_P(r_{\beta,\NC})$ where $r_{\beta,\NC} = o(1)$ and
\begin{align}								\label{eq-betadagger}
	\beta^{\dagger,(-k)}
	&	
	\in
	\argmin_{\beta \in \mathcal{B} }
	\EXP
	\Bigg[
	\begin{array}{l}
	w(\bC_i)^2 \bI (\bA_i, \bX_i , \widehat{\pi}^{(-k)} ) \T  B\big(  \beta(\bC_i) \big) 
	\\
	\times
	\big\{ \bY_i - \widehat{\bg}^{(-k)} (\bA_i, \bX_i) \big\}^{\otimes 2} 
	B \big(  \beta(\bC_i) \big) \bI (\bA_i, \bX_i ,  \widehat{\pi}^{(-k)}  )
	\end{array}
	\, \Bigg| \,
	\II_k^c
	\Bigg] \ .
\end{align}		
	\item[\hypertarget{(EN4)}{(EN4)}] (\textit{Convergence Rate of Outcome Covariance Model}) We have $\NC^{1/2} r_{g,\NC} r_{\beta,\NC}$ is $o(1)$.
\end{itemize}
At a high level, Assumptions \hyperlink{(EN1)}{(EN1)} and \hyperlink{(EN2)}{(EN2)} are similar to assumptions about nuisance functions in modern, cross-fitted estimators of treatment effects in i.i.d. settings. In contrast, Assumption \hyperlink{(EN3)}{(EN3)} and \hyperlink{(EN4)}{(EN4)} impose conditions on the outcome covariance model 
where the estimated outcome covariance model $\widehat{\beta}^{(k)}$ based on empirical risk minimization (i.e., equation \eqref{eq-genbeta}) converges to the minimizer of the population objective function (i.e., equation \eqref{eq-betadagger}), denoted as $\beta^{\dagger,(-k)}$. If we only desire consistency of $\widehat{\tau}$ without asymptotic Normality, we only need Assumption \hyperlink{(EN3)}{(EN3)}. On the contrary, if we desire asymptotic Normality of $\widehat{\tau}$, Assumption \hyperlink{(EN4)}{(EN4)} places additional restrictions on $r_{\beta,\NC}$ where the product of the convergence rate of $\widehat{\beta}^{(k)}$ and the outcome regression $\widehat{g}^{(-k)}$ is of order $\NC^{-1/2}$. Assumptions \hyperlink{(EN3)}{(EN3)} and \hyperlink{(EN4)}{(EN4)} are satisfied if investigators choose a sufficiently ``nice'' $\mathcal{B}$. 
 For example, we can choose the outcome covariance model as a singleton set, say $\mathcal{B} = \{0\}$, and Assumptions \hyperlink{(EN3)}{(EN3)} and \hyperlink{(EN4)}{(EN4)} will automatically hold with $r_{\beta,\NC}=0$. Also, our simple function class $\paraB$ in equation \eqref{eq-ParaB} will satisfy these rate condition with $r_{\beta,\NC}=\NC^{-1/2}$; see Lemma \ref{Supp-lemma:consistency2} 
in the Supplementary Material for a formal proof.

We make two remarks about the minimizer of the population objective function $\beta^{\dagger,(-k)}$. First, the population objective function fixes the nuisance functions $\widehat{\pi}^{(-k)}, \widehat{g}^{(-k)}$,  because the outcome covariance model conditions on the estimated $(\widehat{\pi}^{(-k)}, \widehat{g}^{(-k)})$ from the auxiliary sample and uses the data from the main sample only to estimate $\widehat{\beta}^{(k)}$.
Second, $\beta^{\dagger,(-k)}$ does not have to be the ``true'' outcome covariance model; rather, $\beta^{\dagger,(-k)}$ is simply the best possible outcome covariance model when we solve the population objective function over some function class $\mathcal{B}$.


Theorem \ref{thm:3} formally characterizes the property of $\widehat{\tau}$.
\begin{theorem} \label{thm:3} 
Suppose Assumptions \hyperlink{(A1)}{(A1)}-\hyperlink{(A3)}{(A3)}, \hyperlink{(M1)}{(M1)}-\hyperlink{(M2)}{(M2)}, \hyperlink{(EN1)}{(EN1)} and \hyperlink{(EN3)}{(EN3)} hold. Then the estimator $\widehat{\tau}$ in equation \eqref{eq-tauhat} satisfies the following properties.
\begin{itemize}
	\item[(1)] (\textit{Double Robustness}) If either \hyperlink{(EN2.PS)}{(EN2.PS)} or \hyperlink{(EN2.OR)}{(EN2.OR)} holds, $\widehat{\tau}$ is consistent for the treatment effect $\tau^*$, i.e., $\widehat{\tau} = \tau^* + o_P(1)$.
	\item[(2)] (\textit{Asymptotic Normality}) If \hyperlink{(EN2.Both)}{(EN2.Both)} and  \hyperlink{(EN4)}{(EN4)} holds, $\widehat{\tau}$ has an asymptotically Normal distribution centered at the true treatment effect, i.e., $\sqrt{\NC} \big( \widehat{\tau} - \tau^* \big)$ weakly converges to \\
	$ N \big( 0 ,  \VAR \big\{ \phi(\bO_i, \pi^*, g^*, \beta^* ) \big\} \big)$ where
	\begin{align*}
	\beta^*
	\in
	\argmin_{\beta \in \mathcal{B} }
	\EXP
	\left[
	\begin{array}{l}
	w(\bC_i)^2 \bI (\bA_i, \bX_i , \pi^*) \T  B\big(  \beta(\bC_i) \big) 
	\\
	\hspace*{1.5cm}
	\times
	\big\{ \bY_i - \bg^*(\bA_i, \bX_i) \big\}^{\otimes 2} 
	B \big(  \beta(\bC_i) \big) \bI (\bA_i, \bX_i , \pi^*)
	\end{array}
	\right]
	 \ .
		\end{align*}
		
		Also, a consistent estimator of the variance of $\widehat{\tau}$ is
	\begin{align*}
		&
		\widehat{\sigma}^2
		=
		\frac{1}{\NC}	\sum_{k=1}^2  \sum_{i \in \II_k} 
		\Big\{
			\phi(\bO_i, \widehat{\pi}^{(-k)}, \widehat{g}^{(-k)}, \widehat{\beta}^{(k)})- \widehat{\tau}_{k}
		\Big\}^2 
		\ , \
		&& \widehat{\tau}_{k} = 
			\frac{2}{\NC} \sum_{i \in \II_k}
	\phi(\bO_i, \widehat{\pi}^{(-k)}, \widehat{g}^{(-k)}, \widehat{\beta}^{(k)}) \ .
	\end{align*}
	\item[(3)] (\textit{Efficiency Gain Under Known Treatment Assignment Mechanism}) Suppose \hyperlink{(EN2.Both)}{(EN2.Both)} and  \hyperlink{(EN4)}{(EN4)} hold. Additionally, if the same condition in part 2 of Lemma \ref{thm:2} concerning the treatment assignment mechanism holds, the asymptotic relative efficiency (ARE) between $\widehat{\tau}$ and $\overline{\tau}$ is ${\rm ARE}(\widehat{\tau},\overline{\tau}) = \VAR\big\{\varphi (\bO_i, e^*, g^*) \big\} / \VAR \big\{ \phi(\bO_i, \pi^*, g^*, \beta^*) \big\} \geq 1$.
\end{itemize}
\end{theorem}
Theorem \ref{thm:3} shows that the proposed estimator $\widehat{\tau}$ is consistent, doubly robust, and asymptotically Normal. Critically, we do not need our estimated outcome covariance model $\widehat{\beta}$ to converge to some underlying ``true'' model to guarantee double robustness, consistency, and asymptotic Normality of $\widehat{\tau}$; as mentioned earlier, this is similar to GEE estimators where any, arbitrary weighting matrix leads to consistent and asymptotically Normal estimates so long as the weights converges to some function. Also, similar to Lemma \ref{thm:2}, under some assumptions, the proposed estimator $\widehat{\tau}$ is theoretically guaranteed to be at least as efficient than the existing estimator $\overline{\tau}$.

\section{Simulation}									\label{sec:sim}
\subsection{Finite-Sample Performance of $\widehat{\tau}$ and $\overline{\tau}$}								\label{sec:varyingN}
We study the finite-sample performances of our proposed estimator $\widehat{\tau}$ and the existing estimator $\overline{\tau}$ through a simulation study.  For pre-treatment covariates $\bX_{ij}$, we use six variables $(W_{1ij}, W_{2ij}, W_{3ij}, C_{1i}, C_{2i}, \NI_i)$ where $W_{1ij}$, $W_{2ij}$, and $C_{1i}$ are from the standard Normal distribution and  $W_{3ij}$ and $C_{2i}$ are from the Bernoulli distribution with mean parameter 0.3 and 0.7, respectively. The number of clusters varies from $\NC \in \{25,50,100,250,500\}$ and the cluster size takes on values between $2000/\NC$ and $3000/\NC$ so that the expected total number of individuals in the simulated dataset is 2500. 

The treatment follows a mixed effects model with non-linear terms, i.e., 
\begin{align*}			
	&
	\hspace*{-0.25cm}
	P \big( A_{ij} = 1 \cond \bX_{ij} , V_i)
	=
	\text{expit} \Bigg\{
	\begin{array}{l}
	-0.5 + 0.5 W_{1ij} - \ind \big( W_{2ij} > 1 \big) \quad
	\\
	\quad
	+ 0.5 W_{3ij} - 0.25 C_{1i} + C_{2i} + V_i
	\end{array}
	\Bigg\},
\end{align*}
where $V_i \sim N(0,\sigma_V^2)$. When $\sigma_V=0$, the random effect $V_i$ in the propensity score model vanishes and each element of $\bA_{i}$ are conditionally independent from each other given $\bX_i$.  But, when $\sigma_V \neq 0$, treatments in cluster $i$ are correlated with each other. 

For the outcome, we use a mixed effects model with the following non-linear terms
\begin{align}										\label{eq-Sim-OR}
	&
	Y_{ij} \cond (A_{ij}, \bX_{ij}, U_i)
	 \sim
	N \big(
	3 + (2.1 + W_{2ij}^2 + 3W_{3ij} ) A_{ij}
	+
	2 W_{1ij} - C_{1i}^2 + W_{2ij} C_{2i} + U_i
	\ , \ 1
	\big),
\end{align}
where $U_i$ follows $N(0,\sigma_U^2)$. Similar to the treatment model, individuals' outcomes in cluster $i$ are correlated via $\sigma_U$. 
The target estimand is the average treatment effect with $w(\bC_i)=1$, which equals $4$ under the models above, i.e., $\tau^* = 4$. 

We consider four different values of $(\sigma_V,\sigma_U)$: (i) no treatment correlation and weak outcome correlation where $(\sigma_V,\sigma_U)=(0,0.5)$; 
(ii) no treatment correlation and strong outcome correlation where $(\sigma_V,\sigma_U)=(0,1.5)$; 
(iii) strong treatment correlation and weak outcome correlation where $(\sigma_V,\sigma_U)=(1.5,0.5)$; 
and (iv) strong treatment correlation and strong outcome correlation where $(\sigma_V,\sigma_U)=(1.5,1.5)$. To estimate the nuisance functions in both $\widehat{\tau}$ and $\overline{\tau}$,  we use ensembles of multiple ML methods via the super learner algorithm \citep{SL2007, Polley2010}; see Section \ref{Supp-sec:Sim} 
of the Supplementary Materials for the exact ML methods used in the super learner and other implementation details. For estimating the outcome covariance model $\beta$, we choose the function space $\paraB$ in equation \eqref{eq-ParaB} where we use 1, 2, 3, 5, and 3 strata parameters (i.e., $\text{dim}(\bm{\gamma}) \in \{ 1,2,3,5,3 \}$) for cluster sizes $\NC = 25,50,100,250$, and $500$, respectively. 
We repeat the simulation 200 times and report the empirical bias, empirical standard error, and the coverage rate of 95\% confidence intervals. 

Also, to make the terms $\sigma_V$ and $\sigma_U$ governing the correlation structures more interpretable, we compute the intra-cluster correlation coefficient (ICC), a commonly used measure in multilevel studies to assess correlation between study units in the same cluster. For the outcome ICC, denoted as ${\rm ICC}_{Y}(\sigma_U)$, it  varies from 0 to 0.8. 
 For the treatment ICC, denoted as ${\rm ICC}_A(\sigma_V)$, it varies varies from 0.05 to 0.36. 
 For the exact details on computing ICCs, especially for binary variables, see Section \ref{Supp-sec:Sim} 
of the Supplementary Materials.

Table \ref{tab:0} summarizes the result. Broadly speaking, the proposed estimator $\widehat{\tau}$ achieve the smallest standard errors compared to the existing estimator $\overline{\tau}$ across all simulation scenarios, with $\overline{\tau}$ becoming less efficient under strong outcome correlation. Also, the coverage rates of confidence intervals based on $\widehat{\tau}$ are closer to the nominal coverage than those based on $\overline{\tau}$. Finally, even with small number of clusters where the bounded cluster condition in Assumption \hyperlink{(M1)}{(M1)} may be suspect, say $\NC = 25$ clusters with roughly 100 individuals per cluster, we see that $\widehat{\tau}$ has small bias and nominal coverage. Based on the result, we believe $\widehat{\tau}$ can be used in settings where each cluster has more study units compared to the total number of clusters, say if households are the study units and U.S. states are clusters.

\begin{table}[!htp]
	\hspace*{-0.5cm}
		\renewcommand{\arraystretch}{1.1} \centering
		\footnotesize
		\setlength{\tabcolsep}{2.5pt}
\begin{tabular}{|c|c|c|c|c|c|c|c|c|c|c|c|c|c|c|c|c|c|}
\hline
\multirow{2}{*}{Est.}           & \multirow{2}{*}{$(\sigma_V,\sigma_U)$} & \multirow{2}{*}{\begin{tabular}[c]{@{}c@{}}($\text{ICC}_A$,\\ \ $\text{ICC}_Y$)\end{tabular}} & \multicolumn{3}{c|}{\begin{tabular}[c]{@{}c@{}}$\NC=25$\\ $\EXP(\NI_i)=100$\end{tabular}}                 & \multicolumn{3}{c|}{\begin{tabular}[c]{@{}c@{}}$\NC=50$\\ $\EXP(\NI_i)=50$\end{tabular}}                & \multicolumn{3}{c|}{\begin{tabular}[c]{@{}c@{}}$\NC=100$\\ $\EXP(\NI_i)=25$\end{tabular}}                & \multicolumn{3}{c|}{\begin{tabular}[c]{@{}c@{}}$\NC=250$\\ $\EXP(\NI_i)=10$\end{tabular}}                & \multicolumn{3}{c|}{\begin{tabular}[c]{@{}c@{}}$\NC=500$\\ $\EXP(\NI_i)=5$\end{tabular}}               \\ \cline{4-18} 
                                     &                                        &                                              & Bias & SE & {\tiny Cover.} & Bias & SE &  {\tiny Cover.} & Bias & SE & {\tiny Cover.} & Bias & SE & {\tiny Cover.} & Bias & SE & {\tiny Cover.} \\ \hline
 \multirow{4}{*}{$\overline{\tau}$} & (0.0,0.5) & (0.05,0.20)  & 0.40   & 8.50 & 0.995  & 0.39  & 7.48 & 0.965  & 0.18  & 7.52 & 0.940  & 0.98  & 6.53 & 0.960  & 0.19  & 6.54 & 0.955 \\ \cline{2-18}
                                         & (0.0,1.5) & (0.05,0.69)  & 1.72  & 27.10 & 0.990 & -0.48 & 12.07 & 0.985  & 1.01 & 12.71 & 0.950 & -0.17 & 10.69 & 0.965  & 1.00 & 11.00 & 0.950 \\ \cline{2-18}
                                         & (1.5,0.5) & (0.27,0.20)  & 7.37  & 88.19 & 0.990  & 3.12 & 31.34 & 0.980  & 1.35 & 16.92 & 0.975  & 0.69  & 8.59 & 0.975  & 0.48  & 7.11 & 0.960 \\ \cline{2-18}
                                         & (1.5,1.5) & (0.27,0.69)  & 7.64 & 222.3 & 0.990  & 7.31 & 85.20 & 0.965  & 0.14 & 45.70 & 0.945  & 1.00 & 19.79 & 0.970  & 0.75 & 13.42 & 0.950       \\ \hline
  \multirow{4}{*}{$\widehat{\tau}$} & (0.0,0.5) & (0.05,0.20) & -0.43   & 6.22 & 0.920 & -0.12  & 6.13 & 0.965  & 0.18  & 6.65 & 0.920  & 0.55  & 6.25 & 0.935  & 0.20  & 6.20 & 0.950 \\ \cline{2-18}
                                         & (0.0,1.5) & (0.05,0.69) & -0.08   & 6.58 & 0.945  & 0.01  & 6.21 & 0.955  & 0.07  & 6.19 & 0.960  & 0.27  & 6.59 & 0.970  & 1.05  & 7.58 & 0.950 \\ \cline{2-18}
                                         & (1.5,0.5) & (0.27,0.20) & -0.38   & 6.63 & 0.930 & -0.15  & 7.38 & 0.905 & -0.29  & 7.15 & 0.925 & -0.17  & 7.11 & 0.920 & -0.06  & 6.14 & 0.980 \\ \cline{2-18}
                                         & (1.5,1.5) & (0.27,0.69) & -0.51   & 7.09 & 0.940 & -0.26  & 7.10 & 0.935 & -0.08  & 7.09 & 0.955 & -0.09  & 7.23 & 0.950  & 0.84  & 7.79 & 0.965       \\ \hline
\end{tabular}
\caption{Finite-sample performance of $\widehat{\tau}$ and $\overline{\tau}$ under different number of clusters. Each row represents the values of $\sigma_V$ and $\sigma_U$. Each column shows the empirical biases, empirical standard errors, and coverages of 95\% confidence intervals. $\NC$ is number of clusters and the average number of individual per cluster (i.e., cluster size) is 2500/$\NC$. Bias and SE columns are scaled by 100.}
\label{tab:0}
\end{table}

%

\subsection{Comparison to Existing Nonparametric Methods}					\label{sec:comparison}
Next, we compare the performance of $\widehat{\tau}$ to existing nonparametric estimators of treatment effects in the literature. The rationale behind this comparison is inspired by the discussion in \citet{Carvalho2019} where a practitioner may naively apply existing i.i.d.-based nonparametric methods in multilevel settings due to its simplicity. Comparing these estimators to the proposed estimator $\widehat{\tau}$, which is specifically tailored for multilevel settings, may shed some light on the cost of naively applying existing methods in practice.

For comparison, we use causal forests \citep{WA2018,grf} implemented in the \texttt{grf} R-package \citep{grfpackage}, and R/U-learners \citep{NieWager2020} with gradient boosting and lasso via the \texttt{rlearner} R package. Again, we remark that these competing ML methods were initially developed for i.i.d. data. But, \texttt{grf} provides some ways to accommodate clustering by using the options \texttt{weight.vector} and \texttt{clusters} options in the function \texttt{average\_treatment\_effect} \citep{grfvig}. We use these recommended tuning procedures for both causal forests and R/U-learners. The simulation model is the same as Section \ref{sec:varyingN} except we fix $\NC = 500$.

Table \ref{tab:1} summarizes the result.  
Among competing ML methods initially developed for i.i.d. data, GRF with the recommended modifications for clustering seems to get close to the targeted nominal coverage, ranging from 91\% to 95.5\%. This suggests that the suggestions in \citet{grfvig} to tweak the original GRF to accommodate clustering has some merit and is worth theoretically investigating in the future. In contrast, R- and U-Learners' coverage rates are often much lower, ranging from 70\% to 92\%. Finally, our estimator $\widehat{\tau}$ maintain nominal coverage and achieves the smallest standard error among all estimators. 

\begin{table}[!htp]
		\renewcommand{\arraystretch}{1.1} \centering
		\footnotesize
		\setlength{\tabcolsep}{2pt}
\begin{tabular}{|c|c|c|c|c|c|c|c|c|c|c|c|c|}
\hline
$(\sigma_V,\sigma_U)$       & \multicolumn{3}{c|}{$(0.0,0.5)$}   & \multicolumn{3}{c|}{$(0.0,1.5)$}   & \multicolumn{3}{c|}{$(1.5,0.5)$}   & \multicolumn{3}{c|}{$(1.5,1.5)$}   \\ \hline
$({\rm ICC}_A,{\rm ICC}_Y)$ & \multicolumn{3}{c|}{$(0.05,0.20)$} & \multicolumn{3}{c|}{$(0.05,0.69)$} & \multicolumn{3}{c|}{$(0.27,0.20)$} & \multicolumn{3}{c|}{$(0.27,0.69)$} \\ \hline
Statistic & Bias     & SE    & Cover.    & Bias     & SE    & Cover.    & Bias     & SE     & Cover.     & Bias     & SE     & Cover.     \\ \hline 
           R-Learner-B   & -1.61  & 7.25  & 0.875   & -1.86   & 9.90  & 0.920  & -5.30   & 7.12  & 0.805   & -9.88  & 12.43  & 0.725 \\ \hline
           R-Learner-L  & -11.45  & 9.62  & 0.750  & -11.43  & 12.78  & 0.780  & -8.88  & 10.67  & 0.885   & -9.30  & 15.09  & 0.885 \\ \hline
           U-Learner-B   & -3.28  & 7.37  & 0.860   & -3.66   & 9.96  & 0.900  & -6.38   & 7.31  & 0.755  & -11.13  & 12.59  & 0.700 \\ \hline
           U-Learner-L  & -11.40  & 9.44  & 0.810  & -11.41  & 12.79  & 0.800  & -8.85  & 10.74  & 0.900   & -9.30  & 15.11  & 0.885 \\ \hline
                   GRF   & -0.10  & 7.58  & 0.930    & 0.81  & 11.18  & 0.915   & 0.07   & 7.61  & 0.955    & 0.18  & 12.97  & 0.910 \\ \hline
 $\overline{\tau}$    & 0.19  & 6.54  & 0.955    & 1.00  & 11.00  & 0.950   & 0.48   & 7.11  & 0.960    & 0.75  & 13.42  & 0.950 \\ \hline
  $\widehat{\tau}$    & 0.20  & \textbf{6.20}  & 0.950    & 1.05   & \textbf{7.58}  & 0.950  & -0.06   & \textbf{6.14}  & 0.980    & 0.84   & \textbf{7.79}  & 0.965 \\ \hline
\end{tabular}	
\caption{Comparison of different estimators under varying correlation strength. Each row represents different estimators for the average treatment effect. Each column shows the empirical biases, empirical standard errors, and coverages of 95\% confidence intervals under different values of $\sigma_V$ and $\sigma_U$.  Bias and SE columns are scaled by 100. The values in boldface are the smallest standard errors under each data generating process.}
\label{tab:1}
\end{table}

We extend the above comparisons by considering non-normal outcome regression random effect $U_i$. Specifically, we use the same outcome regression as equation \eqref{eq-Sim-OR} but the distribution of $U_i$ is generated from the mixture of the following four distributions: $N(0,0.5^2)$, $0.5 \cdot t(5)$, ${\rm Laplace}(0,0.5)$, and ${\rm Unif}(-0.25,0.25)$. The ANOVA-type estimate of the ICC of the outcome is 0.23. We keep the same propensity score model under $\sigma_V \in \{0,1.5\}$ and 
Table \ref{tab:1-1} summarizes the result. Similar to Table \ref{tab:1}, under all settings, $\widehat{\tau}$ has the smallest standard error. 
Moreover, $\widehat{\tau}$ achieves closer to nominal coverage compared to existing nonparametric estimators. Overall, the additional simulation supports the robustness and the efficiency gains of $\widehat{\tau}$ under different correlation structures among individuals.

\begin{table}[!htp]
		\renewcommand{\arraystretch}{1.1} \centering
		\footnotesize
		\setlength{\tabcolsep}{3pt}
\begin{tabular}{|c|c|c|c|c|c|c|}
\hline
$\sigma_V$                  & \multicolumn{3}{c|}{$0.0$}         & \multicolumn{3}{c|}{$1.5$}   \\ \hline
$({\rm ICC}_A,{\rm ICC}_Y)$ & \multicolumn{3}{c|}{$(0.05,0.23)$} & \multicolumn{3}{c|}{$(0.27,0.23)$} \\ \hline
Statistic                   & Bias      & SE      & Cover.     & Bias      & SE      & Cover.     \\ \hline
           R-Learner-B   & -2.01   & 6.86  & 0.910  & -5.56   & 8.40  & 0.745 \\ \hline
           R-Learner-L  & -11.50  & 10.00  & 0.755  & -8.97  & 11.75  & 0.850 \\ \hline
           U-Learner-B   & -3.82   & 7.02  & 0.855  & -6.54   & 8.47  & 0.730 \\ \hline
           U-Learner-L  & -11.41  & 10.01  & 0.795  & -8.91  & 11.84  & 0.845 \\ \hline
                   GRF   & -0.41   & 7.80  & 0.910  & -0.09   & 8.69  & 0.905 \\ \hline
 $\overline{\tau}$    & 0.04   & 6.37  & 0.965   & 0.47   & 7.61  & 0.965 \\ \hline
  $\widehat{\tau}$   & -0.14   & \textbf{5.90}  & 0.965  & -0.17   & \textbf{7.48}  & 0.935 \\ \hline
\end{tabular}
\caption{Comparison of different estimators under non-normal $U_i$. Each row represents different estimators for the average treatment effect. Each column shows the empirical biases, empirical standard errors, and coverages of 95\% confidence intervals under different values of $\sigma_V$ and $\sigma_U$.  Bias and SE columns are scaled by 100. The values in boldface are the smallest standard errors under each data generating process.}
\label{tab:1-1}
\end{table}



\section{Applications: Early Childhood Longitudinal Study}									\label{sec:application}				


We apply our method to estimate average treatment effects of center-based pre-school programs on children's reading score in kindergarten from the Early Childhood Longitudinal Study's Kindergarten Class of 1998-1999 (ECLS-K) dataset \citep{ECLSK2009}; note that a similar question with respect to children's math score was asked in \citet{Lee2021}. Briefly, the dataset consists of children followed from kindergarten through eighth grade in the United States. Children define the individual study units, kindergarten classrooms define clusters, and the subscript ${ij}$ indicates the $j$th child in the $i$th kindergarten. We let $A_{ij}=1$ if the child received center-based care before kindergarten and $A_{ij}=0$ otherwise (i.e., parental care); note that the treatment was assigned at the individual-level. The outcome $Y_{ij} \in [22,94]$ is the standardized reading score of each child, measured during the Fall semester and after treatment assignment. For cluster-level pre-treatment covariates, we include cluster size, region (northeast/midwest/south/west), kindergarten location (central city/urban/rural), and kindergarten type (public/private). For individual-level pre-treatment covariates, we include the child's gender, age, race, family type, parental education, and economic status.
We restrict our analysis 
 to children with complete data on the outcome, treatment, and pre-treatment covariates. 
This results in 15,980 children in 942 kindergartens, which corresponds to 16.96 children per kindergarten. Cluster size varies between 1 and 25 children; the first, second, and third quartiles of cluster size are 14, 19, and 21, respectively; 
see Section \ref{Supp-sec:Supp-ECLSK} 
of the Supplementary Materials for additional details.



To assess the outcome and the treatment correlations, we compute the ICCs as follows. For the outcome ICC, we fit a linear mixed effects regression model where   $Y_{ij}$ is the independent variable, $(A_{ij}, \bX_{ij})$ are the dependent variables, and kindergarten classrooms serve as random effects. For the treatment ICC, we fit a logistic mixed effect regression model where $A_{ij}$ is the independent variable, $ \bX_{ij}$ are the dependent variables, and kindergarten classrooms serve as random effects. Using these models, we find that the ICCs of the outcome and the treatment variables are 0.103 and 0.063, respectively, suggesting a moderate amount of correlation/dependencies between children in the same kindergarten classroom.

We focus on $\tau_{\text{ATE}}^*$, the overall average treatment effect with $w(\bC_i) = 1$ (i.e., the average treatment effect across all clusters), and $\{ \tau_1^*,\ldots,\tau_{12}^*\}$, 12 weighted average treatment effects defined by region and kindergarten location with $w_1(\bC_i) = \ind \{$kindergarten $i$ is in a central city in northeast region$\}$, $\ldots$, $w_{12}(\bC_i) = \ind \{$kindergarten $i$ is a rural area in west region$\}$. The choice of the 12 weighted average treatment effects are based on the previous works \citep{ECLSK1, ECLSK2} indicating that educational attainment varies by region and location. We compute 
the proposed estimator $\widehat{\tau}$ using the same procedure in Section \ref{sec:sim} except we repeat cross-fitting 100 times 
and use 24 strata parameters (i.e., $\text{dim}(\bm{\gamma})=24$) to estimate the outcome covariance model $\beta$ in $\paraB$. We remark that by Theorem \ref{thm:3}, as long as either the outcome model or the conditional propensity score is correctly specified, $\widehat{\tau}$ is consistent, irrespectively of the estimated outcome covariance model. Also, if the convergence rates are sufficiently fast, $\widehat{\tau}$ is asymptotically Normal. 

\begin{figure}[!htb]
	\centering
	\includegraphics[width=1\textwidth]{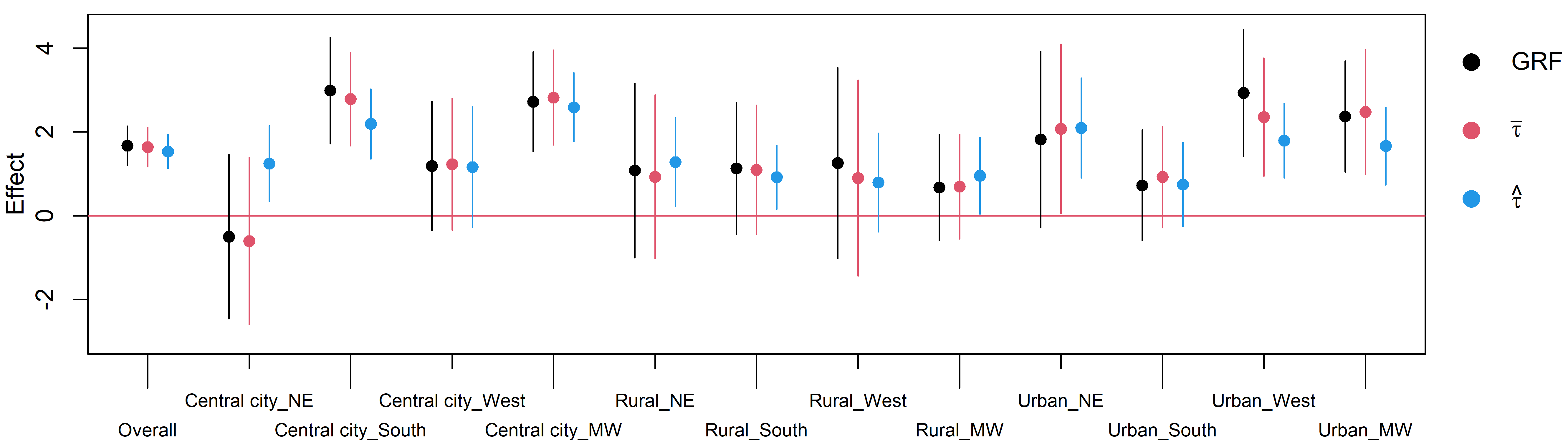}
	\caption{Summary of the data analysis from the ECLS-K dataset. 
	Each column shows the overall average treatment effect (labeled as overall) and the subgroup average treatment effect (labeled as Central city\_NE, $\ldots$, Urban\_MW) where NE and MW are shorthand for northeast and midwest, respectively. The black ($\bullet$), red ({\color{red}$\bullet$}), and blue ({\color{myblue}$\bullet$}) dots show the point estimates  of the causal forest-based estimator (denoted as GRF), the existing doubly robust estimator (denoted as $\overline{\tau}$), and our proposed estimator (denoted as $\widehat{\tau}$), respectively.  The vertical lines show 95\% confidence intervals.
	}
	\label{fig:1}
\end{figure}

Figure \ref{fig:1} summarizes the result where the 12 weighted average treatment effects are divided by the empirical proportion of each subgroup, i.e., $\{ \sum_{i=1}^\NC w_t(\bC_i) / \NC \}^{-1} \widehat{\tau}_t $ for $t=1,\ldots,12$. As a consequence, the latter 12 estimates in the figure present the estimated subgroup average treatment effects, i.e. $\EXP \{ \pot{ Y_{ij} }{1} - \pot{ Y_{ij} }{0} \cond w_t(\bC_i) = 1 \}$; note that since the empirical proportions $\{ \sum_{i=1}^\NC w_t(\bC_i) / \NC \}$ converge to the true proportions of $P(w_t(\bC_i) = 1)$, by Slutsky's theorem, the estimator for the subgroup average treatment effect enjoys the same properties as the estimator for $\tau^*$. For comparison, we use the causal forest-based estimator and the existing estimator $\overline{\tau}$ in Section \ref{sec:comparison}. First, all three estimators show the similar overall effects around 1.6 and indicate that the effect is significant at level 0.05. Second, the subgroup average treatment effects seem to be heterogeneous across census blocks and locations, and the significance of the effect estimates is different depending on the estimators. In particular, for the subgroup average treatment effect among city schools in the northeast census region, the causal forest-based estimator and $\overline{\tau}$ show negative, but insignificant at level 0.05, effect estimates whereas the proposed estimator $\widehat{\tau}$ show a positive, significant at level 0.05 effect estimate. Similarly, the effect estimates in rural northeast, south, and midwest regions based on the causal forest-based estimator and $\overline{\tau}$ are insignificant at level 0.05 in whereas those based on the proposed estimator $\widehat{\tau}$ are significant at level 0.05. Lastly, the proposed estimator $\widehat{\tau}$ always yield smaller variances compared to the competing estimators. Specifically, our 
 proposed estimator $\widehat{\tau}$ yields 6.6 to 54.9 percent shorter confidence intervals and has 12.8 and 79.7 percent smaller variance compared to the other estimators.

\section{Conclusion} 							\label{sec:Conclude}

This paper presents a new estimator $\widehat{\tau}$ to infer treatment effects in multilevel studies. The new estimator is doubly robust, asymptotically Normal, and does not require parametric specifications of the outcome model, the treatment model, and the correlation structure. The new estimator $\widehat{\tau}$ leverages recent ML-based methods in causal inference in i.i.d. settings to estimate, among other things, two non-standard functionals in causal inference. These two functionals are designed to capture the correlation between the treatment and the outcome among study units in the same cluster. Also, the new estimator $\widehat{\tau}$ can be more efficient than existing estimators in multilevel studies; we present both theoretical conditions that guarantee efficiency improvements and numerical studies showing $\widehat{\tau}$ performing better than existing, nonparametric estimators.


We end by re-iterating some important limitations of the new estimator $\widehat{\tau}$ and, in light of these limitations, offer some guidelines on how to use $\widehat{\tau}$ in practice. First, while $\widehat{\tau}$ performs well in data and all the simulation studies we considered, we do not show whether $\widehat{\tau}$ is optimal in terms of semiparametric efficiency. As hinted in Sections \ref{sec:M1} and \ref{sec:nfunc}, 
in nonparametric, multilevel settings, the semiparametric efficiency bound of $\tau^*$ depends on both the smoothness of the conditional mean of the outcome as well as the covariance of the outcomes in a cluster as each part can have non-diminishing first-order terms in a locally linear, asymptotic expansion of the estimator around $\tau^*$. In light of this, we conjecture that $\widehat{\tau}$ can achieve the semiparametric efficiency bound under more stronger assumptions about the exchangeability of the elements of the outcome covariance matrix. Second, while our theory currently relies on the bounded cluster size assumption, as demonstrated in Section \ref{sec:varyingN}, our methods seem to perform well even if the cluster size is larger than the number of clusters. Of course, we do not expect this phenomena to hold in settings where the cluster is very large and there are a couple number of clusters. 
Despite these limitations, we believe the proposed estimator provides a new approach to robustly and efficiently estimate treatment effects in multilevel settings where the study units exhibit dependence within clusters and can be a valuable tool for practitioners working with multilevel observational studies.



\newpage

\appendix

\section{Details of the Main Paper}									\label{sec:Detail}

\subsection{Details of Section \ref{Main-sec:Setup}: Properties of Existing Estimators in Multilevel Studies}						\label{sec:ExistingMethod}

We consider a simple modification of the recent cross-fitting estimator by \citet{Victor2018}. At a high level, we randomly split the observed data into non-overlapping folds, say two folds $\II_1$ and $\II_2$, at the cluster level instead of at the individual-level; we remark that the original proposal by \citet{Victor2018} randomly splits the data at the individual-level and assumes that the data was i.i.d. One fold, referred to as the auxiliary data and denoted as $\II_k^c$, is used to estimate the nuisance functions and the other fold, referred to as the main data and denoted as $\II_k$, evaluates the estimated nuisance functions obtained from the first fold. We then switch the roll of the two split samples to fully use the data for treatment effect estimation.

 Let $\widehat{e}^{(-k)}$ and $\widehat{g}^{(-k)}$ be the evaluation of estimated propensity score (i.e. $\EXP (A_{ij} \cond \bX_{ij})$) and the outcome regression (i.e. $\EXP (Y_{ij} \cond A_{ij}, \bX_{ij})$) based on the auxiliary data $\II_k^c$ in the main data $\II_k$. Then, we use  $\widehat{e}^{(-k)}$ and $\widehat{g}^{(-k)}$  as plug-ins to the augmented inverse probability weighted (AIPW) estimator \citep{Robins1994, Scharfstein1999}, denoted as $\overline{\tau}$ below.
\begin{align*}			
	& \overline{\tau} = \overline{\tau} (\widehat{e}^{(-k)}, \widehat{g}^{(-k)}) 
	= \frac{1}{2} \sum_{k=1}^2 \frac{1}{\NC/2}  \sum_{i \in \II_k} 	{\varphi}(\bO_i, \widehat{e}^{(-k)}, \widehat{g}^{(-k)}) \ , \
		\\ \nonumber 
		&
		\varphi(\bO_i, e , g) =  \frac{w(\bC_i)}{n_i} \sum_{j =1}^{\NI_i} \bigg[  \frac{A_{ij} \big\{ Y_{ij} - g^*(1, \bX_{ij}) \big\} }{e(1 \cond \bX_{ij})} + g^*(1, \bX_{ij}) 
		-   \frac{(1 - A_{ij}) \big\{ Y_{ij} - g^*(0, \bX_{ij}) \big\}  }{e(0 \cond \bX_{ij})} - g^*(0, \bX_{ij})   \bigg]. \nonumber 
\end{align*}
Similar to the original cross-fitting estimator in \citet{Victor2018}, we make a parallel set of assumptions about the properties of the estimated nuisance functions.
\begin{itemize}
	\item[\hypertarget{(E1)}{(E1)}] (\textit{Estimated Nuisance Functions}) For all $k=1,2$, and $(A_{ij},\bX_{ij})$, we have	$ \widehat{e}^{(-k)} \big( 1 \cond \bX_{ij}) \in [c_e, 1-c_e]$, and $\big| \widehat{g}^{(-k)}(A_{ij}, \bX_{ij}) - g^*(A_{ij}, \bX_{ij}) \big| \leq c_g$ for some positive constants $c_e$ and $c_g$. 
	\item[\hypertarget{(E2)}{(E2)}] (\textit{Convergence Rate of Estimated Nuisance Functions})
	For all $k=1,2$, consider one of the following three conditions. 	
	\begin{itemize}
	\item[\hypertarget{(E2.PS)}{(E2.PS)}] We have $\big\| \widehat{e}^{(-k)} \big( 1 \cond \bX_{ij} \big) - e^* \big( 1 \cond \bX_{ij} \big) \big\|_{P,2} = O_P(r_{e,\NC})$ where $r_{e,\NC}$ is $o(1)$ and $
	\big\| \widehat{g}^{(-k)}(A_{ij}, \bX_{ij}) - g(A_{ij},\bX_{ij}) \big\|_{P,2} = O_P(r_{g,\NC})$ where $g$ is some function satisfying $\big| g^*(A_{ij},\bX_{ij}) - g(A_{ij},\bX_{ij}) \big| \leq c_g $ for all $(A_{ij},\bX_{ij})$ and $r_{g,\NC}$ is $o(1)$.	
	\item[\hypertarget{(E2.OR)}{(E2.OR)}] We have $\big\| \widehat{g}^{(-k)}(A_{ij}, \bX_{ij}) - g^*(A_{ij},\bX_{ij}) \big\|_{P,2} = O_P(r_{g,\NC})$ where $r_{g,\NC}$ is $o(1)$ and $\big\| \widehat{e}^{(-k)} \big( 1 \cond \bX_{ij} \big) - e \big( 1 \cond \bX_{ij} \big) \big\|_{P,2} = O_P(r_{e,\NC})$ where $e$ is some function satisfying $e(1 \cond \bX_{ij}) \in [c_e,1-c_e]$ for all $\bX_{ij}$ and $r_{e,\NC}$ is $o(1)$.
	\item[\hypertarget{(E2.Both)}{(E2.Both)}] We have $
		\big\| \widehat{e}^{(-k)} \big( 1 \cond \bX_{ij} \big) - e^* \big( 1 \cond \bX_{ij} \big) \big\|_{P,2} = O_P(r_{e,\NC})$
		and $
	\big\| \widehat{g}^{(-k)}(A_{ij}, \bX_{ij}) - g^*(A_{ij},\bX_{ij}) \big\|_{P,2} = O_P(r_{g,\NC})$
	where $r_{e,\NC}$ 
	and $r_{g,\NC}$ are $o(1)$ and $r_{e,\NC}r_{g,\NC}$ is $o(\NC^{-1/2})$. 
	\end{itemize}
\end{itemize}
In words, both Assumption \hyperlink{(E1)}{(E1)} and variants of \hyperlink{(E2)}{(E2)} state that the estimated nuisance functions $\widehat{e}^{(-k)}$ and $\widehat{g}^{(-k)}$ are well-behaved estimators. In particular, \hyperlink{(E2.PS)}{(E2.PS)} states that $\widehat{e}^{(-k)}$ is a consistent estimator of the true propensity score $e^*$, \hyperlink{(E2.OR)}{(E2.OR)} states that  $\widehat{g}^{(-k)}$ is a consistent estimator of the true outcome regression model $g^*$, and \hyperlink{(E2.Both)}{(E2.Both)} state that both $\widehat{e}^{(-k)}$ and $\widehat{g}^{(-k)}$ are consistent estimators of their true counterparts $e^*$ and $g^*$, respectively. Also, \hyperlink{(E2.PS)}{(E2.PS)} and \hyperlink{(E2.OR)}{(E2.OR)} do not require $\sqrt{\NC}$ rates of convergence whereas \hyperlink{(E2.Both)}{(E2.Both)} place restrictions on the convergence rates; these restrictions are satisfied if the estimated propensity score $\widehat{e}^{(-k)}$ and the estimated outcome regression $\widehat{g}^{(-k)}$ are estimated at rates faster than $\NC^{-1/4}$. 

Theorem \ref{thm:1} shows that the proposed adjustment to the original cross-fitting estimator leads to a consistent, doubly robust, and asymptotically Normal estimate of the average treatment effect in multilevel studies. 
\begin{theorem}						\label{thm:1}
Suppose Assumptions \hyperlink{Main-(A1)}{(A1)}-\hyperlink{Main-(A3)}{(A3)} and \hyperlink{Main-(M2)}{(M2)} in the main paper, \hyperlink{(M1)'}{(M1)'} and \hyperlink{(E1)}{(E1)} hold where \hyperlink{(M1)'}{(M1)'} is given below. \\

\noindent \hypertarget{(M1)'}{(M1)'} (\textit{Nonparametric Model}) The observed data satisfies the following, nonparametric restrictions for some constants $\NT$ and $c_e>0$.
{\small \begin{align*}
 &
P(\bO_1,\ldots,\bO_\NC)  =
\prod_{i=1}^{\NC}   P (\bO_i) \ , \
P (\bO_i) \in \left\{ P  \left|  
\begin{array}{l}
  \NI_i \leq \NT, \
 \EXP \big( Y_{ij} \cond A_{ij} =a, \bX_{ij} = \bx \big) =  
g(a,\bx) , \\
\EXP  \big( A_{ij} =a \cond \bX_{ij} = \bx \big) = e( a \cond \bx)
\end{array} \right.
\right\} \ . \
\end{align*}}

Then, the cross-fitting estimator $\overline{\tau}$ in \eqref{eq-bartau} satisfies the following properties. 
\begin{enumerate}

\item[(1)](\textit{Double Robustness}) If either \hyperlink{(E2.PS)}{(E2.PS)} or \hyperlink{(E2.OR)}{(E2.OR)} holds,  the estimator $\overline{\tau}$ is consistent for the true treatment effect, i.e. $\overline{\tau} = \tau^* + o_P(1)$.
\item[(2)](\textit{Asymptotic Normality}) If \hyperlink{(E2.Both)}{(E2.Both)} holds, the estimator $\overline{\tau}$ has an asymptotically Normal distribution centered at the true treatment effect, i.e. $\sqrt{\NC} \big( \overline{\tau} - \tau^* \big)
		\stackrel{D}{\rightarrow}
		N \big( 0 , \VAR\big\{\varphi (\bO_i, e^*, g^*) \big\}  \big)$.
Also, the asymptotic variance of the estimator can be consistently estimated by the estimator $\overline{\sigma}^2$,
	\begin{align*}
		\overline{\sigma}^2
		=
		\frac{1}{\NC}	\sum_{k=1}^2  \sum_{i \in \II_k} 
		\Big\{
			\varphi(\bO_i, \widehat{e}^{(-k)}, \widehat{g}^{(-k)}) - \overline{\tau}_{k}
		\Big\}^2 \ , \
		\overline{\tau}_k
		=
		\frac{1}{\NC/2}  \sum_{i \in \II_k} 	{\varphi}(\bO_i, \widehat{e}^{(-k)}, \widehat{g}^{(-k)})\ .
	\end{align*}
\end{enumerate}
\end{theorem}
In words, Theorem \ref{thm:1} states the original cross-fitting estimator, with a minor tweak on how the samples are split,  retains its consistency, double robustness, and asymptotic Normality in multilevel studies where the data is no longer i.i.d. However, a critical difference between the traditional i.i.d. setting and the multilevel setting is that $\overline{\tau}$ is not efficient when \hyperlink{(E2.Both)}{(E2.Both)} holds. Roughly speaking, the efficiency loss is attributed to $\overline{\tau}$ not accounting for the intra-cluster correlation of the outcomes and the treatment variables. For the outcome variable, even though there maybe intra-cluster correlation among the residualized outcomes $Y_{ij} - \widehat{g}^{(-k)}(A_{ij},\bX_{ij})$ in $\overline{\tau}$, the estimator simply takes an equally-weighted average of these terms rather than an unequally weighted average where the weights depend on the correlation structure. For the treatment, the term $\widehat{e}^{(-k)}(a \cond \bX_{ij}) = P(A_{ij} = a \cond \bX_{ij})$ in $\overline{\tau}$ integrates over peers' treatment assignments and covariates in the same cluster, i.e. $(\bA_\eij, \bX_\eij)$. But, if there was intra-cluster correlation among the treatment variables, the propensity score $P(A_{ij} = a \cond \bX_{ij})$ does not capture this correlation, leading to potential losses in efficiency. 

\subsection{Details of Section \ref{Main-sec:result}}

\subsubsection{Discussion on Model \protect\hyperlink{Main-(M1)}{(M1)}}		\label{sec:modelM1}

We show that the examples in Section \ref{Main-sec:M1} in the main paper satisfy Model \hyperlink{Main-(M1)}{(M1)}. 

\begin{itemize}
	\item[1-1.] \emph{GLMM in Section \ref{Main-sec:2-3}}: \\
	This is a special case of 3. \emph{Locally independent, nonparametric latent variable models of \citep{Lord1980,Holland1986}}.
	\item[1-2.]  \emph{GEE models in Section \ref{Main-sec:2-3}}: \\
	Suppose $\NI_i \stackrel{i.i.d.}{\sim} P_\NI$ where $\text{supp}(\NI_i) \subset \{ 1, \ldots, \overline{M} \}$ and the following models hold:
\begin{align*}	
	&
	\EXP ( Y_{ij} \cond A_{ij} , \bX_{ij} ) 
	=
	g(A_{ij},\bX_{ij}) = 
	\mathcal{L}_g^{-1} \Big(
	\big( 1, A_{ij}, \bX_{ij} \T \big) \bm{\beta} \Big) , \quad{}
	\\
	&
	\pi(1 \cond \bA_\eij, \bX_{iij}, \bX_\eij ) 
	=
	P ( A_{ij} = 1 \cond \bA_\eij, \bX_{ij}, \bX_\eij) = \mathcal{L}_\pi^{-1}
	\Big( \big( \bA_\eij\T, \bX_{ij} \T , \bX_\eij \T \big)  \bm{\gamma} \Big)  \ ,
\end{align*}
where $\mathcal{L}_g$ and $\mathcal{L}_\pi$ are appropriate link functions. Then, it is trivial that Model \hyperlink{Main-(M1)}{(M1)} holds with $P(\bO_i) = \sum_{\ell = 1}^{\NT} P(\NI_i = \ell) P ( \bO_i \cond \NI_i = \ell )$.

	\item[2.] \emph{Nonparametric, individual-level i.i.d. models}: \\
	Suppose $\bC_i \stackrel{i.i.d.}{\sim} P (\bC_i) $ where $\text{supp}(\NI_i) \subset \{ 1, \ldots, \overline{M} \}$. Then, we find $\EXP ( Y_{ij} \cond A_{ij} , \bX_{ij} ) $ is identical across all units as follows.
	\begin{align*}
		\EXP ( Y_{ij} \cond A_{ij} = a , \bX_{ij} = \bx ) 
		& = 
		\EXP ( Y_{ij} \cond A_{ij} = a , \bX_{ij} = \bx , \bC_i = \bm{c} ) 
		\\
		&
		=
		\int y \frac{
			P( Y_{ij} = y, A_{ij} = a , \bX_{ij} = \bx \cond \bC_i = \bm{c} )
		}{
			\int P( Y_{ij} = t, A_{ij} = a , \bX_{ij} = \bx \cond \bC_i = \bm{c} ) \,  dP(t)
		}
		\, dP(y)
		\\
		&
		=
		\int y \frac{
			P( Y_{i'j'} = y, A_{i'j'} = a , \bX_{i'j'} = \bx \cond \bC_{i'} = \bm{c} )
		}{
			\int P( Y_{i'j'} = t, A_{i'j'} = a , \bX_{i'j'} = \bx \cond \bC_{i'} = \bm{c} ) \,  dP(t)
		}
		\, dP(y)
		\\
		& = 
		\EXP ( Y_{i'j'} \cond A_{i'j'} = a , \bX_{i'j'} = \bx , \bC_{i'} = \bm{c} ) 
		\\
		& = 
		\EXP ( Y_{i'j'} \cond A_{i'j'} = a , \bX_{i'j'} = \bx ) 
		\ , \ ij \neq i'j'
		\ .
	\end{align*}	
	Additionally, we find
	\begin{align*}
		&
		P( A_{ij} = a , \bA_\eij = \ba' , \bX_{ij} = \bx , \bX_\eij = \bx' \cond \bC_i = \bm{c} )
		\\
		&
		=
		\int P( Y_{ij} = y, \bY_\eij = \by', A_{ij} = a , \bA_\eij = \ba' , \bX_{ij} = \bx , \bX_\eij = \bx' \cond \bC_i = \bm{c} ) \, dP(y,\by')
		\\
		&
		=
		\int \prod_{j=1}^{\NI_i} P( \bO_{ij} = \bo \cond \bC_i = \bm{c} ) \, dP(y,\by') \ .
	\end{align*}
	Since $\prod_{j=1}^{\NI_i} P( \bO_{ij} = \bo \cond \bC_i = \bm{c} )$ has the same form across all units, we find 
	\begin{align*}
		&
		P( A_{ij} = a , \bA_\eij = \ba' , \bX_{ij} = \bx , \bX_\eij = \bx' \cond \bC_i = \bm{c} )
		\\
		&
		=
		P( A_{i'j'} = a , \bA_{i'(-j')} = \ba' , \bX_{i'j'} = \bx , \bX_{i'(-j')} = \bx' \cond \bC_{i'} = \bm{c} ) \ , \ ij \neq i'j' \ .
	\end{align*}
	As a result, we have 
	\begin{align*}
		&
		P( \bA_\eij = \ba' , \bX_{ij} = \bx , \bX_\eij = \bx' \cond \bC_i ) \\
		&
		= P (\bA_{i'(-j')} = \ba' , \bX_{i'j'} = \bx , \bX_{i'(-j')} = \bx' \cond \bC_{i'} = \bm{c} ) \ , \ ij \neq i'j' \ ,
	\end{align*}
	and
	\begin{align*}
		&
		P( A_{ij} = a \cond \bA_\eij = \ba' , \bX_{ij} = \bx , \bX_\eij = \bx' )
		\\
		&
		P( A_{ij} = a \cond \bA_\eij = \ba' , \bX_{ij} = \bx , \bX_\eij = \bx' , \bC_i = \bm{c} )
		\\
		&
		=
		\frac{
		P( A_{ij} = a , \bA_\eij = \ba' , \bX_{ij} = \bx , \bX_\eij = \bx' \cond \bC_i=\bm{c} )
		}{ P( \bA_\eij = \ba' , \bX_{ij} = \bx , \bX_\eij = \bx' \cond \bC_i=\bm{c} ) }
		\\
		&
		=
		\frac{
		P( A_{i'j'} = a , \bA_{i'(-j')} = \ba' , \bX_{i'j'} = \bx , \bX_{i'(-j')} = \bx' \cond \bC_{i'}=\bm{c} )
		}{P (\bA_{i'(-j')} = \ba' , \bX_{i'j'} = \bx , \bX_{i'(-j')} = \bx' \cond \bC_{i'}=\bm{c} ) } 
		\\
		&
		=
		P( A_{i'j'} = a \cond \bA_{i'(-j')} = \ba' , \bX_{i'j'} = \bx , \bX_{i'(-j')} = \bx' , \bC_{i'} = \bm{c} )
		\\
		& = 
		P( A_{i'j'} = a \cond \bA_{i'(-j')} = \ba' , \bX_{i'j'} = \bx , \bX_{i'(-j')} = \bx' )
		\ , \ ij \neq i'j' \ .
	\end{align*}
	This implies $P( A_{ij} = a \cond \bA_\eij = \ba' , \bX_{ij} = \bx , \bX_\eij = \bx' )$ is identical across all units, and this conditional probability satisfies the restriction in Model \hyperlink{Main-(M1)}{(M1)}.

	\item[3.] \emph{Locally independent, nonparametric latent variable models of \citep{Lord1980,Holland1986}}: \\
	Suppose $\NI_i \stackrel{i.i.d.}{\sim} P_\NI$ where $\text{supp}(\NI_i) \subset \{ 1, \ldots, \overline{M} \}$. We first study the density of $(A_{ij}, \bX_{ij} , \bU_i)$:
	\begin{align*}
		&
		f (A_{ij} = a, \bX_{ij} = \bx, \bU_i = \bu )
		\\
		& =
		\int
		f (A_{ij} = a, \bA_{\eij} = \ba' , \bX_{ij} = \bx, \bX_{\eij} = \bx', \bU_i = \bu , \bV_i = \bv ) \, d(\ba' \bx' \bv)
		\\
		& = 
		\int
		f (A_{ij} = a, \bA_{\eij} = \ba' \cond \bX_{ij} = \bx, \bX_{\eij} = \bx', \bU_i = \bu , \bV_i = \bv )
		\\
		& \hspace*{3cm} \times
		f (\bX_{ij} = \bx, \bX_{\eij} = \bx') f(\bU_i = \bu) f (\bV_i = \bv )
		 \, d(\ba' \bx' \bv)
		 \\
		& = 
		\bigg[
		\int
		f (A_{ij} = a \cond \bX_{ij} = \bx , \bV_i = \bv )
		\bigg\{
		\prod_{k \neq j}
		f (A_{ik} = a_k' \cond \bX_{ik} = \bx_k' , \bV_i = \bv ) \bigg\}
		\\
		& \hspace*{3cm} \times
		f (\bX_{ij} = \bx, \bX_{\eij} = \bx') 
		f (\bV_i = \bv )
		 \, d(\ba' \bx' \bv) \bigg] \times
		 f(\bU_i = \bu) 
		 \\
		 & =
		 f (A_{ij} = a, \bX_{i} = \bx) f(\bU_i = \bu ) \ .
	\end{align*}
	As a consequence, we find $f(\bU_i = \bu \cond A_{ij} = a, \bX_{ij} = \bx) = f(\bU_i = \bu)$. From the assumption, we have $f(\bV_i = \bv \cond \bX_{ij} = \bx) = f(\bV_i = \bv)$.

	Using the established result, we study the outcome regression:
	\begin{align*}
		&
		\EXP (Y_{ij} \cond A_{ij} = a , \bX_{ij} = \bx)
		\\
		& = 
		\EXP_{\bU | A, \bX} \big\{ \EXP (Y_{ij} \cond A_{ij} = a , \bX_{ij} = \bx, \bU_i) \cond
		 A_{ij} = a , \bX_{ij} = \bx \big\}
		\\
		& = 
		\int \EXP (Y_{ij} \cond A_{ij} = a , \bX_{ij} = \bx, \bU_i = \bu) f( \bU_i = \bu \cond A_{ij} = a, \bX_{ij} = \bx) \, d\bu
		\\
		& = 
		\int \EXP (Y_{ij} \cond A_{ij} = a , \bX_{ij} = \bx, \bU_i = \bu) f( \bU_i = \bu)  \, d\bu
		\\
		& = 
		\int \EXP (Y_{i'j'} \cond A_{i'j'} = a , \bX_{i'j'} = \bx, \bU_{i'} = \bu) f( \bU_{i'} = \bu) \, d\bu
		\\
		& = \cdots  =
		\EXP (Y_{i'j'} \cond A_{i'j'} = a , \bX_{i'j'} = \bx) \ .
	\end{align*}
	The fourth and sixth lines hold from the established result. The fifth line holds because the distributions of $Y_{ij} \cond (A_{ij}, \bX_{ij}, \bU_i)$ and $\bU_i$ is identical across $ij$, respectively. Thus, we find the outcome regression is identical across $ij$. 
	
	Second, we show that the conditional propensity score is identical across $ij$. To begin, we re-write $P(A_{ij} =a_j , \bA_{\eij} = \ba_{(-j)} \cond \bX_{ij} = \bx_j, \bX_{\eij} = \bx_{(-j)})$:
		\begin{align*}
			&
			P( A_{ij} =a_j , \bA_{\eij} = \ba_{(-j)} \cond \bX_{ij} = \bx_j, \bX_{\eij} = \bx_{(-j)} )
			\\
			&
			=
			\EXP \big\{ 
				\ind (A_{ij} =a_j , \bA_{\eij} = \ba_{(-j)} ) \cond \bX_{ij} = \bx_j, \bX_{\eij} = \bx_{(-j)}
			\big\}
			\\
			&
			=
			\EXP_{\bV| \bX}
			\Big[
			\EXP \big\{ 
				\ind (A_{ij} =a_j , \bA_{\eij} = \ba_{(-j)} ) \cond \bX_{ij} = \bx_j, \bX_{\eij} = \bx_{(-j)} , \bV_i
			\big\} \, \Big| \,
			\bX_{ij} = \bx_j, \bX_{\eij} = \bx_{(-j)} \Big]
			\\
			&
			=
			\int 
			\EXP \big\{ 
				\ind ( A_{ij} =a_j , \bA_{\eij} = \ba_{(-j)} ) \cond \bX_{ij} = \bx_j, \bX_{\eij} = \bx_{(-j)} , \bV_i = \bv
			\big\}
			\\
			& \hspace*{3cm} \times
			f(\bV_i = \bv \cond \bX_{ij} = \bx_j, \bX_{\eij} = \bx_{(-j)} ) \, d\bv
			\\
			&
			=
			\int 
			\prod_{k=1}^{\NI_i = \NI}
			\EXP \big\{ 
				\ind (A_{ik} =a_k ) \cond \bX_{ik} = \bx_k , \bV_i = \bv
			\big\}
			f (\bV_i=\bv) \, d\bv
			\\
			&
			=
			\int 
			\prod_{k'=1}^{\NI_{i'} = \NI}
			\EXP \big\{ 
				\ind (A_{i'k'} =a_k ) \cond \bX_{i'k'} = \bx_k , \bV_{i'} = \bv
			\big\}
			f (\bV_{i'}=\bv) \, d\bv
			\\
			& = \cdots = 
			P( A_{i'j'} =a_{j} , \bA_{i'(-j')} = \ba_{(-j')} \cond \bX_{i'j'} = \bx_{j'}, \bX_{i'(-j')} = \bx_{(-j')} ) \ .
		\end{align*}
		As a result, we find the conditional propensity score is identical across $ij$:
		\begin{align*}
			&
			 P(A_{ij} = a_{j} \cond \bA_{i(-j)} = \ba_{(-j)}, \bX_{ij} = \bx_{j} , \bX_{i(-j)} = \bx_{(-j)} ) 
			\\
			& = 
			\frac{ P(A_{ij} = a_{j} , \bA_{i(-j)} = \ba_{(-j)} \cond \bX_{ij} = \bx_{j} , \bX_{i(-j)} = \bx_{(-j)} ) }
			{P(\bA_{i(-j)} = \ba_{(-j)} \cond \bX_{ij} = \bx_{j} , \bX_{i(-j)} = \bx_{(-j)} )}
			\\
			& = 
			\frac{ P(A_{ij} = a_{j} , \bA_{i(-j)} = \ba_{(-j)} \cond \bX_{ij} = \bx_{j} , \bX_{i(-j)} = \bx_{(-j)} ) }
			{
			\left\{
			\begin{array}{l}
			P(A_{ij} = 1 , \bA_{i(-j)} = \ba_{(-j)} \cond \bX_{ij} = \bx_{j} , \bX_{i(-j)} = \bx_{(-j)} ) \\
			\quad + P(A_{ij} = 0 , \bA_{i(-j)} = \ba_{(-j)} \cond \bX_{ij} = \bx_{j} , \bX_{i(-j)} = \bx_{(-j)} ) 
			\end{array}
			\right\}
			}
			\\
			& = 
			\frac{ P(A_{i'j'} = a_{j} , \bA_{i'(-j')} = \ba_{(-j)} \cond \bX_{i'j'} = \bx_{j'} , \bX_{i'(-j')} = \bx_{(-j)} ) }
			{
			\left\{
			\begin{array}{l}
			P(A_{i'j'} = 1 , \bA_{i'(-j')} = \ba_{(-j)} \cond \bX_{i'j'} = \bx_{j} , \bX_{i'(-j')} = \bx_{(-j)} ) \\
			\quad + P(A_{i'j'} = 0 , \bA_{i'(-j')} = \ba_{(-j)} \cond \bX_{i'j'} = \bx_{j} , \bX_{i'(-j')} = \bx_{(-j)} ) 
			\end{array}
			\right\}
			}
			\\
			& = \cdots = P(A_{i'j'} = a_{j} \cond \bA_{i'(-j')} = \ba_{(-j)}, \bX_{i'j'} = \bx_{j} , \bX_{i'(-j')} = \bx_{(-j)} ) \ .
		\end{align*}
	Then, it is trivial that Model \hyperlink{Main-(M1)}{(M1)} holds with $P(\bO_i) = \sum_{\ell = 1}^{\NT} P(\NI_i = \ell) P ( \bO_i \cond \NI_i = \ell )$.


	\item[4.] \emph{Nonparaemtric mixture models}: \\
	The support of $\NI_i$ is the union of the support of $\NI_i \cond \bm{Z}_i = \bm{z}$. Since the support of $\NI_i \cond \bm{Z}_i = \bm{z}$ belongs to $\{ 1,\ldots,\NT \}$, we have $\text{supp} (\NI_i) \subset \{ 1,\ldots, \NT\}$. 
	
	Next, we show the identical outcome regression and the conditional propensity score across $ij$. Instead of assuming the identical conditional distribution $\bm{Z}_i \cond (h_1(\bO_{ij}), h_2(\bO_{\eij}) )$ for any fixed functions $h_1$ and $h_2$, it suffices to have the condition for only two pairs of $(h_1, h_2)$: (i) $h_1 (\bO_{ij}) = (A_{ij}, \bX_{ij})$ and $h_2 (\bO_{\eij}) = 1$ and (ii) $h_1(\bO_{ij}) = \bX_{ij}$ and $h_2(\bO_{\eij}) = \bX_{\eij}$. Then, we find $P(\bm{Z}_{i} = \bm{z} \cond A_{ij} = a, \bX_{ij} = \bx)$ and $P(\bm{Z}_{i} = \bm{z} \cond \bX_{ij} = \bx , \bX_{\eij} = \bx')$ have the same form across all $(i,j)$. Under this relaxed assumption on $\bm{Z}_i$, we show that the outcome regression is identical across $ij$. Let $f_{\bm{Z}|A,\bX}$ be the density of $\bm{Z}_i \cond (A_{ij}, \bX_{ij})$. Then, we find
	\begin{align*}
		&
		\EXP (Y_{ij} \cond A_{ij} = a, \bX_{ij} = \bx) 
		\\
		& 
		=
		\EXP_{\bm{Z} | A, \bX} \Big\{ \EXP (Y_{ij} \cond A_{ij} = a, \bX_{ij} = \bx, \bm{Z}_i)  \, \Big| \, A_{ij} = a, \bX_{ij} =\bx \Big\}		
		\\
		& 
		=
		\int \EXP (Y_{ij} \cond A_{ij} = a, \bX_{ij} = \bx, \bm{Z}_i = \bm{z} ) f_{\bm{Z}|A,\bX}(\bm{Z}_i = \bm{z} \cond A_{ij}=a, \bX_{ij} = \bx ) \, d\bm{z}
		\\
		& 
		=
		\int \EXP (Y_{i'j'} \cond A_{i'j'} = a, \bX_{i'j'} = \bx, \bm{Z}_{i'} = \bm{z} ) f_{\bm{Z}|A,\bX}(\bm{Z}_{i'} = \bm{z} \cond A_{i'j'}=a, \bX_{i'j'} = \bx ) \, d\bm{z}
		\\
		& = \cdots =
		\EXP (Y_{i'j'} \cond A_{i'j'} = a, \bX_{i'j'} = \bx) \ .
	\end{align*}
	Lastly, we show the conditional propensity score is identical across $ij$. Let $f_{\bm{Z}| \bX}$ be the density of $\bm{Z}_i \cond (\bX_{ij}, \bX_{\eij})$. Then, we find
	\begin{align*}
		&
		\EXP (A_{ij} = a , \bA_{\eij} = \ba' \cond \bX_{ij} = \bx, \bX_{\eij} = \bx') 
		\\
		& 
		=
		\EXP_{\bm{Z} | \bX} \Big\{ \EXP (A_{ij} = a , \bA_{\eij} = \ba' \cond \bX_{ij} = \bx, \bX_{\eij} = \bx' , \bm{Z}_i)  \, \Big| \, \bX_{ij} =\bx , \bX_{\eij} = \bx' \Big\}		
		\\
		& 
		=
		\int \EXP (A_{ij} = a , \bA_{\eij} = \ba' \cond \bX_{ij} = \bx, \bX_{\eij} = \bx' , \bm{Z}_i = \bm{z} ) 
		\\
		& \hspace*{4cm} \times
		 f_{\bm{Z}|\bX}(\bm{Z}_i = \bm{z} \cond \bX_{ij} = \bx , \bX_{\eij} = \bx' ) \, d\bm{z}
		\\
		& 
		=
		\int \EXP (A_{i'j'} = a , \bA_{i'(-j')} = \ba' \cond \bX_{i'j'} = \bx, \bX_{i'(-j')} = \bx' , \bm{Z}_{i'} = \bm{z} ) 
		\\
		& \hspace*{4cm} \times
		 f_{\bm{Z}|\bX}(\bm{Z}_{i'} = \bm{z} \cond \bX_{i'j'} = \bx , \bX_{i'(-j')} = \bx' ) \, d\bm{z}
		\\
		& = \cdots =
		\EXP (A_{i'j'} = a , \bA_{i'(-j')} = \ba' \cond \bX_{i'j'} = \bx, \bX_{i'(-j')} = \bx') 
		\ .
	\end{align*}
	This show that Model \hyperlink{Main-(M1)}{(M1)} holds with $P(\bO_i) = \int P(\bO_i \cond \bm{Z}_i = \bm{z} ) \, dP(\bm{z})$.

\end{itemize}

\subsubsection{Practical Considerations}								\label{sec:practical}
This section highlights some practical implementation details concerning the proposed estimator $\widehat{\tau}$, especially for investigators who are more familiar with using parametric or semiparametric methods like GLMM and GEE. Much of these tips already exist in the literature and the references below contain additional details.

First, because the estimator $\widehat{\tau}$ is constructed using cross-fitting,  the finite sample performance of the estimators may depend on the particular random split. To mitigate the impact associated with the random split, Section 3.4 of \citet{Victor2018} suggests repeating the sample splitting $S$ times and obtain the estimator $\widehat{\tau}^{(s)}$ $(s=1,\ldots,S)$ and the corresponding variance estimator $\widehat{\sigma}^{2,(s)}$. Afterwards, the medians is computed , i.e.
	$\widehat{\tau}^{\text{med}}
	=
	\median_{s=1,\ldots,S}
	\widehat{\tau}^{(s)} $.
along with its standard error, i.e. $ \widehat{\sigma}^{2,\text{med}} =
	\median_{s=1,\ldots,S}
	\Big[
	 \big\{ \widehat{\tau}^{(s)} - \widehat{\tau}^{\text{med}} \big\}^2
	 +
	 \widehat{\sigma}^{2,(s)}
	 \Big]
$.
As shown in Corollary 3.3 of \citet{Victor2018}, the median-adjusted estimators can replace the established results about $\widehat{\tau}$.

Second, for estimating $g$, we can use existing nonparametric estimation methods. Specifically, we can treat $Y_{ij}$ as the dependent variable and $(A_{ij}, \bX_{ij})$ as the independent variables.

Third, to estimate the conditional propensity score $\pi$, naively using the raw variables $(\bA_\eij, \bX_{ij}, \bX_\eij)$ as independent variables in existing nonparametric regression estimator may not work because each cluster may have different cluster size, leading to different number of independent variables. For example, without covariates, a cluster of size $2$ will have 1 independent variable while a cluster of size 10 will have 9 independent variables when we use $\bA_\eij$ as independent variables in a nonparametric regression estimator. To resolve the issue, \citet{vvLaan2014}, \citet{Sofrygin2016}, and \citet{Ogburn2020arxiv} suggested making structural assumptions on $\pi$ where the investigator uses fixed, dimension-reducing summaries of peers' data $(\bA_\eij, \bX_\eij)$ as independent variables. Some popular examples of these dimension-reducing summaries include the proportion of treated peers, $\overline{A}_\eij = \sum_{\ell \neq j} A_{i\ell}/(\NI_i-1)$, and the average of peers' covariates, $\overline{\bX}_\eij = \sum_{\ell \neq j} \bX_{i\ell}/(\NI_i-1)$. 
With the additional structural assumptions, nonparametric estimation of $\pi$ is similar to nonparametric estimation of $e$ where $A_{ij}$ is the dependent variable and $(\overline{A}_\eij, \bX_{ij}, \overline{\bX}_\eij) $ are the independent variables. Note that investigators can use additional summaries of peers' data as independent variables or use more flexible models of contagion in network settings  \citep{Salathe2010, Keeling2011,Lorna2013,Imai2015}

Fourth, in some multilevel studies, there may be few ``outliers'' clusters where the clusters are either very large (or very small) compared to other clusters in the study. In such cases, one may remove these outlying clusters or reweigh them by tweaking the weights $w(\bC_i)$ in the average treatment effect. 
Or, for large outlying clusters, one may randomly remove observations so that the number of individuals per cluster used in the estimation procedures are roughly similar to each other; 
this adjustment is often referred to as undersampling \citep[Chapter 5.2]{Fernandez2018}. One can also take the median of multiple conditional propensity score estimates that are obtained from repeated undersampling procedures to improve numerical stability.

\subsection{Details of Section \ref{Main-sec:sim}}					\label{sec:Sim}

\subsubsection{Details of the Superlearner Libraries}		\label{sec:Sim-SL}

We include the following methods and the corresponding R packages in our super learner library: linear regression via \texttt{glm}, lasso/elastic net via \texttt{glmnet} \citep{glmnet}, spline via \texttt{earth} \citep{earth} and \texttt{polspline} \citep{polspline}, generalized additive model via \texttt{gam} \citep{gam}, boosting  via \texttt{xgboost} \citep{xgboost} and \texttt{gbm} \citep{gbm}, random forest via \texttt{ranger} \citep{ranger}, and neural net via \texttt{RSNNS} \citep{RSNNS}. For estimating the conditional propensity score $\pi$, we follow the details in Section \ref{sec:practical} where we use $\overline{A}_\eij$ and $\overline{\bX}_\eij$ as the summary statistics of $\bA_\eij$ and $\bX_\eij$, respectively, perform undersampling five times by randomly choosing $\min \NI_i$ (i.e. size of the smallest cluster) individuals from each cluster, and use the median adjustment for $\widehat{\tau}$ based on five random splits. 

\subsubsection{Details of the Intra-cluster Correlation Coefficient}

First, we presented the closed form of the ICC used in the main paper. The ICC of continuous $Y$ is defined as ${\rm ICC}_{Y} (\sigma_U) = \sigma_U^2/(1+\sigma_U^2)$. For a binary $A_i$, we obtain the ANOVA-type estimate of the ICC from \citet{Ridout1999} and \citet{ICC2019}, i.e.
\begin{align*}
	{\rm ICC}_{A} (\sigma_V) = \frac{{\rm MS}_{A,b} - {\rm MS}_{A,w}}{{\rm MS}_{A,b} + (\NI_0 - 1) {\rm MS}_{A,w}}
	\ , \
	\NI_0 = \frac{1}{\NC-1}\bigg( \sum_{i=1}^\NC \NI_i - \sum_{i=1}^\NC \frac{\NI_i^2}{\sum_{i=1}^\NC \NI_i} \bigg)
\end{align*}
where ${\rm MS}_{A,b}$ and ${\rm MS}_{A,w}$ are the between-group and within-group mean squares of the residuals $r_{ij} = A_{ij} - e^*(A_{ij} \cond \bX_{ij})$: 
\begin{align*}
	&
	{\rm MS}_{A,w}
	=
	\frac{1}{\sum_{i=1}^\NC \NI_i - \NC}\bigg\{
	\sum_{i=1}^\NC \sum_{j=1}^{\NI_i} r_{ij}^2 - 
	\sum_{i=1}^\NC
	\frac{ (\sum_{j=1}^{\NI_i} r_{ij})^2} {\NI_i} \bigg\} \ ,
	\\
	&
	{\rm MS}_{A,b}
	=
	\frac{1}{ \NC-1}
	\bigg\{
	\sum_{i=1}^\NC
	\frac{ (\sum_{j=1}^{\NI_i} r_{ij})^2} {\NI_i} - \frac{ (\sum_{i=1}^\NC \sum_{j=1}^{\NI_i} r_{ij} )^2 }{\sum_{i=1}^{\NC} \NI_i  } \bigg\} \ .
\end{align*}


\subsubsection{Finite-Sample Relative Efficiency Between $\widehat{\tau}$ and $\overline{\tau}$} \label{sec:sim-illust}

We study the finite-sample efficiency characteristics of $\widehat{\tau}$  and $\overline{\tau}$ for different values of $(\sigma_V,\sigma_U)$. 
Specifically, we use the simulation models in Section \ref{Main-sec:varyingN} in the main paper when $\NC = 500$. We vary both $(\sigma_V,\sigma_U)$ from a two-dimensional grid $\{0,0.1,\ldots,2\}^{\otimes 2}$. We then compare the efficiency between $\overline{\tau} (e^*,g^*)$ and $\widehat{\tau}(\pi^*,g^*, \widehat{\beta})$. Here, we use the true nuisance functions $\pi$, $g$, and $e$ to better highlight the differences in efficiency without the variability arising from estimating them. Specifically, for each $(\sigma_V,\sigma_U)$ combination, we generate the data 1,000 times and obtain the empirical relative efficiency between the two estimators, i.e. $ \rho (\sigma_V,\sigma_U)  = \texttt{Var}\big\{ \overline{\tau}(e^*,g^*) \big\} / \texttt{Var} \big\{ \widehat{\tau}(\pi^*,g^*, \widehat{\beta}) \big\}$.

\begin{figure}[!htb]
	\centering
	\includegraphics[width=\textwidth]{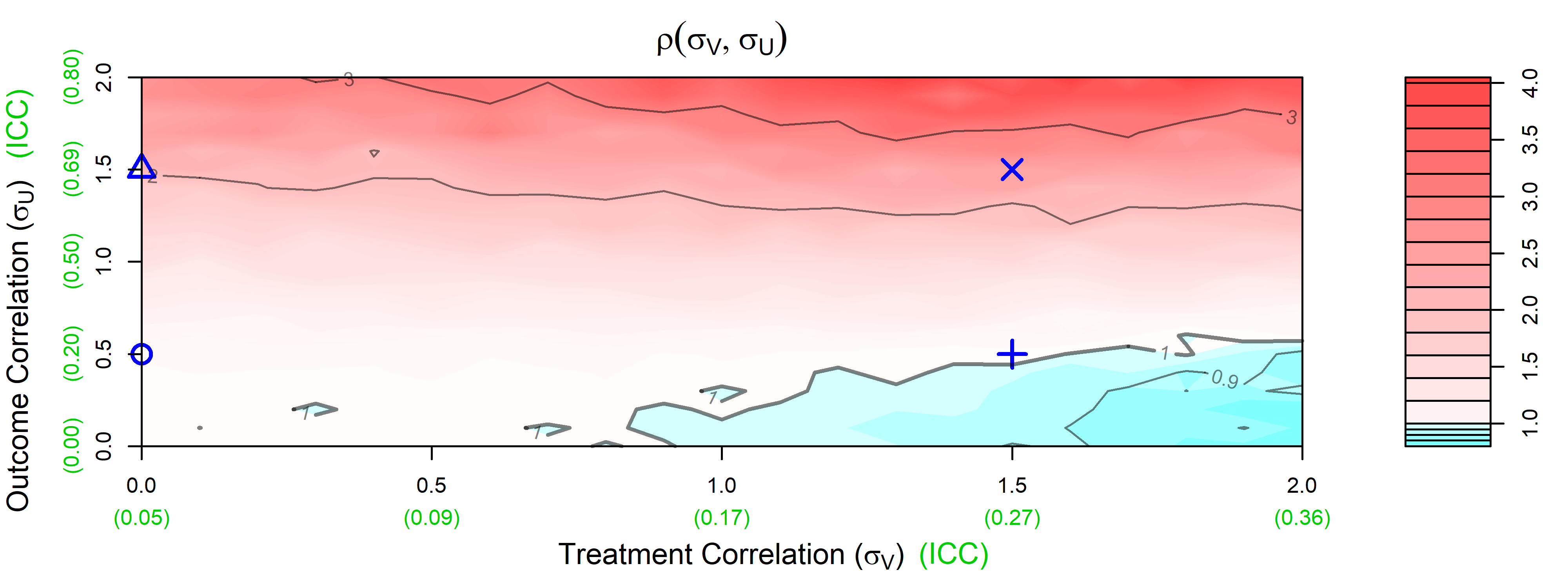} 
	\caption{The empirical relative efficiency between $\widehat{\tau}$ and $\overline{\tau}$ for different values of $(\sigma_V, \sigma_U)$. The black $x$ and $y$ axis are $\sigma_V$ and $\sigma_U$, respectively, and the {\color{green1} green} values are the corresponding ICC values. {\color{red1} Red} regions correspond to settings where $\widehat{\tau}$ is more efficient than $\overline{\tau}$. {\color{cyan} Cyan} regions correspond to settings where $\overline{\tau}$ is more efficient than $\widehat{\tau}$. Blue dots ({\color{blue}$\circ,\triangle,+,\times$}) correspond to the simulation scenario in Section \ref{Main-sec:varyingN} in the main paper.}
	\label{fig-1}
\end{figure}

Figure \ref{fig-1} shows the empirical relative efficiencies between the two estimators $\widehat{\tau}$ and $\overline{\tau}$ for different values of $(\sigma_V, \sigma_U)$. We see that $\widehat{\tau}$ is more efficient than $\overline{\tau}$ when the outcome correlation, as governed by $\sigma_U$, is large. For example, if the outcome ICC is above 0.2, $\widehat{\tau}$ is always more efficient than $\overline{\tau}$, irrespective of the correlation between the treatment values. Also, when there is no correlation between treatments (i.e. when $\sigma_V = 0$), $\widehat{\tau}$ is always more efficient than $\overline{\tau}$, corroborating the result in Lemma \ref{Main-thm:2} in the main paper. Interestingly, $\overline{\tau}$ is more efficient when the treatment correlation, as governed by $\sigma_V$, is large, but not by a large margin compared to when the outcome correlation is large; $\overline{\tau}$  is no more than 20\% efficient compared to $\widehat{\tau}$ when $\sigma_V$ is large, but $\widehat{\tau}$ can be 300\% more efficient compared to $\overline{\tau}$ when $\sigma_U$ is large. In short, we expect that $\widehat{\tau}$ would perform better when (i) the units' treatments in the same cluster are weakly correlated to each other and (ii) the outcomes of units are correlated. 

\newpage

\subsection{Details of Section \ref{Main-sec:application}}								\label{sec:Supp-ECLSK}

We assess covariate balance and overlap in the ECLS-K dataset. We first obtain the median estimates of the propensity score and the conditional propensity score as follows.
\begin{align*}
	&
	\widehat{e}_{\rm med}(a \cond \bX_{ij}) = \median_{k=1,\ldots,100} \widehat{e}_{[k]}(a \cond \bX_{ij})
	\ , \\
	&
	\widehat{\pi}_{\rm med}(a \cond \bA_\eij, \bX_{ij} , \bX_\eij ) = \median_{k=1,\ldots,100} \widehat{\pi}_{[k]}(a \cond \bA_\eij, \bX_{ij} , \bX_\eij ) \ , 
\end{align*}
where $\widehat{e}_{[k]}$ and $\widehat{\pi}_{[k]}$ are the estimated propensity score and conditional propensity score from $k$th cross-fitting procedure, respectively. 

To assess covariate balance after the (conditional) propensity score adjustment, we test the following hypotheses:
\begin{align*}
	& H_{0,{\rm no}} : \EXP \bigg\{ \frac{A_{ij} \bX_i}{P(A_{ij}=1)} - \frac{(1-A_{ij}) \bX_{ij}}{P(A_{ij}=0)} \bigg\} = 0 \ , \\
	& 	H_{0,e} : \EXP \bigg\{ \frac{A_{ij} \bX_{ij}}{e(1 \cond \bX_{ij})} - \frac{(1-A_{ij}) \bX_{ij}}{e(0 \cond \bX_{ij})} \bigg\} = 0 \ , \\
	& 	H_{0,\pi} : \EXP \bigg\{ \frac{A_{ij} \bX_{ij}}{\pi(1 \cond \bA_\eij, \bX_{ij} , \bX_\eij )} - \frac{(1-A_{ij}) \bX_{ij}}{\pi(0 \cond \bA_\eij, \bX_{ij} , \bX_\eij ) } \bigg\} = 0 \ .
\end{align*}
We randomly sample one individual from each cluster 10,000 times to guarantee the independence between observations used in the hypothesis testing. For each randomly chosen samples, we obtain three $t$-statistics associated with the above hypotheses by using $\sum_{ij} A_{ij} / \sum_i \NI_i$, $\widehat{e}_{\rm med}$, $\widehat{\pi}_{\rm med}$ as the estimates of $P(A_{ij} = 1)$, $e$, and $\pi$, respectively. To mitigate the effect of the particular random split, we obtain the median of 10,000 $t$-statistics. 

Table \ref{tab:balance} shows the result. The $t$-statistics regarding $H_{0,e}$ suggest that covariate balance is achieved in all covariates. Similarly, the $t$-statistics regarding $H_{0,\pi}$ show that covariate balance is achieved in all covariates except socioeconomic status ($t$-statistic$=2.390$). However, compared to the $t$-statistic regarding $H_{0,{\rm no}}$ in socioeconomic status ($t$-statistic$=5.295$), we find that adjustment with $\pi$ improved the balance of socioeconomic status. Moreover, given the number of hypotheses of interest is 34, one significant $t$-statistic may occur even though covariate balance is not violated. Therefore, we conclude that covariate balance is achieved for both $e$ and $\pi$.

\begin{table}[!htp]
		\footnotesize
		\renewcommand{\arraystretch}{1.1} \centering
		\setlength{\tabcolsep}{4pt}
\begin{tabular}{|c|c|c|c||c|c|c|c|}
\hline
\multirow{2}{*}{Covariate} & \multicolumn{3}{c||}{Test statistic}                        & \multirow{2}{*}{Covariate} & \multicolumn{3}{c|}{Test statistic}                         \\ \cline{2-4} \cline{6-8} 
                          & $H_{0,{\rm no}}$ & $H_{0,e}$ & $H_{0,\pi}$ &     & $H_{0,{\rm no}}$ & $H_{0,e}$ & $H_{0,\pi}$ \\ \hline
                       Cluster size & -0.548 & -0.237 & -0.346 &
            Census region=Northeast & -0.717 & -0.623 & -0.542 \\ \hline
                Census region=South & -0.736 & -0.773 & -0.826 &
                 Census region=West & -1.969  & 0.039 & -0.425 \\ \hline
                      Location=City & -0.891 & -0.082 & -0.432 &
                     Location=Rural & -1.386 & -0.047 & -0.091 \\ \hline
                Public kindergarten & -2.517 & -0.674 & -1.162 &
                               Male  & 0.189  & 0.822  & 0.942 \\ \hline
                                Age & -1.064 & -0.424 & -0.499 &
                         Race=Asian & -0.288  & 0.212 & -0.363 \\ \hline
                         Race=Black  & 1.766  & 0.229  & 0.567 &
                      Race=Hispanic & -3.213 & -1.019 & -1.500 \\ \hline
                         Race=White  & 0.048 & -0.190  & 0.081 &
                      Intact family & -1.065 & -0.316 & -0.526 \\ \hline
 Parental education $\geq$ College  & 2.071  & 0.197  & 0.542 &
               Socioeconomic status  & 5.876  & 1.371  & 2.373 \\ \hline
\end{tabular}
\caption{Covariate Balance Assessment: Each column shows $t$-statistics under $H_{0,{\rm no}}$, $H_{0,e}$, and $H_{0,\pi}$, respectively. Each row shows pre-treatment covariates.} 
\label{tab:balance}
\end{table}

\begin{figure}[!htb]
	\centering
	\includegraphics[width=\textwidth]{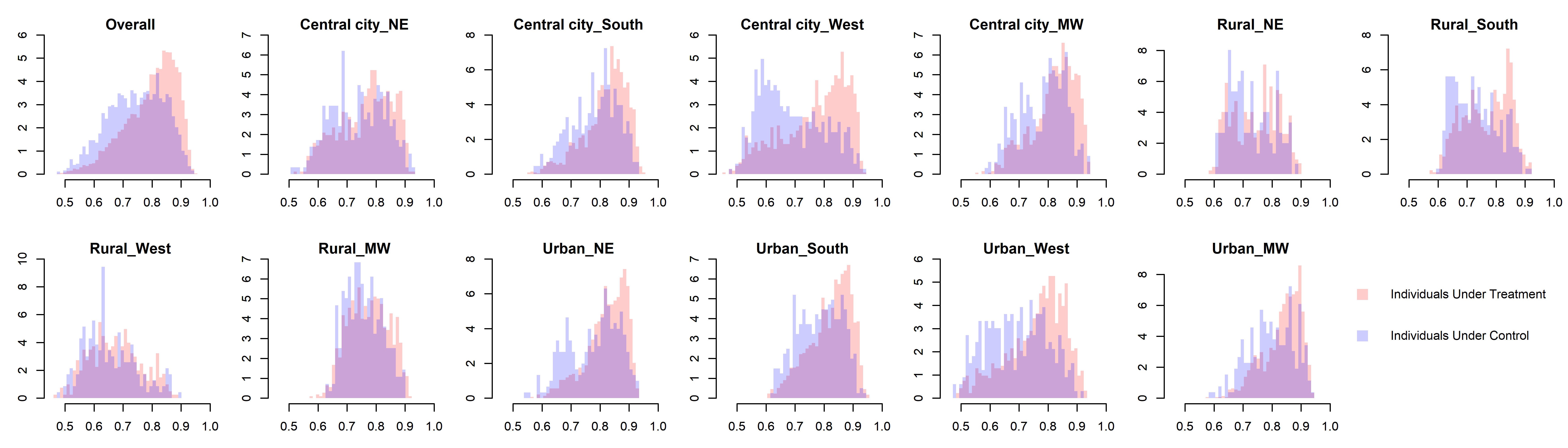} 
	\caption{Overlap Assessment: histograms of $\widehat{e}_{\rm med}(1 \cond \bX_{ij})$.}
	\label{fig-overlap1}
	\vspace*{-0.2cm}
\end{figure}

\begin{figure}[!htb]
	\centering
	\includegraphics[width=\textwidth]{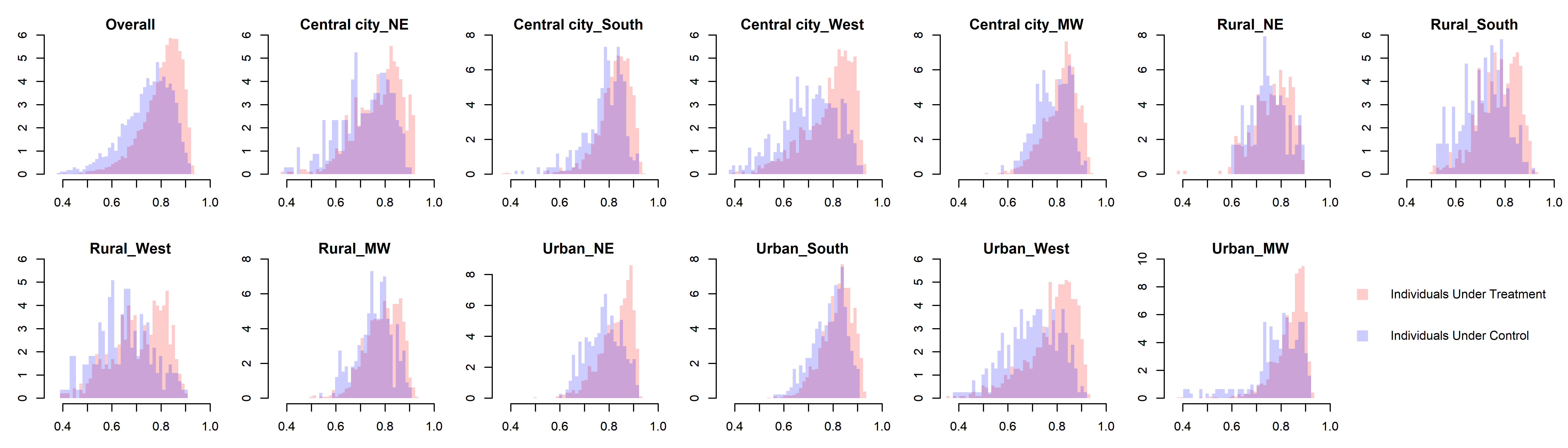} 
	\caption{Overlap Assessment: histograms of $\widehat{\pi}_{\rm med}(1 \cond \bA_{\eij}, \bX_{ij}, \bX_{\eij})$.}
	\label{fig-overlap2} 
	\vspace*{-4cm}
\end{figure}

Next, we assess overlap for $e$ via histrograms. As shown in Figures \ref{fig-overlap1} and \ref{fig-overlap2}, we find the overlap is not violated for both $e$ and $\pi$ across all subgroups.

\newpage
\section{Lemma}							\label{sec:lemma}

In this section, we introduce four lemmas that are used in the proof in Section \ref{sec:proof}.

\begin{lemma} \label{lemma:taubar}
	Let $\bI(\bA_i, \bX_i \con e)$ be
\begin{align*}
		\bI (\bA_i, \bX_i \con e)
	=
	\frac{1}{\NI_i}
	\begin{bmatrix}
	\ind(A_{i1} = 1) / e(1 \cond \bX_{i1}) 
	-
	\ind(A_{i1} = 0) / e(0 \cond \bX_{i1})
		\\
		\vdots
		\\
	\ind(A_{i\NI_i} = 1) / e(1 \cond \bX_{i\NI_i})
	-
	\ind(A_{i\NI_i} = 0) / e(0 \cond \bX_{i\NI_i})
	\end{bmatrix}
	 \ .
\end{align*}
Under conditions \hyperlink{Main-(A1)}{(A1)}-\hyperlink{Main-(A3)}{(A3)} and \hyperlink{Main-(M2)}{(M2)} of the main paper, and \hyperlink{(M1)'}{(M1)'} and \hyperlink{(E1)}{(E1)} in Section \ref{sec:ExistingMethod}, the following conditions hold.
\begin{itemize}
	\item[(a)] 
	The cluster size is bounded as $\NI_i \leq \NT$ for some integer $\NT$. The weight $w(\bC_i)$ satisfies $| w(\bC_i) | \leq C_w$ for all $\bC_i$ with some positive constant $C_w$.
	\item[(b)] We have $\EXP \big\{ \big\| \bY_i - \bg^*(\bA_i, \bX_i) \big\|_2^4 \cond \bA_i, \bX_i \big\} \leq C_4$ for all $\bA_i$, $\bX_i$, and for some some constant $C_4$. Furthermore, we have $ \big\| \Sigma(\bA_i, \bX_i) \big\|_2 \leq C_2$. Moreover, for all $\bX_i$, the smallest eigenvalue of $  \Sigma(\bA_i, \bX_i) $ is lower bounded by $C_2^{-1}$ for some $\bA_i$.
	\item[(c)] For all $\bA_i$, $\bX_i$, we have $ \big\| \widehat{\bg}^{(-k)}(\bA_i, \bX_i) - \bg^*(\bA_i, \bX_i) \big\|_2 \leq C_g$, $\big\| \bI(\bA_i, \bX_i \con e^*) \big\|_2 \leq C_e$, and $\big\| \bI(\bA_i, \bX_i \con \widehat{e}^{(-k)}) \big\|_2 \leq C_e$ for some constants $C_g$ and $C_e$. 

\end{itemize}
Additionally, the following conditions hold for all $k=1,2$ under \hyperlink{Main-(E2.PS)}{(E2.PS)}, \hyperlink{Main-(E2.OR)}{(E2.OR)}, and \hyperlink{Main-(E2.Both)}{(E2.Both)} of the main paper, respectively.
\begin{itemize}
	\item[(d.PS)] $ \big\| \bI(\bA_i, \bX_i \con \widehat{e}^{(-k)}) - \bI(\bA_i, \bX_i \con e^*) \big\|_ {P,2} = O_P(r_{e,\NC})$ with $r_{e,\NC} = o(1)$, and $ \big\| \widehat{\bg}^{(-k)}(\bA_i, \bX_i) - \bg'(\bA_i, \bX_i) \big\|_ {P,2} = O_P(r_{g,\NC})$ for some function $g'$ satisfying $| \bg^*(\bA_i, \bX_i) - \bg'(\bA_i, \bX_i) | \leq C_g$ with $r_{g,\NC} = o(1)$.
 	\item[(d.OR)] $ \big\| \bI(\bA_i, \bX_i \con \widehat{e}^{(-k)}) - \bI(\bA_i, \bX_i \con e') \big\|_ {P,2} = O_P(r_{e,\NC})$ for some function $e'$ satisfying $\| \bI(\bA_i, \bX_i \con e') \|_2 \leq C_e$ with $r_{e,\NC} = o(1)$, and $ \big\| \widehat{\bg}^{(-k)}(\bA_i, \bX_i) - \bg^*(\bA_i, \bX_i) \big\|_ {P,2} = O_P(r_{g,\NC})$ with $r_{g,\NC} = o(1)$.
 	\item[(d.Both)] $ \big\| \bI(\bA_i, \bX_i \con \widehat{e}^{(-k)}) - \bI(\bA_i, \bX_i \con e^*) \big\|_ {P,2} = O_P(r_{e,\NC})$ with $r_{e,\NC} = o(1)$, and $ \big\| \widehat{\bg}^{(-k)}(\bA_i, \bX_i) - \bg^*(\bA_i, \bX_i) \big\|_ {P,2} = O_P(r_{g,\NC})$ with $r_{g,\NC} = o(1)$ and $r_{e,\NC} r_{g,\NC} = o(\NC^{-1/2})$.
\end{itemize}

\end{lemma}

\newpage

\begin{lemma} \label{lemma:tauhat}
	Let $\bI(\bA_i, \bX_i \con \pi)$ be
\begin{align*}
	\bI (\bA_i, \bX_i \con \pi) &=
	\frac{1}{\NI_i}
	\begin{bmatrix}
	\ind(A_{i1} = 1) / \pi(1 \cond \bA_{i(-1)}, \bX_{i1}, \bX_{i(-1)}) 
	-
	\ind(A_{i1} = 0) / \pi(0 \cond \bA_{i(-1)}, \bX_{i1}, \bX_{i(-1)}) 
		\\
		\vdots
		\\
	\ind(A_{i\NI_i} = 1) / \pi(1 \cond \bA_{i(-\NI_i)}, \bX_{i \NI_i} , \bX_{i(-\NI_i)}) 
	-
	\ind(A_{i\NI_i} = 0) / \pi(0 \cond \bA_{i(-\NI_i)}, \bX_{i \NI_i} , \bX_{i(-\NI_i)}) 
	\end{bmatrix}
	 \ .
\end{align*}
Under conditions \hyperlink{Main-(A1)}{(A1)}-\hyperlink{Main-(A3)}{(A3)}, \hyperlink{Main-(M1)}{(M1)}-\hyperlink{Main-(M2)}{(M2)}, and \hyperlink{Main-(EN1)}{(EN1)} of the main paper, the following conditions hold.
\begin{itemize}
	\item[(a)] 
	The cluster size is bounded as $\NI_i \leq \NT$ for some integer $\NT$. The weight $w(\bC_i)$ satisfies $| w(\bC_i) | \leq C_w$ for all $\bC_i$ with some positive constant $C_w$.
	\item[(b)] We have $\EXP \big\{ \big\| \bY_i - \bg^*(\bA_i, \bX_i) \big\|_2^4 \cond \bA_i, \bX_i \big\} \leq C_4$ for all $\bA_i$, $\bX_i$, and for some some constant $C_4$. Furthermore, we have $ \big\| \Sigma(\bA_i, \bX_i) \big\|_2 \leq C_2$. Moreover, for all $\bX_i$, the smallest eigenvalue of $  \Sigma(\bA_i, \bX_i) $ is lower bounded by $C_2^{-1}$ for some $\bA_i$.
	\item[(c)] For all $\bA_i$, $\bX_i$, we have $ \big\| \widehat{\bg}^{(-k)}(\bA_i, \bX_i) - \bg^*(\bA_i, \bX_i) \big\|_2 \leq C_g$, $\big\| \bI(\bA_i, \bX_i \con \pi^*) \big\|_2 \leq C_\pi$, and $\big\| \bI(\bA_i, \bX_i \con \widehat{\pi}^{(-k)}) \big\|_2 \leq C_\pi$ for some constants $C_g$ and $C_\pi$. 

\end{itemize}
Additionally, the following conditions hold for all $k=1,2$ under \hyperlink{Main-(EN2.PS)}{(EN2.PS)}, \hyperlink{Main-(EN2.OR)}{(EN2.OR)}, and \hyperlink{Main-(EN2.Both)}{(EN2.Both)} of the main paper, respectively.
\begin{itemize}
	\item[(d.PS)] $ \big\| \bI(\bA_i, \bX_i \con \widehat{\pi}^{(-k)}) - \bI(\bA_i, \bX_i \con \pi^*) \big\|_ {P,2} = O_P(r_{\pi,\NC})$ with $r_{\pi,\NC} = o(1)$, and $ \big\| \widehat{\bg}^{(-k)}(\bA_i, \bX_i) - \bg'(\bA_i, \bX_i) \big\|_ {P,2} = O_P(r_{g,\NC})$ for some function $g'$ satisfying $\| \bg^*(\bA_i, \bX_i) - \bg'(\bA_i, \bX_i) \|_2 \leq C_g$ with $r_{g,\NC} = o(1)$.
 	\item[(d.OR)] $ \big\| \bI(\bA_i, \bX_i \con \widehat{\pi}^{(-k)}) - \bI(\bA_i, \bX_i \con \pi') \big\|_ {P,2} = O_P(r_{\pi,\NC})$ for some function $\pi'$ satisfying $\| \bI(\bA_i, \bX_i \con \pi') \|_2 \leq C_\pi$ with $r_{\pi,\NC} = o(1)$, and $ \big\| \widehat{\bg}^{(-k)}(\bA_i, \bX_i) - \bg^*(\bA_i, \bX_i) \big\|_ {P,2} = O_P(r_{g,\NC})$ with $r_{g,\NC} = o(1)$.
 	\item[(d.Both)] $ \big\| \bI(\bA_i, \bX_i \con \widehat{\pi}^{(-k)}) - \bI(\bA_i, \bX_i \con \pi^*) \big\|_ {P,2} = O_P(r_{\pi,\NC})$ with $r_{\pi,\NC} = o(1)$, and $ \big\| \widehat{\bg}^{(-k)}(\bA_i, \bX_i) - \bg^*(\bA_i, \bX_i) \big\|_ {P,2} = O_P(r_{g,\NC})$ with $r_{g,\NC} = o(1)$. Furthermore,  $r_{\pi,\NC} r_{g,\NC} = o(\NC^{-1/2})$.
\end{itemize}

Additionally, under \hyperlink{Main-(EN3)}{(EN3)}, we have 
\begin{itemize}
	\item[(e)] $\big\| \widehat{\beta}^{(k)}(\bC_i) - \beta^{\ddagger,(-k)}(\bC_i) \big\|_{P,2} = O_P(r_{\beta,\NC})$ where $r_{\beta,\NC} = o(1)$ with some function $\beta^{\ddagger,(-k)} \in \mathcal{B}$ that is fixed given $\II_k^c$. Moreover, there exists a fixed function $\beta'^{,(-k)}  \in \mathcal{B}$ so that $\big\| \beta^{\ddagger,(-k)} (\bC_i) - \beta'^{,(-k)}(\bC_i) \big\|_{P,2} = O_P(r_{\pi,\NC}) + O_P(r_{g,\NC})$. 
\end{itemize}

Lastly, under \hyperlink{Main-(EN4)}{(EN4)}, we have 
\begin{itemize}
		\item[(f)] $r_{\beta,\NC} r_{g,\NC} = o(\NC^{-1/2})$.
\end{itemize}

\end{lemma}

\newpage

\begin{lemma}						\label{lemma:consistency1}
	Suppose that conditions (a)-(d.Both) in Lemma \ref{lemma:tauhat} hold. For given $\mathcal{B}$, let $\beta^{\dagger,(-k)}$ and $\beta^*$ be the solutions to the following minimization problems \eqref{eq-defbeta0}  and \eqref{eq-303}, respectively.  
	\begin{align}
	\beta^{\dagger,(-k)}
	&	
	\in
	\argmin_{\beta \in \mathcal{B} }
	\EXP \Big\{ w(\bC_i)^2 \widehat{\bI}_i^{(-k),\intercal} B(\bC_i \con \beta) \widehat{\bm{S}}_i^{(-k)} B(\bC_i \con \beta) \widehat{\bI}_i^{(-k)}
	\, \Big| \, \II_k^c
	 \Big\} \ ,
	\label{eq-defbeta0}
	\\
		 	\beta^*
	&
	\in
	\argmin_{\beta \in \mathcal{B} }
	\EXP \Big\{ w(\bC_i)^2 \bI_i^{*,\intercal} B(\bC_i \con \beta) \bm{S}_i B(\bC_i \con \beta) \bI_i^* \Big\}
		\label{eq-303}
\end{align}	
where $\bI_i^* = \bI (\bA_i, \bX_i \con \pi^* )$, $ \widehat{\bI}_i^{(-k)} = \bI(\bA_i,\bX_i \con \widehat{\pi}^{(-k)})$, $\bg_i^*(\bA_i) = \bg^*(\bA_i, \bX_i)$, $\widehat{\bg}_i^{(-k)}(\bA_i) = \widehat{\bg}^{(-k)}(\bA_i,\bX_i)$, $\bm{S}_i = \big\{ \bY_i - \bg_i^*(\bA_i) \big\}^{\otimes 2}$, and $\widehat{\bm{S}}_i^{(-k)} = \big\{ \bY_i - \widehat{\bg}_i^{(-k)} (\bA_i) \big\}^{\otimes 2}$. Then, $\beta^{\ddagger,(-k)}$ and $\beta'^{,(-k)}$ in  condition (e) in Lemma \ref{lemma:tauhat} can be replaced with  $\beta^{\dagger,(-k)}$ and $\beta^*$, respectively, i.e., $\big\|
		\beta^{\dagger,(-k)} (\bC_i) -
			\beta^*(\bC_i) 
		\big\|_{P,2}
		= 
		O_P(r_{\pi,\NC}) + O_P(r_{g,\NC})$.
\end{lemma}

\begin{lemma}						\label{lemma:consistency2}
	Suppose that conditions (a)-(d.Both) in Lemma \ref{lemma:tauhat} hold. Let $\beta^{\dagger,(-k)}(\bC_i) = \sum_{\ell=1}^\MT \ind \{ L(\bC_i) = \ell \} \gamma_\ell^{\dagger,(-k)}$ be the solution to the minimization problem \eqref{eq-defbeta0} under $\paraB$. Additionally, suppose that $\bm{\gamma}^{\dagger,(-k)}$ belongs to the interior of $[-B_0,B_0]^{\otimes \MT}$. Then, condition (f) in Lemma \ref{lemma:tauhat} holds with $r_{\beta,\NC}=\NC^{-1/2}$, i.e., $\big\|
		\widehat{\beta}^{(k)} (\bC_i) - \beta^{\dagger,(-k)} (\bC_i)
	\big\|_{P,2} = O_P(\NC^{-1/2})$.
\end{lemma}

\newpage

\section{Proof of Lemmas and Theorems in the Main Paper}									\label{sec:proof}

\subsection{Proof of Theorem \ref{thm:1}}							\label{proof-thm:1}

We prove the theorem in order of (2) Asymptotic Normality and (1) Double Robustness. Note that the assumptions in Theorem \ref{thm:1} implies the conditions in Lemma \ref{lemma:taubar}. 

\subsubsection{Proof of (2) Asymptotic Normality}						\label{proof-thm:1-(2)}

				We use the following forms of $\widehat{\varphi}_k$ and $\varphi^*$ in the proof.
	\begin{align*}
		\widehat{\varphi}_k(\bO_i)
		&		
		=
		w(\bC_i) 
		\bigg[
		\bI(\bA_i, \bX_i \con \widehat{e}^{(-k)}) \big\{ \bY_i - \widehat{\bg}^{(-k)}(\bA_i, \bX_i) \big\}
		+
		\frac{\bm{1}\T}{\NI_i} 	\big\{ \widehat{\bg}^{(-k)}(\bm{1}, \bX_i) - \widehat{\bg}^{(-k)} (\bm{0}, \bX_i) \big\} 
		\bigg]
		\\
		&
		=
		w(\bC_i) 
		\bigg[
		\widehat{\bI}_i^{(-k),\intercal} \big\{ \bY_i - \widehat{\bg}_i^{(-k)}(\bA_i) \big\}
		+
		\frac{\bm{1}\T}{\NI_i} 	\big\{ \widehat{\bg}_i^{(-k)}(\bm{1}) - \widehat{\bg}_i^{(-k)} (\bm{0}) \big\} 
		\bigg]
		\ ,
		\\
		\varphi^*(\bO_i)
		&		
		=
		w(\bC_i) 
		\bigg[
		\bI(\bA_i, \bX_i \con e^*) \big\{ \bY_i - \bg^* (\bA_i, \bX_i) \big\}
		+
		\frac{\bm{1}\T}{\NI_i} 	\big\{ \bg^* (\bm{1}, \bX_i) - \bg^* (\bm{0}, \bX_i) \big\} 
		\bigg]
		\\
		& = 
		w(\bC_i) 
		\bigg[
		\bI_i^{*,\intercal} \big\{ \bY_i - \bg_i^* (\bA_i) \big\}
		+
		\frac{\bm{1}\T}{\NI_i} 	\big\{ \bg_i^* (\bm{1}) - \bg_i^* (\bm{0}) \big\} 
		\bigg]
		 \ .
	\end{align*}		
		We find $\sqrt{\NC}( \overline{\tau}_{\PARASUB} - \tau_{\PARASUB}^* )$ is decomposed as follows
	\begin{align}						\label{proof-AN-old-001}
		\sqrt{\NC} \big(  \overline{\tau}_{\PARASUB} - \tau_{\PARASUB}^* \big)
		& =
		\frac{1}{\sqrt{2} } \sum_{k=1}^2 \sqrt{ \frac{\NC}{2}} \big(  \overline{\tau}_{\PARASUB k} - \tau_{\PARASUB}^* \big)
		\nonumber
		\\
		&
		=
		\frac{1}{\sqrt{2}}
		\sum_{k=1}^2
		\frac{1}{\sqrt{\NC/2}} \sum_{i \in \II_k} \Big\{
			\widehat{\varphi}_k(\bO_i) - \tau_{\PARASUB}^*
		\Big\}
		\nonumber
		\\
		& = 
		\frac{1}{\sqrt{2}}
		\sum_{k=1}^2
		\Bigg[
		\frac{1}{\sqrt{\NC/2}} \sum_{i \in \II_k} \Big\{ \widehat{\varphi}_k(\bO_i) - \varphi^*(\bO_i) \Big\}
		+
		\frac{1}{\sqrt{\NC/2}} \sum_{i \in \II_k} \Big\{ \varphi^*(\bO_i)  - \tau_{\PARASUB}^* \Big\} 
		\Bigg]  \ .
	\end{align}
	The second term in the bracket is trivially satisfies the asymptotic Normality from the central limit theorem so it suffices to show the first term the bracket is $o_P(1)$, which is decomposed into $[B]$ and $[C]$ as follows.
	\begin{align}								\label{proof-AN-old-002}
		&
		\frac{1}{\sqrt{\NC/2}} \sum_{i \in \II_k} \Big\{ \widehat{\varphi}_k(\bO_i) - \varphi^*(\bO_i) \Big\}
		\\
		& =
		\underbrace{
		\frac{1}{\sqrt{\NC/2}} \sum_{i \in \II_k} \bigg[ 
		\Big\{ \widehat{\varphi}_k(\bO_i) - \varphi^*(\bO_i) \Big\}
		-
		\EXP \Big\{ \widehat{\varphi}_k(\bO_i) - \varphi^*(\bO_i) \, \Big| \, \II_k^c \Big\}
		\bigg]
		}_{[B]}
		+
		\sqrt{ \frac{\NC}{2} } \underbrace{
		\EXP \Big\{ \widehat{\varphi}_k(\bO_i) - \varphi^*(\bO_i) \, \Big| \, \II_k^c \Big\}
		}_{[C]} \ . 
		\nonumber		
	\end{align}	

	We show that $[B] = O_P( r_{e,\NC}^2)  + O_P(r_{g,\NC}^2 ) $ and $[C] =O_P(r_{e,\NC}r_{g,\NC})$.
	
	\begin{itemize}[leftmargin=0.5cm]
		\item \textbf{(Rate of $[B]$)}:  The squared expectation of $[B]$ is
	\begin{align}																	\label{proof-AN-old-003}
		\EXP \big\{ [B]^2 \cond \II_k^c \big\}
		&
		=
		\frac{1}{\NC/2} \sum_{i \in \II_k} \EXP \Big[ \big\{  \widehat{\varphi}_k(\bO_i) - \varphi^*(\bO_i) \big\}^2 \, \Big| \, \II_k^c \Big]
		=
		 \EXP \Big[ \big\{ \widehat{\varphi}_k(\bO_i) - \varphi^*(\bO_i) \big\}^2  \, \Big| \, \II_k^c \Big]  \ .
	\end{align}
	We find $ \widehat{\varphi}_k(\bO_i) - \varphi^*(\bO_i)$ is represented as
	\begin{align*}
		&
		 \widehat{\varphi}_k(\bO_i) - \varphi^*(\bO_i)
		\\
		& = 
		w(\bC_i) \bigg[
		\widehat{\bI}_i^{(-k),\intercal} \big\{ \bY_i - \widehat{\bg}_i^{(-k)}(\bA_i) \big\}
		-
		\bI_i^{*,\intercal} \big\{ \bY_i - \bg_i^*(\bA_i) \big\}
		+ \frac{\bm{1}\T}{\NI_i} \Big\{ \widehat{\bg}_i^{(-k)} (\bm{1}) -  \widehat{\bg}_i^{(-k)} (\bm{0})  - \bg_i^* (\bm{1}) + \bg_i^* (\bm{0}) \Big\}
		\bigg]
		\\
		& = 
		\frac{w(\bC_i)}{2} 
		\Big\{ \widehat{\bI}_i^{(-k)} - \bI_i^* \Big\}\T
		\Big\{ 2\bY_i - \widehat{\bg}_i^{(-k)}(\bA_i) - \bg_i^*(\bA_i) \Big\}
		+
		\frac{w(\bC_i)}{2} 
		\Big\{ \widehat{\bI}_i^{(-k)} + \bI_i^* \Big\}\T
		\Big\{ \widehat{\bg}_i^{(-k)}(\bA_i) - \bg_i^*(\bA_i)  \Big\}
		\\
		& \hspace*{1cm}
		+ \frac{w(\bC_i) \bm{1}\T}{\NI_i} \Big\{ \widehat{\bg}_i^{(-k)} (\bm{1})  - \bg_i^* (\bm{1})  \Big\}
		-  \frac{w(\bC_i) \bm{1}\T}{\NI_i} \Big\{ \widehat{\bg}_i^{(-k)} (\bm{0})  - \bg_i^* (\bm{0})  \Big\} \ .
	\end{align*}
	We use the inequality $(a_1+\ldots+a_4)^2 \leq 8 (a_1^2 + \ldots+a_4^2)$ and $\big\| A B \big\|_2 \leq \big\|A \big\|_2 \big\|B\big\|_2$ to obtain an upper bound of $\big\{  \widehat{\varphi}_k(\bO_i) - \varphi^*(\bO_i) \big\}^2$ as follows.
	\begin{align}							\label{proof-AN-old-004}
		\big\{  \widehat{\varphi}_k(\bO_i) - \varphi^*(\bO_i) \big\}^2
		& \leq
		2 w(\bC_i)^2
		\Bigg[
		\Big\| \widehat{\bI}_i^{(-k)} - \bI_i^* \Big\|_2^2
		\Big\| 2\bY_i - \widehat{\bg}_i^{(-k)}(\bA_i) - \bg_i^*(\bA_i) \Big\|_2^2
		\\
		& \hspace*{3cm}
		+
		\Big\| \widehat{\bI}_i^{(-k)}  + \bI_i^*  \Big\|_2^2
		\Big\| \widehat{\bg}_i^{(-k)}(\bA_i) - \bg_i^*(\bA_i) \Big\|_2^2
		\nonumber
		\\
		& \hspace*{3cm}
		+ 
		\frac{4}{\NI_i}
		\Big\| \widehat{\bg}_i^{(-k)} (\bm{1})  - \bg_i^* (\bm{1}) \Big\|_2^2
		+ 
		\frac{4}{\NI_i}
		\Big\| \widehat{\bg}_i^{(-k)} (\bm{0})  - \bg_i^* (\bm{0}) \Big\|_2^2
		\Bigg] \ .
		\nonumber
	\end{align}
	Since $|w(\bC_i)|\leq C_w$, it is sufficient to study the asymptotic behavior of the following terms.
	\begin{align}
		&
		\EXP \bigg\{
			\Big\| \widehat{\bI}_i^{(-k)} - \bI_i^* \Big\|_2^2
		\Big\| 2\bY_i - \widehat{\bg}_i^{(-k)}(\bA_i) - \bg_i^*(\bA_i) \Big\|_2^2
		\, \bigg| \, \II_k^c \bigg\}
		\label{proof-old-gap1}
		\\
		&
		\EXP \bigg\{
		\Big\| \widehat{\bI}_i^{(-k)}  + \bI_i^*  \Big\|_2^2
		\Big\| \widehat{\bg}_i^{(-k)}(\bA_i) - \bg_i^*(\bA_i) \Big\|_2^2
		\, \bigg| \, \II_k^c \bigg\}
		\label{proof-old-gap2}
		\\
		&
		\EXP \bigg\{
		\Big\| \widehat{\bg}_i^{(-k)} (\ba)  - \bg_i^* (\ba) \Big\|_2^2
		\, \bigg| \, \II_k^c \bigg\}
		\ , \ba=\bm{1},\bm{0} \ .
		\label{proof-old-gap3}
	\end{align}
	
	First, we find $\EXP \big\{ \| \widehat{\bI}_i^{(-k)} - \bI_i^*\|_2^2 \cond \II_k^c \big\} = O_P(r_{e,\NC}^2)$ and $\| \widehat{\bI}_i^{(-k)} + \bI_i^*\|_2^2 \leq 4C_e^2$ from conditions (c) and (d) of Lemma \ref{lemma:taubar}. 
	
	Second, we have $\EXP\big\{ \big\| \widehat{\bg}_i^{(-k)}(\bA_i) - \bg_i^*(\bA_i) \big\|_2^2 \cond \II_k^c \big\} = O_P(r_{g,\NC}^2)$ from condition (d) of Lemma \ref{lemma:taubar}. Moreover, $  \EXP \big\{ \big\|2\bY_i - \widehat{\bg}_i^{(-k)}(\bA_i) - \bg_i^*(\bA_i) \big\|_2^2 \cond \II_k^c \big\} $ is upper bounded by a constant as follows.
	\begin{align*}
		&
	\EXP \big\{  \big\|2\bY_i - \widehat{\bg}_i^{(-k)}(\bA_i) - \bg_i^*(\bA_i) \big\|_2^2 \cond \II_k^c \big\} 
	\\
	&
	=
	\EXP \Big[ \EXP \big\{  \big\|2\bY_i - \widehat{\bg}_i^{(-k)}(\bA_i) - \bg_i^*(\bA_i) \big\|_2^2 \cond \bA_i , \bX_i ,  \II_k^c \big\} \, \Big| \, \II_k^c \Big]
	\\
	&
	= \EXP \Big[ 4 \bm{\epsilon}_i\T \bm{\epsilon}_i  + 4 \bm{\epsilon}_i\T \big\{ \widehat{\bg}_i^{(-k)}(\bA_i) - \bg_i^*(\bA_i) \big\} + \big\| \widehat{\bg}_i^{(-k)}(\bA_i) - \bg_i^*(\bA_i) \big\|_2^2
	\, \Big| \, \II_k^c \Big]
	\\
	&
	= \EXP \Big[ 4 \bm{1}\T \Sigma(\bA_i, \bX_i) \bm{1}  + \big\| \widehat{\bg}_i^{(-k)}(\bA_i) - \bg_i^*(\bA_i) \big\|_2^2
	\, \Big| \, \II_k^c \Big]
	\\
	& 
	\leq \NT C' + C_g^2 \ .
	\end{align*}
	The three equalities are straightforward and the last inequality is from conditions (b) and (c) of Lemma \ref{lemma:taubar}. 
	
	Finally, under condition (d), we find 
	\begin{align*}
		&
		\EXP \Big\{
		\Big\| \widehat{\bg}_i^{(-k)} (\bA_i)  - \bg_i^* (\bA_i) \Big\|_2^2
		\, \Big| \, \II_k^c \Big\}
		\\
		&
		=
		\sum_{m=1}^\NT
		\Bigg[
		P(\NI_i = m) 
		\EXP \Big\{
		\Big\| \widehat{\bg}_i^{(-k)} (\bA_i)  - \bg_i^* (\bA_i) \Big\|_2^2
		\, \Big| \, 
		\NI_i = m , 
		\II_k^c
		\Big\}
		\Bigg]
		\\
		& 
		= 
		\sum_{m=1}^\NT
		\Bigg[
		P(\NI_i = m) 
		\EXP \bigg[
			\sum_{\ba_i \in \{0,1\}^m }
			P(\bA_i = \ba_i \cond \bX_i, \NI_i = m)
			\\
			&
			\hspace*{4cm}
			\times
			\EXP \Big\{
		\Big\| \widehat{\bg}_i^{(-k)} (\ba_i)  - \bg_i^* (\ba_i) \Big\|_2^2
		\, \Big| \, 
		\bA_i = \ba_i ,
		\bX_i ,
		\NI_i = m , 
		\II_k^c
		\Big\}
		\, \bigg| \, \NI_i=m, \II_k^c
		\bigg]
		\Bigg] \ .
	\end{align*}
	From the positivity assumption, $P(\bA_i = \ba_i \cond \bX_i, \NI_i = m) \geq \delta$ for some constant $\delta$, especially at $\ba_i=\bm{1}$. Therefore, $\EXP \big\{
		\big\| \widehat{\bg}_i^{(-k)} (\bA_i)  - \bg_i^* (\bA_i) \big\|_2^2
		\cond \II_k^c \big\}$ is lower bounded by
		\begin{align*}
		&
			\EXP \Big\{
		\Big\| \widehat{\bg}_i^{(-k)} (\bA_i)  - \bg_i^* (\bA_i) \Big\|_2^2
		\, \Big| \, \II_k^c \Big\}
		\\
		& 
		\geq 
		\delta
		\sum_{m=1}^\NT
		\Bigg[
		P(\NI_i = m) 
		\EXP \bigg[			
			\EXP \Big\{
		\Big\| \widehat{\bg}_i^{(-k)} (\bm{1})  - \bg_i^* (\bm{1}) \Big\|_2^2
		\, \Big| \, 
		\bA_i = \bm{1} ,
		\bX_i ,
		\NI_i = m , 
		\II_k^c
		\Big\}
		\, \bigg| \, \NI_i=m, \II_k^c
		\bigg]
		\Bigg] 
		\\
		&
		=
		\delta
		\sum_{m=1}^\NT
		\Bigg[
		P(\NI_i = m) 
		\EXP \bigg\{			\Big\| \widehat{\bg}_i^{(-k)} (\bm{1})  - \bg_i^* (\bm{1}) \Big\|_2^2
		\, \bigg| \, \NI_i=m, \II_k^c
		\bigg\}
		\Bigg]
		\\
		&
		=
		\delta
		\EXP \Big\{			\Big\| \widehat{\bg}_i^{(-k)} (\bm{1})  - \bg_i^* (\bm{1}) \Big\|_2^2
		\, \Big| \, \II_k^c
		\Big\} \ .
		\end{align*}
		Since $\EXP \big\{
		\big\| \widehat{\bg}_i^{(-k)} (\bA_i)  - \bg_i^* (\bA_i) \big\|_2^2
		\cond \II_k^c \big\} = O_P(r_{g,\NC}^2)$, we obtain $\EXP \big\{ \big\| \widehat{\bg}_i^{(-k)} (\bm{1})  - \bg_i^* (\bm{1}) \big\|_2^2	\cond \II_k^c 		\big\} =O_P(r_{g,\NC}^2)$. Similarly, $\EXP \big\{ \big\| \widehat{\bg}_i^{(-k)} (\bm{0})  - \bg_i^* (\bm{0}) \big\|_2^2	\cond \II_k^c 		\big\} =O_P(r_{g,\NC}^2)$.

	Using the established results, we find the convergence rates of \eqref{proof-old-gap1}-\eqref{proof-old-gap3}. First, the rate of \eqref{proof-old-gap1} is $O_P(r_{e,\NC}^2)$.
	\begin{align*}
		&
		\EXP \bigg\{
			\Big\| \widehat{\bI}_i^{(-k)} - \bI_i^* \Big\|_2^2
		\Big\| 2\bY_i - \widehat{\bg}_i^{(-k)}(\bA_i) - \bg_i^*(\bA_i) \Big\|_2^2
		\, \bigg| \, \II_k^c \bigg\}
		\\
		&
		=
		\EXP \Bigg[
			\Big\| \widehat{\bI}_i^{(-k)} - \bI_i^* \Big\|_2^2
		\EXP \bigg\{
		\Big\| 2\bY_i - \widehat{\bg}_i^{(-k)}(\bA_i) - \bg_i^*(\bA_i) \Big\|_2^2
		\, \bigg| \, \bA_i, \bX_i, \II_k^c \bigg\}
		\, \Bigg| \, \II_k^c \Bigg]
		\\
		& 
		\leq
		4C_B^2 ( \NT C' + C_g^2 ) \EXP \Big\{ \Big\| \widehat{\bI}_i^{(-k)} - \bI_i^* \Big\|_2^2 \, \Big| \,  \II_k^c \Big\} 
		= O_P(r_{e,\NC}^2) \ .
	\end{align*}
	Second, the rate of \eqref{proof-old-gap2} is $O_P(r_{g,\NC}^2)$.
	\begin{align*}
		&
		\EXP \bigg\{
		\Big\| \widehat{\bI}_i^{(-k)}  + \bI_i^*  \Big\|_2^2
		\Big\| \widehat{\bg}_i^{(-k)}(\bA_i) - \bg_i^*(\bA_i) \Big\|_2^2
		\, \bigg| \, \II_k^c \bigg\}
		\leq
		4C_e^2 C_B^2
		\EXP \bigg\{
		\Big\|  \widehat{\bg}_i^{(-k)}(\bA_i) - \bg_i^*(\bA_i)  \Big\|_2^2
		\, \bigg| \, \II_k^c \bigg\} = 
		O_P(r_{g,\NC}^2)  \ .
	\end{align*}
	Lastly, \eqref{proof-old-gap3} is $O_P(r_{g,\NC}^2)$ from the established result. As a consequence, by plugging in the rate in \eqref{proof-AN-old-004}, we have $\EXP ( [B]^2 \cond \II_k^c \big) =O_P( r_{e,\NC}^2)  + O_P(r_{g,\NC}^2 ) $. Moreover, this implies $[B] = O_P( r_{e,\NC}^2)  + O_P(r_{g,\NC}^2 ) $ from Lemma 6.1 of \citet{Victor2018}. 
	
	\item \textbf{(Rate of $[C]$)}: $[C]$ is represented as 
		\begin{align*}
		[C]
		&
		=
		\EXP \Big\{ \widehat{\varphi}_k(\bO_i) - \varphi^*(\bO_i) \, \Big| \, \II_k^c \Big\}
		\\
		&
		=
		\EXP \Bigg[ w(\bC_i) \bigg[
		\widehat{\bI}_i^{(-k),\intercal}  \Big\{ \bY_i - \widehat{\bg}_i^{(-k)}(\bA_i) \Big\}
	+
	\frac{\bm{1}\T}{\NI_i} \Big\{ \widehat{\bg}_i^{(-k)}(\bm{1}) - \widehat{\bg}_i^{(-k)}(\bm{0}) \Big\} 		
	-
	\frac{\bm{1}\T}{\NI_i} \Big\{ \bg_i^*(\bm{1}) - \bg_i^*(\bm{0}) \Big\} 		
	\bigg]
	\, \Bigg| \, \II_k^c \Bigg] \ .
		\end{align*}	
	The first term in the second line is decomposed into 
	\begin{align}								\label{proof-AN-old-005}
	\EXP \Big[
	w(\bC_i) \widehat{\bI}_i^{(-k),\intercal}  \big\{ \bY_i - \widehat{\bg}_i^{(-k)}(\bA_i) \big\}  \, \Big| \, \II_k^c \Big]
	& = 
	\EXP \Big[ w(\bC_i) \bI_i^{*,\intercal} \big\{ \bg_i^*(\bA_i) - \widehat{\bg}_i^{(-k)}(\bA_i) \big\} \, \Big| \, \II_k^c \Big]
	\\
	& \hspace*{1cm}
	+
	\EXP \Big[ w(\bC_i) \big\{ \widehat{\bI}_i^{(-k)} -\bI_i^* \big\} \T \big\{ \bg_i^*(\bA_i) - \widehat{\bg}_i^{(-k)}(\bA_i) \big\} \, \Big| \, \II_k^c \Big] \ .
	\nonumber
	\end{align}
	From some algebra, the first term of \eqref{proof-AN-old-005} is equivalent as
	\begin{align*}
		&
		\EXP \Big[ w(\bC_i) \bI_i^{*,\intercal} \big\{ \bg_i^*(\bA_i) - \widehat{\bg}_i^{(-k)}(\bA_i) \big\} \, \Big| \, \II_k^c \Big]
		=
		\EXP \Bigg[
		\frac{w(\bC_i)  \bm{1}\T}{\NI_i} \Big\{ \bg_i^*(\bm{1}) - \bg_i^*(\bm{0}) 
		- \widehat{\bg}_i^{(-k)}(\bm{1}) + \widehat{\bg}_i^{(-k)}(\bm{0}) \Big\} 	
	\, \Bigg| \, \II_k^c \Bigg] \ .
	\end{align*}
	The second term of \eqref{proof-AN-old-005} is 
	\begin{align*}
		&
		\bigg| 
		\EXP \Big[ w(\bC_i) \big\{ \widehat{\bI}_i^{(-k)} -\bI_i^* \big\}\T \big\{ \bg_i^*(\bA_i) - \widehat{\bg}_i^{(-k)}(\bA_i) \big\} \, \Big| \, \II_k^c \Big]
		\bigg|
		\\
		&
		\leq		
		C_w
		\big\| \widehat{\bI}_i^{(-k)} -\bI_i^* \big\|_{P,2}  \big\| \bg_i^*(\bA_i) - \widehat{\bg}_i^{(-k)}(\bA_i) \big\|_{P,2}
		= O_P(r_{e,\NC}r_{g,\NC}) \ .
	\end{align*}
	The inequality holds from the H\"older's inequality. This implies $[C] =O_P(r_{e,\NC}r_{g,\NC}) $.
	\end{itemize}

We use the established rates of $[B]$ and $[C]$ in \eqref{proof-AN-old-002} which leads to the following result.
\begin{align}							\label{proof-AN-old-006}
	\frac{1}{\sqrt{\NC/2}} \sum_{i \in \II_k} \Big\{  \widehat{\varphi}_k(\bO_i) - \varphi^*(\bO_i) \Big\}
	&
	=
	[B]
	+
	\sqrt{\frac{\NC}{2}} [C]
	\nonumber
	\\
	&
	=
	O_P( r_{e,\NC} ) + O_P( r_{g,\NC} )
	+
	\sqrt{\NC} O_P(r_{e,\NC} r_{g,\NC} ) 
	=
	o_P(1)  \ .
\end{align}
From \eqref{proof-AN-old-001}, we find
\begin{align}							\label{proof-AN-old-007}
	\sqrt{\NC} \big(  \overline{\tau}_{\PARASUB} - \tau_{\PARASUB}^* \big)
	&=
	\frac{1}{\sqrt{\NC}} \Big\{ \varphi^*(\bO_i)  - \tau_{\PARASUB}^* \Big\}  + o_P(1)
	\stackrel{D}{\rightarrow}
	N \Big( 0 , \VAR\big\{ \varphi^*(\bO_i) \big\} \Big)  \ .
\end{align}

Lastly, we show that the variance estimator is consistent. Let $\overline{\sigma}_k^2 = (\NC/2)^{-1} \sum_{i \in \II_k} \{
			\widehat{\varphi}_k(\bO_i) - \overline{\tau}_{\PARASUB k} \}^2 $. We decompose $\overline{\sigma}_k^2 -  \VAR\{\varphi^*(\bO_i) \} $ as $\overline{\sigma}_k^2 - S_k^2 + S_k^2 - \VAR\{\varphi^*(\bO_i) \}$ where $S_k^2 =
		(2/\NC)  \sum_{i \in \II_k} \big\{ \varphi^*(\bO_i)  - \tau_{\PARASUB}^* \big\}^2$.
	From the law of large numbers, we have $S_k^2 - \VAR\{\varphi^*(\bO_i) \} = o_P(1)$ so it is sufficient to show $\overline{\sigma}_k^2 - S_k^2 = o_P(1)$ which is represented as follows.
	\begin{align*}
		&
		\overline{\sigma}_k^2 - S_k^2
		\\
		&
		=
		\frac{1}{\NC/2} \sum_{i \in \II_k} 
		\Big\{
			\widehat{\varphi}_k(\bO_i) - \overline{\tau}_{\PARASUB k}
		\Big\}^2
		-
		\frac{1}{\NC/2} \sum_{i \in \II_k} \Big\{  \varphi^*(\bO_i)  - \tau_{\PARASUB}^* \Big\}^2
		\\
		& = 
		\frac{1}{\NC/2} \sum_{i \in \II_k} 
		\Big\{
			\widehat{\varphi}_k(\bO_i) - \overline{\tau}_{\PARASUB k}
			-
			\varphi^*(\bO_i)  + \tau_{\PARASUB}^* 
		\Big\}
		\Big\{
			\widehat{\varphi}_k(\bO_i) - \overline{\tau}_{\PARASUB k}
			-
			\varphi^*(\bO_i)  + \tau_{\PARASUB}^* 
			 +
			2  	\varphi^*(\bO_i)  -2 \tau_{\PARASUB}^* 
		\Big\}
		\\
		& =
		\underbrace{ 
		\frac{1}{\NC/2} \sum_{i \in \II_k} 
		\Big\{
			\widehat{\varphi}_k(\bO_i) - \overline{\tau}_{\PARASUB k}
			-
			\varphi^*(\bO_i)  + \tau_{\PARASUB}^* 
		\Big\}^2
		}_{V_\NC}
		+
		\frac{2}{\NC/2} \sum_{i \in \II_k} 
		\Big\{
		\widehat{\varphi}_k(\bO_i) - \overline{\tau}_{\PARASUB k}
			-
			\varphi^*(\bO_i)  + \tau_{\PARASUB}^* 
		\Big\} \Big\{ \varphi^*(\bO_i) - \tau_{\PARASUB}^* \Big\}
		\\
		& 
		\leq V_\NC + 2 \sqrt{V_\NC} S_k^2 \ .
	\end{align*}
	The inequality holds from the H\"older's inequality. Since $S_k^2 = \sigma^2 + o_P(1) = O_P(1)$, it suffices to show that $V_\NC = o_P(1)$.
	
	We observe that $V_\NC$ is upper bounded by
	\begin{align}								\label{proof-AN-old-101}
		V_\NC
		\leq 
		\frac{2}{\NC/2} \sum_{i \in \II_k} \Big\{ \widehat{\varphi}_k(\bO_i ) - \varphi^* (\bO_i  ) \Big\}^2
		+
		\Big(  \overline{\tau}_{\PARASUB k} - \tau_{\PARASUB}^*  \Big)^2 \ .
	\end{align}
		From \eqref{proof-AN-old-006}, the first term of \eqref{proof-AN-old-101} is $o_P(1)$. The second term is also $o_P(1)$ from \eqref{proof-AN-old-007}. Therefore, $S_k^2 - \VAR\{\varphi^*(\bO_i) \} = o_P(1)$ and $\overline{\sigma}_k^2 -  \VAR\{\varphi^*(\bO_i) \} = o_P(1)$. This concludes the proof.

\subsubsection{Proof of (1) Double Robustness}						\label{proof-thm:1-(1)}

The outline of the proof is given as follows. In \textbf{[Case PS]} and \textbf{[Case OR]}, we assume \hyperlink{Main-(E2.PS)}{(E2.PS)} (i.e., $e' = e^*$) and \hyperlink{Main-(E2.OR)}{(E2.OR)} (i.e., $g'=g^*$), respectively. The proof is similar to the proof of Theorem  \ref{thm:1}-(b) in Section \ref{proof-thm:1-(2)} except we consider the empirical mean instead of $\sqrt{\NC}$-scaled empirical mean.

\begin{itemize}[leftmargin=0.5cm]
	\item \textbf{[Case PS]}: We assume $e' = e^*$. Let $\varphi'(\bO_i)$ be
	\begin{align*}
		\varphi' (\bO_i)
		=
		w(\bC_i) 
		\bigg[
		\bI_i^{*,\intercal} \big\{ \bY_i - \bg_i' (\bA_i) \big\}
		+
		\frac{\bm{1}\T}{\NI_i} 	\big\{ \bg_i' (\bm{1}) - \bg_i' (\bm{0}) \big\} 
		\bigg]
		  \ .
	\end{align*}
	To study $\EXP \big\{ \varphi'(\bO_i) \big\}$, we first find that the expectation of $w(\bC_i) \bI_i^{*,\intercal} \big\{ \bY_i - \bg_i' (\bA_i) \big\}$.
	\begin{align}							\label{proof-DR-1-001}
	\nonumber
		\EXP \Big[ w(\bC_i) \bI_i^{*,\intercal} \big\{ \bY_i - \bg_i' (\bA_i) \big\} \Big]
		&
		=
		\EXP
		\Bigg[
			\frac{w(\bC_i)}{\NI_i}
			\sum_{j=1}^{\NI_i}
			\bigg\{
			\frac{ \ind(A_{ij} = 1) }{ e^*(1 \cond \bX_{ij} )} - \frac{ \ind(A_{ij} = 0) }{e^*(0 \cond \bX_{ij} )}
	\bigg\}
			\Big\{ g^*(A_{ij}, \bX_{ij} ) - g'(A_{ij}, \bX_{ij} ) \Big\}
		\Bigg]
		\\
		\nonumber
		&
		=
		\EXP
		\Bigg[
			\frac{w(\bC_i)}{\NI_i}
			\sum_{j=1}^{\NI_i}
			\bigg[
				\Big\{ g^*(1, \bX_{ij} ) - g^*(0, \bX_{ij} )  \Big\} - \Big\{ g'(1, \bX_{ij} ) - g'(0, \bX_{ij} ) \Big\}
			\bigg]
		\Bigg]
		\\
		&
		=
		\EXP
		\Bigg[
			\frac{w(\bC_i) \bm{1}\T }{\NI_i}
			\bigg[ \Big\{ \bg_i^* (\bm{1}) - \bg_i^* (\bm{0})  \Big\} - \Big\{ \bg_i' (\bm{1}) - \bg_i' (\bm{0})  \Big\} 
			\bigg]
		\Bigg] \ .
	\end{align}
	Therefore, the expectation of $\varphi'(\bO_i)$ is $\tau_{\PARASUB}^*$.
	\begin{align*}
		\EXP \big\{ \varphi'(\bO_i) \big\}
		&
		=
		\EXP 
		\Bigg[
			w(\bC_i) 
		\bigg[
		\bI_i^{*,\intercal} \big\{ \bY_i - \bg_i' (\bA_i) \big\}
		+
		\frac{\bm{1}\T}{\NI_i} 	\big\{ \bg_i' (\bm{1}) - \bg_i' (\bm{0}) \big\} 
		\bigg]
		\Bigg]
		=
		\EXP
		\Bigg[
			\frac{w(\bC_i) \bm{1}\T }{\NI_i}
			\Big\{ \bg_i^* (\bm{1}) - \bg_i^* (\bm{0})  \Big\}
		\Bigg]  = \tau_{\PARASUB}^*
		\ . 
	\end{align*}

	We find that the proof in Section \ref{proof-thm:1-(2)} now involves with $\varphi'(\bO_i)$ instead of $\varphi^*(\bO_i)$. Specifically, \eqref{proof-AN-old-001} divided by $\sqrt{\NC}$ becomes
	\begin{align*}
		\overline{\tau}_{\PARASUB} - \tau_{\PARASUB}^*
		& = 
		\frac{1}{2}
		\sum_{k=1}^2
		\frac{1}{\NC/2} \sum_{i \in \II_k} \Big\{ \widehat{\varphi}_k(\bO_i) - \varphi'(\bO_i) \Big\}
		+
		\frac{1}{2} \sum_{k=1}^2
		\frac{1}{\NC/2} \sum_{i \in \II_k} \Big\{ \varphi'(\bO_i)  - \tau_{\PARASUB}^* \Big\} 
		\\
		& =
		\frac{1}{2}
		\sum_{k=1}^2
		\EXP \Big\{ \widehat{\varphi}_k(\bO_i) - \varphi'(\bO_i) \, \Big| \, \II_k^c \Big\}
		+ o_P(1)  \ . 
	\end{align*}
	The second equality is from the law of large numbers. The first term of the right hand side is represented as 
		\begin{align*}
		&
		\EXP \Big\{ \widehat{\varphi}_k(\bO_i) - \varphi'(\bO_i) \, \Big| \, \II_k^c \Big\}
		\\
		&
		=
		\EXP \Bigg[ w(\bC_i) \bigg[
		\widehat{\bI}_i^{(-k),\intercal}  \Big\{ \bY_i - \widehat{\bg}_i^{(-k)}(\bA_i) \Big\}
	+
	\frac{\bm{1}\T}{\NI_i} \Big\{ \widehat{\bg}_i^{(-k)}(\bm{1}) - \widehat{\bg}_i^{(-k)}(\bm{0}) \Big\} 		
	-
	\frac{\bm{1}\T}{\NI_i} \Big\{ \bg_i^*(\bm{1}) - \bg_i^*(\bm{0}) \Big\} 		
	\bigg]
	\, \Bigg| \, \II_k^c \Bigg] \ .
	\nonumber
		\end{align*}	
	The first term in the second line is decomposed into 
	\begin{align}								\label{proof-DR-1-002}
	\EXP \Big[
	w(\bC_i) \widehat{\bI}_i^{(-k),\intercal}  \big\{ \bY_i - \widehat{\bg}_i^{(-k)}(\bA_i) \big\}  \, \Big| \, \II_k^c \Big]
	& = 
	\EXP \Big[ w(\bC_i) \bI_i^{*,\intercal} \big\{ \bg_i^*(\bA_i) - \widehat{\bg}_i^{(-k)}(\bA_i) \big\} \, \Big| \, \II_k^c \Big]
	\\
	& \hspace*{1cm}
	+
	\EXP \Big[ w(\bC_i) \big\{ \widehat{\bI}_i^{(-k)} -\bI_i^* \big\} \T \big\{ \bg_i^*(\bA_i) - \widehat{\bg}_i^{(-k)}(\bA_i) \big\} \, \Big| \, \II_k^c \Big] \ .
	\nonumber
	\end{align}
	From \eqref{proof-DR-1-001}, the first term of \eqref{proof-DR-1-002} is equivalent as
	\begin{align*}
		&
		\EXP \Big[ w(\bC_i) \bI_i^{*,\intercal} \big\{ \bg_i^*(\bA_i) - \widehat{\bg}_i^{(-k)}(\bA_i) \big\} \, \Big| \, \II_k^c \Big]
		=
		\EXP \Bigg[
		\frac{w(\bC_i) \bm{1}\T}{\NI_i} \Big\{ \bg_i^*(\bm{1}) - \bg_i^*(\bm{0}) - \widehat{\bg}_i^{(-k)}(\bm{1}) + \widehat{\bg}_i^{(-k)}(\bm{0}) \Big\} 		
	\, \Bigg| \, \II_k^c \Bigg] \ .
	\end{align*}
	The second term of \eqref{proof-DR-1-002} is 
	\begin{align*}
		\bigg| 
		\EXP \Big[ w(\bC_i) \big\{ \widehat{\bI}_i^{(-k)} -\bI_i^* \big\}\T \big\{ \bg_i^*(\bA_i) - \widehat{\bg}_i^{(-k)}(\bA_i) \big\} \, \Big| \, \II_k^c \Big]
		\bigg|
		&
		\leq		
		C_w
		\big\| \widehat{\bI}_i^{(-k)} -\bI_i^* \big\|_{P,2}  \big\| \bg_i^*(\bA_i) - \widehat{\bg}_i^{(-k)}(\bA_i) \big\|_{P,2}
		\\
		&
		\leq		
		C_w C_g
		\big\| \widehat{\bI}_i^{(-k)} -\bI_i^* \big\|_{P,2} 
		= O_P(r_{e,\NC}) \ .
	\end{align*}
	The inequality holds from the H\"older's inequality. This implies $\EXP \big\{ \widehat{\varphi}_k(\bO_i) - \varphi'(\bO_i) \cond \II_k^c \big\} = O_P(r_{e,\NC})= o_P(1)$, so does $\overline{\tau}_{\PARASUB} - \tau_{\PARASUB}^*$. This concludes the proof. 
	
	\item \textbf{[Case OR]}: We assume $g' = g^*$. Let $\varphi'(\bO_i)$ be
	\begin{align*}
		\varphi' (\bO_i)
		=
		w(\bC_i) 
		\bigg[
		\bI_i'^{,\intercal} \big\{ \bY_i - \bg_i^* (\bA_i) \big\}
		+
		\frac{\bm{1}\T}{\NI_i} 	\big\{ \bg_i^* (\bm{1}) - \bg_i^* (\bm{0}) \big\} 
		\bigg]
		  \ .
	\end{align*}
	It is trivial that $\EXP \big\{ \varphi'(\bO_i) \big\} = \tau_{\PARASUB}^*$.

	We find that the proof in  Section \ref{proof-thm:1-(2)} now involves with $\varphi'(\bO_i)$ instead of $\varphi^*(\bO_i)$. Specifically, \eqref{proof-AN-old-001} divided by $\sqrt{\NC}$ becomes
	\begin{align*}
		\overline{\tau}_{\PARASUB} - \tau^*
		& = 
		\frac{1}{2}
		\sum_{k=1}^2
		\frac{1}{\NC/2} \sum_{i \in \II_k} \Big\{ \widehat{\varphi}_k(\bO_i) - \varphi'(\bO_i) \Big\}
		+
		\frac{1}{2} \sum_{k=1}^2
		\frac{1}{\NC/2} \sum_{i \in \II_k} \Big\{ \varphi'(\bO_i)  - \tau_{\PARASUB}^* \Big\} 
		\\
		& =
		\frac{1}{2}
		\sum_{k=1}^2
		\EXP \Big\{ \widehat{\varphi}_k(\bO_i) - \varphi'(\bO_i) \, \Big| \, \II_k^c \Big\}
		+ o_P(1)  \ . 
	\end{align*}
	The second equality is from the law of large numbers. The first term of the right hand side is represented as 
		\begin{align*}
		&
		\EXP \Big\{ \widehat{\varphi}_k(\bO_i) - \varphi'(\bO_i) \, \Big| \, \II_k^c \Big\}
		\\
		&
		=
		\EXP \Bigg[ w(\bC_i) \bigg[
		\widehat{\bI}_i^{(-k),\intercal}  \Big\{ \bY_i - \widehat{\bg}_i^{(-k)}(\bA_i) \Big\}
	+
	\frac{\bm{1}\T}{\NI_i} \Big\{ \widehat{\bg}_i^{(-k)}(\bm{1}) - \widehat{\bg}_i^{(-k)}(\bm{0}) \Big\} 		
	-
	\frac{\bm{1}\T}{\NI_i} \Big\{ \bg_i^*(\bm{1}) - \bg_i^*(\bm{0}) \Big\} 		
	\bigg]
	\, \Bigg| \, \II_k^c \Bigg] \ .
	\nonumber
		\end{align*}	
	The first term in the second line is decomposed into 
	\begin{align}								\label{proof-DR-2-002}
	\EXP \Big[
	w(\bC_i) \widehat{\bI}_i^{(-k),\intercal}  \big\{ \bY_i - \widehat{\bg}_i^{(-k)}(\bA_i) \big\}  \, \Big| \, \II_k^c \Big]
	& = 
	\EXP \Big[ w(\bC_i) \bI_i^{*,\intercal} \big\{ \bg_i^*(\bA_i) - \widehat{\bg}_i^{(-k)}(\bA_i) \big\} \, \Big| \, \II_k^c \Big]
	\\
	& \hspace*{1cm}
	+
	\EXP \Big[ w(\bC_i) \big\{ \widehat{\bI}_i^{(-k)} -\bI_i^* \big\} \T \big\{ \bg_i^*(\bA_i) - \widehat{\bg}_i^{(-k)}(\bA_i) \big\} \, \Big| \, \II_k^c \Big] \ .
	\nonumber
	\end{align}
	The first term of \eqref{proof-DR-2-002} is equivalent as
	\begin{align*}
		&
		\EXP \Big[ w(\bC_i) \bI_i^{*,\intercal} \big\{ \bg_i^*(\bA_i) - \widehat{\bg}_i^{(-k)}(\bA_i) \big\} \, \Big| \, \II_k^c \Big]
		=
		\EXP \Bigg[
		\frac{w(\bC_i) \bm{1}\T}{\NI_i} \Big\{ \bg_i^*(\bm{1}) - \bg_i^*(\bm{0}) - \widehat{\bg}_i^{(-k)}(\bm{1}) + \widehat{\bg}_i^{(-k)}(\bm{0}) \Big\} 		
	\, \Bigg| \, \II_k^c \Bigg] \ .
	\end{align*}
	The second term of \eqref{proof-DR-1-002} is 
	\begin{align*}
		\bigg| 
		\EXP \Big[ w(\bC_i) \big\{ \widehat{\bI}_i^{(-k)} -\bI_i^* \big\}\T \big\{ \bg_i^*(\bA_i) - \widehat{\bg}_i^{(-k)}(\bA_i) \big\} \, \Big| \, \II_k^c \Big]
		\bigg|
		&
		\leq		
		C_w
		\big\| \widehat{\bI}_i^{(-k)} -\bI_i^* \big\|_{P,2}  \big\| \bg_i^*(\bA_i) - \widehat{\bg}_i^{(-k)}(\bA_i) \big\|_{P,2}
		\\
		&
		\leq		
		2 C_w C_e
		 \big\| \bg_i^*(\bA_i) - \widehat{\bg}_i^{(-k)}(\bA_i) \big\|_{P,2}
		= O_P(r_{g,\NC}) \ .
	\end{align*}
	The inequality holds from the H\"older's inequality. This implies $\EXP \big\{ \widehat{\varphi}_k(\bO_i) - \varphi'(\bO_i) \cond \II_k^c \big\} = O_P(r_{g,\NC})= o_P(1)$, so does $\overline{\tau}_{\PARASUB} - \tau_{\PARASUB}^*$. This concludes the proof.  
\end{itemize}

\subsection{Proof of Lemma \ref{Main-thm:2}}							\label{proof-thm:2}

\subsubsection{Proof of (1) Mean Double-Robustness}						\label{proof-thm:2-1}

	Let $\phi(\bO_i \con \pi, g, \beta)$ be	
	\begin{align*}
		\phi(\bO_i \con \pi, g, \beta)
		=
		w(\bC_i) \bigg[
		\bI(\bA_i, \bX_i \con \pi) B(\bC_i \con \beta) \big\{ \bY_i - \bg(\bA_i, \bX_i) \big\}
	+
	\frac{\bm{1}\T}{\NI_i} \big\{ \bg (\bm{1}, \bX_i) - \bg (\bm{0}, \bX_i) \big\} 
	\bigg]
	\ .
	\end{align*}
	We observe that $\bI$ under the true $\pi^*$ is expressed as
	\begin{align*}
		\bI(\bA_i, \bX_i \con \pi^*)
		=
		\frac{1}{P(\bA_i \cond \bX_i)}
		\times 
		\underbrace{ 
		\frac{1}{\NI_i}
		\begin{bmatrix}
			\big\{ \ind(A_{i1} = 1) - \ind(A_{i1} = 0) \big\} P (\bA_{i(-1)} \cond \bX_i)
			\\	\vdots \\
			\big\{ \ind(A_{i\NI_i} = 1) - \ind(A_{i\NI_i} = 0) \big\} P (\bA_{i(-\NI_i)} \cond \bX_i)
		\end{bmatrix} }_{ \bm{v}(\bA_i \cond \bX_i) } \ .
	\end{align*}
	We find the expectation of $\phi$ under $\pi^*$ and $g'$ is
	\begin{align}
		&
		\EXP \big\{ \phi(\bO_i \con \pi^*, g', \beta) \big\}
		\nonumber
		\\
		&
		=
		\EXP \Bigg[
			w(\bC_i) \bigg[ \bm{I}(\bA_i, \bX_i \con \pi^*)\T B(\bC_i \con \beta) \Big\{ \bY_i - \bg'(\bA_i, \bX_i) \Big\}
	+
	\frac{\bm{1}\T}{\NI_i} \Big\{ \bg' (\bm{1}, \bX_i) - \bg' (\bm{0}, \bX_i) \Big\} 
	\bigg]
			\, \Bigg| \,
			\bX_i
		\Bigg]
		\nonumber
		\\
		&
		=
		\EXP \Bigg[
			w(\bC_i) \bigg[  \bm{I}(\bA_i, \bX_i \con \pi^*)\T B(\bC_i \con \beta) \Big\{ \bg^*(\bA_i, \bX_i) - \bg'(\bA_i, \bX_i) \Big\}
			\, \Bigg| \,
			\bX_i
		\Bigg]
			\nonumber
			\\
			& \hspace*{2cm}
		+
		\EXP \Bigg[ w(\bC_i) \frac{\bm{1}\T}{\NI_i} \Big\{ \bg' (\bm{1}, \bX_i) - \bg' (\bm{0}, \bX_i) \Big\} 	\Bigg]
		\nonumber
		\\
		&
		=
		\EXP \Bigg[
			w(\bC_i) \sum_{\ba_i \in \{0,1\}^{\NI_i} } \bm{v} (\ba_i \cond \bX_i)  \T
			B(\bC_i \con \beta) \Big\{ \bg^*(\ba_i, \bX_i) - \bg'(\ba_i, \bX_i) \Big\}
		\Bigg]
		\label{proof-30001}
		\\
		& \hspace*{2cm}
		+
		\EXP \Bigg[ w(\bC_i) \frac{\bm{1}\T}{\NI_i} \Big\{ \bg' (\bm{1}, \bX_i) - \bg' (\bm{0}, \bX_i) \Big\} 	\Bigg] \ .
		\nonumber
	\end{align}
	The terms in \eqref{proof-30001} is represented as
	\begin{align*}
		&
		w(\bC_i) \bigg[ \sum_{\ba_i \in \{0,1\}^{\NI_i} } \bm{v} (\ba_i \cond \bX_i)\T \Big\{ \bg^* (\bA_i, \bX_i) - \bg' (\bm{0}, \bX_i) \Big\} 
		\\
		& \hspace*{0.25cm}
		+
		 \sum_{\ba_i \in \{0,1\}^{\NI_i} } \bm{v} (\ba_i \cond \bX_i)\T \Big\{ B(\bC_i \con \beta) - I \Big\} \Big\{ \bg' (\bm{1}, \bX_i) - \bg' (\bm{0}, \bX_i) \Big\} 
		 \bigg]
		 \\
		 & = 
		 \frac{w(\bC_i)}{\NI_i}\sum_{\ba_i \in \{0,1\}^{\NI_i} }   \sum_{j=1}^{\NI_i} 
		 \Big\{ \ind(a_{ij} = 1) - \ind(a_{ij} = 0 ) \Big\}
		 P( \bA_\eij = \ba_\eij \cond \bX_i) 
		 \Big\{ g^*(a_{ij}, \bX_{ij} ) - g'(a_{ij}, \bX_{ij})
		 \Big\}
		 \\
		 & \hspace*{0.25cm}
		 - \frac{w(\bC_i) \beta(\bC_i) }{\NI_i}
		\sum_{\ba_i \in \{0,1\}^{\NI_i} }  \sum_{j=1}^{\NI_i} 
		\Big\{ \ind(a_{ij} = 1) - \ind(a_{ij} = 0 ) \Big\}
		  P( \bA_\eij =  \ba_\eij \cond \bX_i) 
		 \sum_{\ell \neq j}\Big\{ g^*(a_{i\ell}, \bX_{i\ell} ) - g'(a_{i\ell}, \bX_{i\ell})
		 \Big\}
		 	 \\
		 & = 
		 \frac{w(\bC_i)}{\NI_i}  \sum_{j=1}^{\NI_i} 
		  \sum_{a_{ij} = 0}^1
		 \Big\{ g^*(a_{ij}, \bX_{ij} ) - g'(a_{ij}, \bX_{ij}) \Big\}
		 \Bigg\{ \begin{matrix} \ind(a_{ij} = 1) \hspace*{0.5cm} \\ \hspace*{0.5cm} - \ind(a_{ij} = 0 ) \end{matrix} \Bigg\}
		 \underbrace{
		 \sum_{ \ba_\eij }  
		 P( \bA_\eij = \ba_\eij \cond \bX_i)  }_{=1}
		 \\
		 & \hspace*{0.25cm}
		 - \frac{w(\bC_i) \beta(\bC_i) }{\NI_i}
		  \sum_{j=1}^{\NI_i} 
		  \underbrace{
		  \sum_{a_{ij} = 0}^1
		 \Bigg\{ \begin{matrix} \ind(a_{ij} = 1) \hspace*{0.5cm} \\ \hspace*{0.5cm} - \ind(a_{ij} = 0 ) \end{matrix} \Bigg\}
		\sum_{\ba_\eij }
		  P( \bA_\eij = \ba_\eij \cond \bX_i) 
		 \sum_{\ell \neq j} \Big\{ g^*(a_{i\ell}, \bX_{i\ell} ) - g'(a_{i\ell}, \bX_{i\ell})
		 \Big\} }_{=0}
		 \\
		  & = 
		 \frac{w(\bC_i)}{\NI_i}  \sum_{j=1}^{\NI_i} \bigg[
		 \Big\{ g^*(1, \bX_{ij} ) - g'(1, \bX_{ij}) \Big\} - 
		 \Big\{ g^*(0, \bX_{ij} ) - g'(0, \bX_{ij}) \Big\}
		 \bigg]
		 \\
		 & =
		 w(\bC_i) \frac{\bm{1}\T}{\NI_i} 
		 \bigg[ \Big\{ \bg^* (\bm{1}, \bX_i) - \bg^* (\bm{0}, \bX_i) \Big\}  - \Big\{ \bg' (\bm{1}, \bX_i) - \bg' (\bm{0}, \bX_i) \Big\} \bigg] \ .
	\end{align*}
	Plugging in the result in $\EXP \big\{ \phi(\bO_i \con \pi^*, g', \beta) \big\}$, we find
	\begin{align*}
		\EXP \big\{ \phi(\bO_i \con \pi^*, g', \beta) \big\}
		=
		\EXP \Bigg[ w(\bC_i) \frac{\bm{1}\T}{\NI_i} \Big\{ \bg^* (\bm{1}, \bX_i) - \bg^* (\bm{0}, \bX_i) \Big\} 	\Bigg]  = \tau_{\PARASUB}^* \ .
	\end{align*}

	Next we find the expectation of $\phi$ under $g^*$ and $\pi'$ is $\tau_{\PARASUB}^*$ as follows.
	\begin{align*}
		&
		\EXP \big\{ \phi(\bO_i \con \pi', g^*, \beta) \big\}
		\\
		&
		=
		\EXP \Bigg[
			w(\bC_i) \bigg[ \bm{I}(\bA_i, \bX_i \con \pi')\T B(\bC_i \con \beta)  \Big\{ \bY_i - \bg^*(\bA_i, \bX_i) \Big\}
	+
	\frac{\bm{1}\T}{\NI_i} \Big\{ \bg^* (\bm{1}, \bX_i) - \bg^* (\bm{0}, \bX_i) \Big\} 
	\bigg]
			\, \Bigg| \,
			\bX_i
		\Bigg]
		\\
		&
		=
		\EXP \Bigg[ w(\bC_i)
	\frac{\bm{1}\T}{\NI_i} \Big\{ \bg' (\bm{1}, \bX_i) - \bg' (\bm{0}, \bX_i) \Big\} 
		\Bigg]
		=
		\tau_{\PARASUB}^*  \ .
	\end{align*}

\subsubsection{Proof of (2) Efficiency Gain in Decomposable Propensity Score}

When $P(\bA_i \cond \bX_i) = \prod_{i=1}^{\NI_i} P(A_{ij} \cond \bX_{ij})$, $\bm{I}(\bA_i, \bX_i \con \pi^*)$ reduces to
	\begin{align*}
		\bI(\bA_i, \bX_i \con \pi^*)
		& =
		\frac{1}{P(\bA_i \cond \bX_i)}
		\frac{1}{\NI_i}
		\begin{bmatrix}
			\big\{ \ind(A_{i1} = 1) - \ind(A_{i1} = 0) \big\} P (\bA_{i(-1)} \cond \bX_i)
			\\	\vdots \\
			\big\{ \ind(A_{i\NI_i} = 1) - \ind(A_{i\NI_i} = 0) \big\} P (\bA_{i(-\NI_i)} \cond \bX_i)
		\end{bmatrix} 
		\\
		& =
		\frac{1}{ \prod_{j=1}^{\NI_i} P(A_{ij} \cond \bX_{ij})}
		\frac{1}{\NI_i}
		\begin{bmatrix}
			\big\{ \ind(A_{i1} = 1) - \ind(A_{i1} = 0) \big\} \prod_{j\neq1} P(A_{ij} \cond \bX_{ij})
			\\	\vdots \\
			\big\{ \ind(A_{i\NI_i} = 1) - \ind(A_{i\NI_i} = 0) \big\} \prod_{j\neq \NI_i} P(A_{ij} \cond \bX_{ij})
		\end{bmatrix} 
		\\
		& =
		\frac{1}{\NI_i}
		\begin{bmatrix}
			\big\{ \ind(A_{i1} = 1) - \ind(A_{i1} = 0) \big\} / P(A_{i1} \cond \bX_{i1})
			\\	\vdots \\
			\big\{ \ind(A_{i\NI_i} = 1) - \ind(A_{i\NI_i} = 0) \big\} / P(A_{i\NI_i} \cond \bX_{i\NI_i})
		\end{bmatrix} 
		\\
		& =
		\frac{1}{\NI_i}
		\begin{bmatrix}
			\ind(A_{i1} = 1) /e^*(1 \cond \bX_{i1}) - \ind(A_{i1} = 0) /e^*(0 \cond \bX_{i1})
			\\	\vdots \\
			\ind(A_{i\NI_i} = 1) /e^*(1 \cond \bX_{i\NI_i}) - \ind(A_{i\NI_i} = 0) /e^*(0 \cond \bX_{i\NI_i})
		\end{bmatrix}  \ .
	\end{align*}
	Therefore, $\overline{\tau}_{\PARASUB}(e^*,g')$ is a special case of $\widehat{\tau}_{\PARASUB}(\pi^*,g',\beta)$ when $\beta=0$. From the definition of $\beta_{\pi^*,g}^*$, we have $\VAR \big\{ \widehat{\tau}_{\PARASUB}(\pi^*,g',\beta_{\pi^*,g}^*) \big\} \leq \VAR \big\{ \widehat{\tau}_{\PARASUB}(\pi^*,g',\beta=0) \big\}  = \VAR \big\{ \overline{\tau}_{\PARASUB}(e^*,g') \big\} $.

\subsection{Proof of Theorem \ref{Main-thm:3}}							\label{proof-thm:3}

We prove the theorem in order of (2) Asymptotic Normality, (1) Double Robustness, and (3) Efficiency Gain Under Known Treatment Assignment Mechanism. Note that the assumptions in Theorem \ref{Main-thm:3} implies the conditions in Lemma \ref{lemma:tauhat}. 

\subsubsection{Proof of (2) Asymptotic Normality}						\label{proof-thm:3-(2)}

It suffices to show that 
\begin{align}				\label{proof2-1}
	& \sqrt{ \frac{2}{\NC} } \sum_{i \in \II_k} 
	\big\{ 
	\phi(\bO_i, \widehat{\pi}^{(-k)}, \widehat{g}^{(-k)}, \widehat{\beta}^{(k)})
	- 
	\tau^* 
	\big\}
	=
	\sqrt{ \frac{2}{\NC} } \sum_{i \in \II_k}
	\big\{
	\phi(\bO_i, \pi^*, g^*,  \beta'^{,(-k)})
	-
	\tau^*
	\big\}
	 + o_P(1)
	 \ .
\end{align}
The right hand side of \eqref{proof2-1} converges to $N \big( 0 ,  \VAR \big\{ \phi(\bO_i, \pi^*, g^*, \beta^*) \big\} \big)$ from the central limit theorem. The left hand side of \eqref{proof2-1} is further decomposed as follows.
\begin{align*}
	& \sqrt{ \frac{2}{\NC} } \sum_{i \in \II_k} 
	\big\{ 
	\phi(\bO_i, \widehat{\pi}^{(-k)}, \widehat{g}^{(-k)}, \widehat{\beta}^{(k)})
	- 
	\tau^* 
	\big\}
	\\
	& = 
	\underbrace{
	\sqrt{ \frac{2}{\NC} } \sum_{i \in \II_k} 
	\Bigg[
	\bigg\{
	\begin{array}{l}
	\phi(\bO_i, \widehat{\pi}^{(-k)}, \widehat{g}^{(-k)}, \widehat{\beta}^{(k)})
	\\
	-
	\phi(\bO_i, \widehat{\pi}^{(-k)}, \widehat{g}^{(-k)}, \beta^{\ddagger,(-k)})
	\end{array}	
	\bigg\}
	-
	\EXP \bigg\{
	\begin{array}{l}
	\phi(\bO_i, \widehat{\pi}^{(-k)}, \widehat{g}^{(-k)}, \widehat{\beta}^{(k)})
	\\
	-
	\phi(\bO_i, \widehat{\pi}^{(-k)}, \widehat{g}^{(-k)}, \beta^{\ddagger,(-k)})
	\end{array}	
	\, \bigg| \, \II_k^c \bigg\}
	\Bigg]
	}_{[A] = o_P(1) }
	\\
	&
	+
	\underbrace{
	\sqrt{ \frac{\NC}{2} } \EXP \bigg\{
	\begin{array}{l}
	\phi(\bO_i, \widehat{\pi}^{(-k)}, \widehat{g}^{(-k)}, \widehat{\beta}^{(k)})
	\\
	-
	\phi(\bO_i, \widehat{\pi}^{(-k)}, \widehat{g}^{(-k)}, \beta^{\ddagger,(-k)})
	\end{array}	
	\, \bigg| \, \II_k^c \bigg\}
	}_{[B] = O_P(\NC^{1/2} r_{g,\NC} r_{\beta,\NC}) }
	\\
	&
	+
	\underbrace{
	\sqrt{ \frac{2}{\NC} } \sum_{i \in \II_k} 
	\Bigg[
	\bigg\{ 
	\begin{array}{l}
	\phi(\bO_i, \widehat{\pi}^{(-k)}, \widehat{g}^{(-k)}, \beta^{\ddagger,(-k)})
	\\
	\hspace*{1cm}
	-	\phi(\bO_i, \pi^*, g^*,  \beta'^{,(-k)})
	\end{array}
	\bigg\}
	-
	\EXP
	\bigg\{ 
	\begin{array}{l}
	\phi(\bO_i, \widehat{\pi}^{(-k)}, \widehat{g}^{(-k)}, \beta^{\ddagger,(-k)})
	\\
	\hspace*{1cm}
	-	\phi(\bO_i, \pi^*, g^*,  \beta'^{,(-k)})
	\end{array}
	\, \bigg| \, \II_k^c
	\bigg\}
	\Bigg]
	}_{[C] = O_P(r_{\pi,\NC}^2) + O_P(r_{g,\NC}^2) }
	\\
	&
	+ 
	\underbrace{
	\sqrt{ \frac{\NC}{2} } \EXP
	\bigg\{ 
	\begin{array}{l}
	\phi(\bO_i, \widehat{\pi}^{(-k)}, \widehat{g}^{(-k)}, \beta^{\ddagger,(-k)})
	\\
	\hspace*{1cm}
	-	\phi(\bO_i, \pi^*, g^*,  \beta'^{,(-k)})
	\end{array}
	\, \bigg| \, \II_k^c
	\bigg\} 
	}_{[D] = \NC^{1/2} r_{\pi,\NC} r_{g,\NC}}
	\\
	&
	+
	\underbrace{
	\sqrt{ \frac{2}{\NC} } \sum_{i \in \II_k} 
	\big\{ \phi(\bO_i, \pi^*, g^*,  \beta'^{,(-k)})
	-
	\tau^*
	\big\}}_{[E] \stackrel{D}{\rightarrow} N ( 0 ,  \VAR \{ \phi(\bO_i, \pi^*, g^*, \beta'^{,(-k)}) \} ) } \ .
\end{align*}
In \textbf{[Step 1]}, we show that $\sqrt{\NC}$-scaled empirical mean of $\widehat{\phi}_k(\bO_i \con \widehat{\beta}^{(k)})$ is asymptotically identical to $\sqrt{\NC}$-scaled empirical mean of $\widehat{\phi}_k(\bO_i \con\beta^{\ddagger,(-k)} )$, i.e., showing that $[A]=o_P(1)$ and $[B]=o_P(1)$. In \textbf{[Step 2]}, we show that $\sqrt{\NC}$-scaled empirical mean of $\widehat{\phi}_k(\bO_i \con\beta^{\ddagger,(-k)} )$ is asymptotically identical to $\sqrt{\NC}$-scaled empirical mean of $\phi^* (\bO_i \con \beta'^{,(-k)} )$, i.e., $[C]=o_P(1)$ and $[D]=o_P(1)$. In \textbf{[Step 3]}, we establish the consistency of the variance estimate.

\begin{itemize}[leftmargin=0.5cm]
	\item \textbf{[Step 1]:} $\sqrt{\NC}$-scaled empirical mean of $\widehat{\phi}_k(\bO_i \con \widehat{\beta}^{(k)})$ is asymptotically identical to $\sqrt{\NC}$-scaled empirical mean of $\widehat{\phi}_k(\bO_i \con\beta^{\ddagger,(-k)} )$. \\

	To show the desired result, we use Example 19.7 and Lemma 19.24 of \citet{Vaart1998}. Let $\Phi := \big\{ \widehat{\phi}_k(\bO_i \con \beta) \cond \beta \in \mathcal{B} \big\}$ be the collection of influence function at function $\beta$ where
	\begin{align}												\label{proof-AN-001}
		\widehat{\phi}_k (\bO_i \con \beta)
		&
		=
		w(\bC_i) \bigg[
		\widehat{\bI}_i^{(-k), \intercal} B (\bC_i \con \beta ) \Big\{ \bY_i - \widehat{\bg}_i^{(-k)} (\bA_i) \Big\}
	+
	\frac{\bm{1}\T}{\NI_i} \Big\{ \widehat{\bg}_i^{(-k)} (\bm{1}) - \widehat{\bg}_i^{(-k)} (\bm{0}) \Big\} 
	\bigg]
	\nonumber
	\\
	&
	=
		w(\bC_i) \bigg[
		\widehat{\bI}_i^{(-k), \intercal} \Big\{
			I + (I - \bm{1}\bm{1}\T ) \beta(\bC_i)
		\Big\} \Big\{ \bY_i - \widehat{\bg}_i^{(-k)} (\bA_i) \Big\}
	+
	\frac{\bm{1}\T}{\NI_i} \Big\{ \widehat{\bg}_i^{(-k)} (\bm{1}) - \widehat{\bg}_i^{(-k)} (\bm{0}) \Big\} 
	\bigg]
	\nonumber
	\\
	&
	=
		\underbrace{
		w(\bC_i) \bigg[
		\widehat{\bI}_i^{(-k), \intercal} \Big\{ \bY_i - \widehat{\bg}_i^{(-k)} (\bA_i) \Big\}
	+
	\frac{\bm{1}\T}{\NI_i} \Big\{ \widehat{\bg}_i^{(-k)} (\bm{1}) - \widehat{\bg}_i^{(-k)} (\bm{0}) \Big\} 
	\bigg]	}_{\widehat{\psi}_{1,k}(\bO_i)}
	\nonumber
	\\
	& \hspace*{3cm}
	+ 
	\underbrace{
	w(\bC_i)
	\widehat{\bI}_i^{(-k), \intercal} \big( I - \bm{1}\bm{1}\T \big) \Big\{ \bY_i - \widehat{\bg}_i^{(-k)} (\bA_i) \Big\} 	}_{\widehat{\psi}_{2,k}(\bO_i)}
	\beta(\bC_i) 
	\ .
	\end{align}
	Therefore, the collection $\Phi$ can be understood as the composition $\rho \, \circ \, \big( \{ \widehat{\psi}_{1,k}\}, \{ \widehat{\psi}_{2,k} \}, \mathcal{B} \big)$ where $\rho:\R^3 \rightarrow \R$ is a fixed map with
	\begin{align*}
		\rho( \widehat{\psi}_{1,k}, \widehat{\psi}_{2,k}, \beta  )
		=
		\widehat{\psi}_{1,k} (\bO_i) + \widehat{\psi}_{2,k} (\bO_i) \beta(\bC_i)
		=
		\widehat{\phi}_k( \bO_i \con \beta) \ .
	\end{align*}
	Note that the classes $\big\{ \widehat{\psi}_{1,k} \big\}$ and $\big\{ \widehat{\psi}_{2,k} \big\}$ are Donsker because they are singleton sets and square-integrable \citep[page 270]{Vaart1998}. For two parameters $\beta_1$ and $\beta_2$, we find
	\begin{align*}
	\Big\{ \rho( \widehat{\psi}_{1,k} , \widehat{\psi}_{2,k} , \beta_1) - \rho( \widehat{\psi}_{1,k} , \widehat{\psi}_{2,k} , \beta_2) \Big\}^2
	=
	\Big\{ \widehat{\phi}_k (\bO_i \con \beta_1) - \widehat{\phi}_k (\bO_i \con \beta_2) \Big\}^2
	&
	=
	\widehat{\psi}_{2,k}(\bO_i)^2 \Big| \beta_1(\bC_i) - \beta_2 (\bC_i) \Big|^2 \ .
	\end{align*}
	Therefore, equation (2.10.19) of \citet{VW1996} is satisfied with $L_{\alpha,1} = 0$, $L_{\alpha,2} = 0$, $L_{\alpha,3} = \widehat{\phi}_{2,k}$, $\alpha_1=\alpha_2=\alpha_3=1$. Moreover, $\{ \widehat{\psi}_{1,k}\}$, $\{ \widehat{\psi}_{2,k} \}$, and $\mathcal{B}$ have envelope functions as $\widehat{\psi}_{1,k}$, $\widehat{\psi}_{2,k}$, and constant $B_0$, respectively. Moreover, we find $(L_\alpha \cdot F^\alpha)^2 = \widehat{\psi}_{2,k}(\bO_i)^2 B_0^2$ which is integrable as follows.
	\begin{align}											\label{proof-AN-002}
		\EXP \Big\{ \widehat{\psi}_{2,k}(\bO_i)^2 \, \Big| \, \II_k^c \Big\} 
		& \leq
		\EXP 
		\Big\{
			w(\bC_i)^2 	\big\| \widehat{\bI}_i^{(-k)}  \big\|_2^2 \big\|  I - \bm{1} \bm{1} \T \big\|_2^2
		\big\|\bY_i - \widehat{\bg}_i^{(-k)} (\bA_i) \big\|_2 ^2
		 \, \Big| \, \II_k^c 
		\Big\}
		\nonumber
		\\
		& \leq
		2C_w^2 C_\pi^2 \NT^2 
		\EXP
			\Big\{
		\big\|\bY_i - \bg_i^*(\bA_i) \big\|_2^2 + \big\| \bg_i^*(\bA_i) - \widehat{\bg}_i^{(-k)} (\bA_i) \big\|_2 ^2
		\, \Big| \, \II_k^c
		\Big\}
		\nonumber
		\\
		& \leq
		2C_w^2 C_\pi^2 \NT^2 
		\Big[
		\EXP 
		\Big\{
		\bm{1}\T \Sigma(\bA_i, \bX_i) \bm{1} \, \Big| \, \II_k^c
		\Big\}
		+ C_g^2
		\Big]
		\nonumber
		\\
		& \leq C'\ . 
	\end{align}
	This implies every function in $\Phi$ is square integrable because $\big\{ \widehat{\phi}_k(\bO_i \con \beta) \big\}^2 \leq \widehat{\psi}_{2,k}(\bO_i)^2 B_0^2$. Therefore, Theorem 2.10.20 of \citet{VW1996} can be applies and, as a consequence, $\Phi = \rho \, \circ \, \big( \{ \widehat{\psi}_{1,k}\}, \{ \widehat{\psi}_{2,k} \}, \mathcal{B} \big)$ is Donsker.

	From conditions (a)-(f) of Lemma \ref{lemma:tauhat}, we have 
	\begin{align}																\label{proof-AN-010}
	&
		\int 
		\Big\{ \widehat{\phi}_k (\bo_i \con \widehat{\beta}^{(k)} ) - \widehat{\phi}_k (\bo_i \con \beta^{\ddagger,(-k)} ) \Big\}^2 \, dP(\bo_i)
		\nonumber
		\\
	&
	=
	\int \widehat{\psi}_{2,k}(\bo_i)^2 \Big\{   \widehat{\beta}^{(k)}(\bc_i) - \beta^{\ddagger,(-k)} (\bc_i) \Big\}^2 \, dP(\bo_i)
	\nonumber
	\\
	& 
	\leq
	\int w(\bc_i)^2 	\big\| \widehat{\bI}_i^{(-k)}  \big\|_2^2 \big\|  I - \bm{1} \bm{1} \T \big\|_2^2
		\big\|\by_i - \widehat{\bg}_i^{(-k)} (\ba_i) \big\|_2 ^2 \Big\{  \widehat{\beta}^{(k)}(\bc_i) - \beta^{\ddagger,(-k)} (\bc_i) \Big\}^2 \, dP(\bo_i)
		\nonumber
		\\
	& 
	\leq
	2 C_w^2 C_\pi^2 \NT^2 \int  \Big\{ 	\big\|\by_i -\bg_i^*(\ba_i) \big\|_2^2 + \big\| \bg_i^*(\ba_i) - \widehat{\bg}_i^{(-k)} (\ba_i) \big\|_2 ^2 \Big\} \Big\{  \widehat{\beta}^{(k)}(\bc_i) - \beta^{\ddagger,(-k)} (\bc_i) \Big\}^2 \, dP(\bo_i)
	\nonumber
		\\
	& 
	\leq
	2 C_w^2 C_\pi^2 \NT^2 \int  \Big\{ \bm{1}\T \Sigma(\ba_i, \bx_i) \bm{1} + \big\| \bg_i^*(\ba_i) - \widehat{\bg}_i^{(-k)} (\ba_i) \big\|_2 ^2 \Big\} \Big\{  \widehat{\beta}^{(k)}(\bc_i) - \beta^{\ddagger,(-k)} (\bc_i) \Big\}^2 \, dP(\ba_i,\bx_i)
	\nonumber
	\\
	& 
	\leq C' \int  \Big\{  \widehat{\beta}^{(k)}(\bc_i) - \beta^{\ddagger,(-k)} (\bc_i) \Big\}^2 \, dP(\bc_i)
	\nonumber
	\\
	& = O_P(r_{\beta,\NC}^2) \ .
	\end{align}

	Therefore, we can apply Lemma 19.24 of \citet{Vaart1998} to show the empirical process of $\widehat{\phi}_k$ at $\widehat{\beta}^{(k)}$ is asymptotically the same as that at $\beta^{\dagger, (-k)}$. 
	\begin{align*}
		&
		\frac{1}{\sqrt{\NC/2}} \sum_{i \in \II_k} \Big[ \widehat{\phi}_k (\bO_i \con \widehat{\beta}^{(k)} ) - m ( \widehat{\beta}^{(k)} ) \Big]
		=
		\frac{1}{\sqrt{\NC/2}} \sum_{i \in \II_k} \Big[ \widehat{\phi}_k(\bO_i \con \beta^{\ddagger,(-k)} ) - m ( \beta^{\ddagger,(-k)} ) \Big]
		+
		o_P(1) 
	\end{align*}
	where 
	\begin{align*}
		m(\beta) 
		= \int \widehat{\phi}_k (\bo_i \con \beta) \, dP(\bo_i)
		= \int \Big\{ \widehat{\phi}_{1,k} (\bo_i) + \widehat{\phi}_{2,k} (\bo_i) \beta(\bc_i) \Big\} \, dP(\bo_i)
		 \ .
	\end{align*}
	Therefore, we find
	\begin{align}								\label{proof-AN-003}
		&
		\frac{1}{\sqrt{\NC/2}} \sum_{i \in \II_k} \Big\{ \widehat{\phi}_k(\bO_i \con \widehat{\beta}^{(k)} )  - \tau_{\PARASUB}^* \Big\}
		\\
		&
		=
		\frac{1}{\sqrt{\NC/2}} \sum_{i \in \II_k} \Big\{ \widehat{\phi}_k (\bO_i \con \beta^{\ddagger,(-k)} ) - \tau_{\PARASUB}^* \Big\}
		+
		\underbrace{ 		
		\sqrt{ \frac{\NC}{2} }
		\Big\{
		m \big( \widehat{\beta}^{(k)} \big)
		-
		m \big( \beta^{\ddagger,(-k)} \big)
		\Big\}
		 }_{[A]}
		 +
		 o_P(1)  \ .
		 \nonumber
	\end{align}
From simple algebra, $[A]$ is $o_P(1)$ from conditions (d) and (e) of Lemma \ref{lemma:tauhat}.
\begin{align*}
	\Big|
	[A]
	\Big|
	&
	=
		\sqrt{ \frac{\NC}{2} }
		\Big|
		m \big( \widehat{\beta}^{(k)} \big)
		-
		m \big( \beta^{\ddagger,(-k)} \big)
		\Big|
		 \\
		 &
		 =
		 \sqrt{ \frac{\NC}{2} }
		 \Bigg|
		 \int \widehat{\phi}_{2,k} (\bo_i) \Big\{ \widehat{\beta}^{(k)}(\bc_i) - \beta^{\ddagger, (-k)}(\bc_i) \Big\} \, dP(\bo_i)
		 \Bigg|
		 \\
		 & = \sqrt{ \frac{ \NC}{2} }
		 \Bigg|
		 \int  \bI (\ba_i, \bx_i \con \widehat{\pi}^{(-k)})\T (I - \bm{1}\bm{1}\T ) \Big\{ \bg^*(\ba_i, \bx_i) - \widehat{\bg}^{(-k)} (\ba_i, \bx_i) \Big\}
		 \Big\{ \widehat{\beta}^{(k)}(\bc_i) - \beta^{\ddagger, (-k)}(\bc_i) \Big\} \, dP(\bo_i)
		 \Bigg|
		  \\
		 & 
		 \leq \sqrt{ \frac{ \NC}{2} } C_\pi \NT \,
		 \Big\| \bg^*(\ba_i, \bx_i) - \widehat{\bg}^{(-k)} (\ba_i, \bx_i) \Big\|_{P,2}
		 \bigg[
 		 \int 
		 \Big\{ \widehat{\beta}^{(k)}(\bc_i) - \beta^{\ddagger, (-k)}(\bc_i) \Big\}^2 \, dP(\bo_i)
		 \bigg]^{1/2}
		   \\
		 & 
		= O(\NC^{1/2}) O_P(r_{g,\NC}) O_P (r_{\beta,\NC}) 
		 = o_P(1) \ . 
\end{align*}	

As a consequence, \eqref{proof-AN-003} reduces to
	\begin{align}					\label{proof-AN-004}
		&
		\frac{1}{\sqrt{\NC/2}} \sum_{i \in \II_k} \Big\{ \widehat{\phi}_k(\bO_i \con \widehat{\beta}^{(k)} )  -\tau_{\PARASUB}^* \Big\}
		=
		\frac{1}{\sqrt{\NC/2}} \sum_{i \in \II_k} \Big\{ \widehat{\phi}_k(\bO_i \con \beta^{\ddagger,(-k)} )  - \tau_{\PARASUB}^* \Big\}
		+
		 o_P(1)  \ .
	\end{align}
	
	\item \textbf{[Step 2]}: $\sqrt{\NC}$-scaled empirical mean of $\widehat{\phi}_k(\bO_i \con\beta^{\ddagger,(-k)} )$ is asymptotically identical to $\sqrt{\NC}$-scaled empirical mean of $\phi^* (\bO_i \con \beta'^{,(-k)} )$. \\

	We decompose the the right hand side of \eqref{proof-AN-004} as follows.
	\begin{align}						\label{proof-AN-005}
		&
		\frac{1}{\sqrt{\NC/2}} \sum_{i \in \II_k} \Big\{ \widehat{\phi}_k (\bO_i \con \beta^{\ddagger,(-k)} ) - \tau_{\PARASUB}^* \Big\}
		\nonumber
		\\
		&
		=
		\frac{1}{\sqrt{\NC/2}} \sum_{i \in \II_k} \Big\{ \widehat{\phi}_k(\bO_i \con \beta^{\ddagger,(-k)} )  - \phi^*(\bO_i \con \beta'^{,(-k)}) \Big\}
		+
		\frac{1}{\sqrt{\NC/2}} \sum_{i \in \II_k} \Big\{ \phi^*(\bO_i \con \beta'^{,(-k)})  - \tau_{\PARASUB}^* \Big\} 
	\end{align}
	where
	\begin{align*}
		\phi^* (\bO_i \con \beta)
		=
		w(\bC_i)
	\bigg[
	\bI (\bA_i, \bX_i \con\pi^* )\T B (\bC_i \con \beta ) \big\{ \bY_i - \bg^* (\bA_i, \bX_i) \big\}
	+
	\frac{\bm{1}\T}{\NI_i} \Big\{ \bg^* (\bm{1}, \bX_i) - \bg^* (\bm{0}, \bX_i) \Big\}
	\bigg]  \ .
	\end{align*}
	The second term is trivially satisfies the asymptotic Normality from the central limit theorem so it suffices to show the first term is $o_P(1)$, which is decomposed into $[B]$ and $[C]$ as follows.
	\begin{align}								\label{proof-AN-006}
		&
		\frac{1}{\sqrt{\NC/2}} \sum_{i \in \II_k} \Big\{ \widehat{\phi}_k(\bO_i \con \beta^{\ddagger,(-k)} )  - \phi^*(\bO_i \con \beta'^{,(-k)}) \Big\}
		\\
		& =
		\underbrace{
		\frac{1}{\sqrt{\NC/2}} \sum_{i \in \II_k} \bigg[ 
		\Big\{ \widehat{\phi}_k(\bO_i \con \beta^{\ddagger,(-k)} ) - \phi^*(\bO_i \con \beta'^{,(-k)}) \Big\}
		-
		\EXP \Big\{ \widehat{\phi}_k(\bO_i \con \beta^{\ddagger,(-k)} ) - \phi^*(\bO_i \con \beta'^{,(-k)}) \, \Big| \, \II_k^c \Big\}
		\bigg]
		}_{[B]}
		\nonumber		
		\\
		& \hspace*{2cm}
		+
		\sqrt{ \frac{\NC}{2} } \underbrace{
		\EXP \Big\{ \widehat{\phi}_k(\bO_i \con \beta^{\ddagger,(-k)} ) - \phi^*(\bO_i \con \beta'^{,(-k)}) \, \Big| \, \II_k^c \Big\}
		}_{[C]} \ . 
		\nonumber		
	\end{align}	
	We show that $[B] = O_P( r_{\pi,\NC}^2)  + O_P(r_{g,\NC}^2 ) $ and $[C]=O_P(r_{\pi,\NC}r_{g,\NC})$.
	
	\begin{itemize}[leftmargin=0.5cm]
		\item  \textbf{(Rate of $[B]$)}
	
	The squared expectation of $[B]$ is
	\begin{align}																	\label{proof-AN-011}
		\EXP \big\{ [B]^2 \cond \II_k^c \big\}
		&
		=
		\frac{1}{\NC/2} \sum_{i \in \II_k} \EXP \Big[ \big\{ \widehat{\phi}_k(\bO_i \con \beta^{\ddagger,(-k)}) - \phi^*(\bO_i \con \beta'^{,(-k)} ) \big\}^2 \, \Big| \, \II_k^c \Big]
		\nonumber
		\\
		&
		=
		 \EXP \Big[  \big\{ \widehat{\phi}_k(\bO_i \con \beta^{\ddagger,(-k)}) - \phi^*(\bO_i \con \beta'^{,(-k)} ) \big\}^2  \, \Big| \, \II_k^c \Big]  \ .
	\end{align}
	We find $\widehat{\phi}_k(\bO_i \con \beta^{\ddagger,(-k)}) - \phi^*(\bO_i \con \beta' ) $ is represented as
	\begin{align*}
		&
		\widehat{\phi}_k(\bO_i \con \beta^{\ddagger,(-k)}) - \phi^*(\bO_i \con \beta'^{,(-k)} ) 
		\\
		& = 
		w(\bC_i) \bigg[
		\widehat{\bI}_i^{(-k),\intercal} B_i^{\ddagger,(-k)} \widehat{\bm{R}}_i^{(-k)} 
		-
		\bI_i^{*,\intercal} B_i'^{,(-k)} \bm{R}_i^*
		+ \frac{\bm{1}\T}{\NI_i} \Big\{ \widehat{\bg}_i^{(-k)} (\bm{1}) -  \widehat{\bg}_i^{(-k)} (\bm{0})  - \bg_i^* (\bm{1}) + \bg_i^* (\bm{0}) \Big\}
		\bigg]
		\\
		& = 
		\frac{w(\bC_i)}{4} 
		\Big\{ \widehat{\bI}_i^{(-k)} - \bI_i^* \Big\}\T
		\Big\{  B_i^{\ddagger,(-k)}  + B_i'^{,(-k)} \Big\}
		\Big\{ \widehat{\bm{R}}_i^{(-k)}  + \bm{R}_i^* \Big\}
		\\
		& \hspace*{1cm}
		+
		\frac{w(\bC_i)}{4} 
		\Big\{ \widehat{\bI}_i^{(-k)} + \bI_i^* \Big\}\T
		\Big\{ B_i^{\ddagger,(-k)} - B_i'^{,(-k)} \Big\}
		\Big\{ \widehat{\bm{R}}_i^{(-k)}  + \bm{R}_i^* \Big\}
		\\
		& \hspace*{1cm}
		+
		\frac{w(\bC_i)}{2} 
		\Big\{ \widehat{\bI}_i^{(-k),\intercal} B_i^{\ddagger,(-k)}  + \bI_i^{*,\intercal} B_i'^{,(-k)} \Big\}
		\Big\{ \widehat{\bm{R}}_i^{(-k)} - \bm{R}_i^* \Big\}
		\\
		& \hspace*{1cm}
		+ \frac{w(\bC_i) \bm{1}\T}{\NI_i} \Big\{ \widehat{\bg}_i^{(-k)} (\bm{1})  - \bg_i^* (\bm{1})  \Big\}
		-  \frac{w(\bC_i) \bm{1}\T}{\NI_i} \Big\{ \widehat{\bg}_i^{(-k)} (\bm{0})  - \bg_i^* (\bm{0})  \Big\} \ .
	\end{align*}
	We use the inequality $(a_1+\ldots+a_5)^2 \leq 16 (a_1^2 + \ldots+a_5^2)$ and $\big\| A B \big\|_2 \leq \big\|A \big\|_2 \big\|B\big\|_2$ to obtain an upper bound of $\big\{ \widehat{\phi}_k(\bO_i \con \beta^{\ddagger,(-k)}) - \phi^*(\bO_i \con \beta'^{,(-k)} )\big\}^2$ as follows.
	\begin{align}							\label{proof-AN-007}
		&
		\big\{  \widehat{\phi}_k(\bO_i \con \beta^{\ddagger,(-k)}) - \phi^*(\bO_i \con \beta'^{,(-k)} )  \big\}^2
		\\
		& \leq
		w(\bC_i)^2
		\Bigg[
		\Big\| \widehat{\bI}_i^{(-k)} - \bI_i^* \Big\|_2^2
		\Big\| B_i^{\ddagger,(-k)}  + B_i'^{,(-k)} \Big\|_2^2
		\Big\| \widehat{\bm{R}}_i^{(-k)}  + \bm{R}_i^* \Big\|_2^2
		\nonumber
		\\
		& \hspace*{2cm}
			+
		\Big\| \widehat{\bI}_i^{(-k)} + \bI_i^* \Big\|_2^2
		\Big\| B_i^{\ddagger,(-k)}  - B_i'^{,(-k)} \Big\|_2^2
		\Big\| \widehat{\bm{R}}_i^{(-k)}  + \bm{R}_i^* \Big\|_2^2
		\nonumber
		\\
		& \hspace*{2cm}
		+
		4
		\Big\| \widehat{\bI}_i^{(-k),\intercal} B_i^{\ddagger,(-k)}  + \bI_i^{*,\intercal} B_i'^{,(-k)} \Big\|_2^2
		\Big\| \widehat{\bm{R}}_i^{(-k)} - \bm{R}_i^* \Big\|_2^2
		\nonumber
		\\
		& \hspace*{2cm}
		+ 
		\frac{16}{\NI_i}
		\Big\| \widehat{\bg}_i^{(-k)} (\bm{1})  - \bg_i^* (\bm{1}) \Big\|_2^2
		+ 
		\frac{16}{\NI_i}
		\Big\| \widehat{\bg}_i^{(-k)} (\bm{0})  - \bg_i^* (\bm{0}) \Big\|_2^2
		\Bigg] \ .
		\nonumber
	\end{align}
	Since $|w(\bC_i)|\leq C_w$, it is sufficient to study the asymptotic behavior of the following terms.
	\begin{align}
		&
		\EXP \bigg\{
			\Big\| \widehat{\bI}_i^{(-k)} - \bI_i^* \Big\|_2^2
		\Big\|  B_i^{\ddagger,(-k)}  + B_i'^{,(-k)} \Big\|_2^2
		\Big\| \widehat{\bm{R}}_i^{(-k)}  + \bm{R}_i^* \Big\|_2^2
		\, \bigg| \, \II_k^c \bigg\}
		\label{proof-gap1}
		\\
		&
		\EXP \bigg\{
			\Big\| \widehat{\bI}_i^{(-k)} + \bI_i^* \Big\|_2^2
		\Big\|  B_i^{\ddagger,(-k)}  - B_i'^{,(-k)} \Big\|_2^2
		\Big\| \widehat{\bm{R}}_i^{(-k)}  + \bm{R}_i^* \Big\|_2^2
		\, \bigg| \, \II_k^c \bigg\}
		\label{proof-gap2}
		\\
		&
		\EXP \bigg\{
		\Big\| \widehat{\bI}_i^{(-k),\intercal} B_i^{\ddagger,(-k)}  + \bI_i^{*,\intercal} B_i'^{,(-k)} \Big\|_2^2
		\Big\| \widehat{\bm{R}}_i^{(-k)} - \bm{R}_i^* \Big\|_2^2
		\, \bigg| \, \II_k^c \bigg\}
		\label{proof-gap3}
		\\
		&
		\EXP \bigg\{
		\Big\| \widehat{\bg}_i^{(-k)} (\ba)  - \bg_i^* (\ba) \Big\|_2^2
		\, \bigg| \, \II_k^c \bigg\}
		\ , \ba=\bm{1},\bm{0} \ .
		\label{proof-gap4}
	\end{align}
	
	First, we find $\EXP \big\{ \| \widehat{\bI}_i^{(-k)} - \bI_i^*\|_2^2 \cond \II_k^c \big\} = O_P(r_{\pi,\NC}^2)$ and $\| \widehat{\bI}_i^{(-k)} + \bI_i^*\|_2^2 \leq 4C_\pi^2$ from conditions (c) and (d) of Lemma \ref{lemma:tauhat}. 
	
	Second, using $B_i = I - (\bm{1}\bm{1}\T-I)\beta_m$ for $\NI_i =m$, we have
	\begin{align*}
		\EXP \bigg\{
		\Big\| B_i^{\ddagger,(-k)}  - B_i'^{,(-k)} \Big\|_2^2
		\, \bigg| \, \II_k^c \bigg\}
		&
		\leq
		\big\| \bm{1}\bm{1}\T-I \big\|_2^2 \EXP \Big\{ \big\|  \beta^{\ddagger,(-k)} - \beta'^{,(-k)} \big\|_2^2 \, \Big| \, \II_k^c \Big\}
		\\
		&
		\leq
		\big\| \bm{1}\bm{1}\T-I \big\|_F^2 \EXP \Big\{ \big\| \beta^{\ddagger,(-k)} - \beta'^{,(-k)} \big\|_2^2 \, \Big| \, \II_k^c \Big\}
		\\
		&
		=
		 O_P( r_{\pi,\NC}^2)  + O_P(r_{g,\NC}^2 )  \ .
	\end{align*}
	The last line holds from $\big\| \bm{1}\bm{1}\T-I \big\|_F^2 \leq \NT^2$ and condition (c) of Lemma \ref{lemma:tauhat}. We also find that $	\big\|  \widehat{B}_i^{\ddagger,(-k)} \big\|_2^2 \leq C_B^2$ for some constant $C_B$ as follows.
	\begin{align*}
		\Big\|  \widehat{B}_i^{\ddagger,(-k)} \Big\|_2^2
		&
		\leq
		\Big\| I  +  (I - \bm{1}\bm{1}\T)  \beta^{\ddagger,(-k)} \Big\|_2^2
		\leq
		2 \| I \|_2^2
		+
		2
		\big\| I - \bm{1}\bm{1}\T \big\|_2^2 \big\|\beta^{\ddagger,(-k)} \big\|_2^2
		\leq
		C_B^2 :=
		8 \NT + 4 \NT^2 B_0^2 \ ,
	\end{align*}
	and $\big\| B_i'^{,(-k)} \big\|_2^2 \leq C_B^2$ from similar manner. Thus, $\big\|  \widehat{B}_i^{\ddagger,(-k)} + B_i'^{,(-k)} \big\|_2^2 \leq 4C_B^2$. Moreover, we have 
	\begin{align*}
		\big\| \widehat{\bI}_i^{(-k),\intercal} \widehat{B}_i^{\ddagger,(-k)} \big\|_2^2 
		\leq 
		\big\| \widehat{\bI}_i^{(-k)} \big\|_2^2 \big\| \widehat{B}_i^{\ddagger,(-k)} \big\|_2^2 
		\leq 
		C_\pi^2 C_B^2 \ , \
		\big\| \bI_i^{*,\intercal} B_i'^{,(-k)} \big\|_2^2 
		\leq 
		\big\| \bI_i^* \big\|_2^2 \big\| B_i'^{,(-k)} \big\|_2^2 
		\leq 
		C_\pi^2 C_B^2 \ .
	\end{align*}
	
	Third, we have $\EXP\big\{ \big\|  \widehat{\bm{R}}_i^{(-k)} - \bm{R}_i^*  \big\|_2^2 \cond \II_k^c \big\} = \EXP\big\{ \big\| \widehat{\bg}_i^{(-k)}(\bA_i) - \bg_i^*(\bA_i) \big\|_2^2 \cond \II_k^c \big\} = O_P(r_{g,\NC}^2)$ from condition (d) of Lemma \ref{lemma:tauhat}. Moreover, $  \EXP \big\{ \big\| \widehat{\bm{R}}_i^{(-k)} + \bm{R}_i^* \big\|_2^2 \cond \II_k^c \big\} $ is upper bounded by a constant as follows.
	\begin{align*}
	\EXP \big\{ \big\| \widehat{\bm{R}}_i^{(-k)} + \bm{R}_i^* \big\|_2^2 \cond \II_k^c \big\} 
	&
	=
	\EXP \Big[ \EXP \big\{ \big\| \widehat{\bm{R}}_i^{(-k)} + \bm{R}_i^* \big\|_2^2 \cond \bA_i , \bX_i , \II_k^c \big\} \, \Big| \, \II_k^c \Big]
	\\
	&
	= \EXP \Big[ 4 \bm{\epsilon}_i\T \bm{\epsilon}_i  + 4 \bm{\epsilon}_i\T \big\{ \widehat{\bg}_i^{(-k)}(\bA_i) - \bg_i^*(\bA_i) \big\} + \big\| \widehat{\bg}_i^{(-k)}(\bA_i) - \bg_i^*(\bA_i) \big\|_2^2
	\, \Big| \, \II_k^c \Big]
	\\
	&
	= \EXP \Big[ 4 \bm{1}\T \Sigma(\bA_i, \bX_i) \bm{1}  + \big\| \widehat{\bg}_i^{(-k)}(\bA_i) - \bg_i^*(\bA_i) \big\|_2^2
	\, \Big| \, \II_k^c \Big]
	\\
	& 
	\leq \NT C' + C_g^2 \ .
	\end{align*}
	The three equalities are straightforward and the last inequality is from conditions (b) and (c) of Lemma \ref{lemma:tauhat}. 
	
	Finally, under condition (d), we find 
	\begin{align*}
		&
		\EXP \Big\{
		\Big\| \widehat{\bg}_i^{(-k)} (\bA_i)  - \bg_i^* (\bA_i) \Big\|_2^2
		\, \Big| \, \II_k^c \Big\}
		\\
		&
		=
		\sum_{m=1}^\NT
		\Bigg[
		P(\NI_i = m) 
		\EXP \Big\{
		\Big\| \widehat{\bg}_i^{(-k)} (\bA_i)  - \bg_i^* (\bA_i) \Big\|_2^2
		\, \Big| \, 
		\NI_i = m , 
		\II_k^c
		\Big\}
		\Bigg]
		\\
		& 
		= 
		\sum_{m=1}^\NT
		\Bigg[
		P(\NI_i = m) 
		\EXP \bigg[
			\sum_{\ba_i \in \{0,1\}^m }
			P(\bA_i = \ba_i \cond \bX_i, \NI_i = m)
			\\
			&
			\hspace*{4cm}
			\times
			\EXP \Big\{
		\Big\| \widehat{\bg}_i^{(-k)} (\ba_i)  - \bg_i^* (\ba_i) \Big\|_2^2
		\, \Big| \, 
		\bA_i = \ba_i ,
		\bX_i ,
		\NI_i = m , 
		\II_k^c
		\Big\}
		\, \bigg| \, \NI_i=m, \II_k^c
		\bigg]
		\Bigg] \ .
	\end{align*}
	From the positivity assumption, $P(\bA_i = \ba_i \cond \bX_i, \NI_i = m) \geq \delta$ for some constant $\delta$, especially at $\ba_i=\bm{1}$. Therefore, $\EXP \big\{
		\big\| \widehat{\bg}_i^{(-k)} (\bA_i)  - \bg_i^* (\bA_i) \big\|_2^2
		\cond \II_k^c \big\}$ is lower bounded by
		\begin{align*}
		&
			\EXP \Big\{
		\Big\| \widehat{\bg}_i^{(-k)} (\bA_i)  - \bg_i^* (\bA_i) \Big\|_2^2
		\, \Big| \, \II_k^c \Big\}
		\\
		& 
		\geq 
		\delta
		\sum_{m=1}^\NT
		\Bigg[
		P(\NI_i = m) 
		\EXP \bigg[			
			\EXP \Big\{
		\Big\| \widehat{\bg}_i^{(-k)} (\bm{1})  - \bg_i^* (\bm{1}) \Big\|_2^2
		\, \Big| \, 
		\bA_i = \bm{1} ,
		\bX_i ,
		\NI_i = m , 
		\II_k^c
		\Big\}
		\, \bigg| \, \NI_i=m, \II_k^c
		\bigg]
		\Bigg] 
		\\
		&
		=
		\delta
		\sum_{m=1}^\NT
		\Bigg[
		P(\NI_i = m) 
		\EXP \bigg\{			\Big\| \widehat{\bg}_i^{(-k)} (\bm{1})  - \bg_i^* (\bm{1}) \Big\|_2^2
		\, \bigg| \, \NI_i=m, \II_k^c
		\bigg\}
		\Bigg]
		\\
		&
		=
		\delta
		\EXP \Big\{			\Big\| \widehat{\bg}_i^{(-k)} (\bm{1})  - \bg_i^* (\bm{1}) \Big\|_2^2
		\, \Big| \, \II_k^c
		\Big\} \ .
		\end{align*}
		Since $\EXP \big\{
		\big\| \widehat{\bg}_i^{(-k)} (\bA_i)  - \bg_i^* (\bA_i) \big\|_2^2
		\cond \II_k^c \big\} = O_P(r_{g,\NC}^2)$, we obtain $\EXP \big\{ \big\| \widehat{\bg}_i^{(-k)} (\bm{1})  - \bg_i^* (\bm{1}) \big\|_2^2	\cond \II_k^c 		\big\} =O_P(r_{g,\NC}^2)$. Similarly, $\EXP \big\{ \big\| \widehat{\bg}_i^{(-k)} (\bm{0})  - \bg_i^* (\bm{0}) \big\|_2^2	\cond \II_k^c 		\big\} =O_P(r_{g,\NC}^2)$.

	Using the established results, we find the convergence rates of \eqref{proof-gap1}-\eqref{proof-gap4}. First, the rate of \eqref{proof-gap1} is $O_P(r_{\pi,\NC}^2)$.
	\begin{align*}
		&
		\EXP \bigg\{
			\Big\| \widehat{\bI}_i^{(-k)} - \bI_i^* \Big\|_2^2
		\Big\|  \widehat{B}_i^{\ddagger,(-k)} + B_i'^{,(-k)} \Big\|_2^2
		\Big\| \widehat{\bm{R}}_i^{(-k)}  + \bm{R}_i^* \Big\|_2^2
		\, \bigg| \, \II_k^c \bigg\}
		\\
		&
		=
		\EXP \Bigg[
			\Big\| \widehat{\bI}_i^{(-k)} - \bI_i^* \Big\|_2^2
		\Big\|  \widehat{B}_i^{\ddagger,(-k)} + B_i'^{,(-k)} \Big\|_2^2
		\EXP \bigg\{
		\Big\| \widehat{\bm{R}}_i^{(-k)}  + \bm{R}_i^* \Big\|_2^2
		\, \bigg| \, \bA_i, \bX_i, \II_k^c \bigg\}
		\, \Bigg| \, \II_k^c \Bigg]
		\\
		& 
		\leq
		4C_B^2 ( \NT C' + C_g^2 ) \EXP \Big\{ \Big\| \widehat{\bI}_i^{(-k)} - \bI_i^* \Big\|_2^2 \, \Big| \,  \II_k^c \Big\} 
		\\
		& 
		= O_P(r_{\pi,\NC}^2) \ .
	\end{align*}
	Second, the rate of \eqref{proof-gap2} is $O_P( r_{\pi,\NC}^2)  + O_P(r_{g,\NC}^2 ) $.
	\begin{align*}
		&
		\EXP \bigg\{
			\Big\| \widehat{\bI}_i^{(-k)} + \bI_i^* \Big\|_2^2
		\Big\|  \widehat{B}_i^{\ddagger,(-k)} - B_i'^{,(-k)} \Big\|_2^2
		\Big\| \widehat{\bm{R}}_i^{(-k)}  + \bm{R}_i^* \Big\|_2^2
		\, \bigg| \, \II_k^c \bigg\}
		\\
		& 
		\leq
		\EXP \Bigg[
			\Big\| \widehat{\bI}_i^{(-k)} + \bI_i^* \Big\|_2^2
		\Big\|  \widehat{B}_i^{\ddagger,(-k)} - B_i'^{,(-k)} \Big\|_2^2
		\EXP \bigg\{
		\Big\| \widehat{\bm{R}}_i^{(-k)}  + \bm{R}_i^* \Big\|_2^2
		\, \bigg| \, \bA_i, \bX_i, \II_k^c \bigg\}
		\, \Bigg| \, \II_k^c \Bigg]
		\\
		& 
		\leq
		4C_\pi^2 ( \NT C' + C_g^2 ) \EXP \bigg\{
		\Big\|  \widehat{B}_i^{*\ddagger,(-k)} - B_i'^{,(-k)} \Big\|_2^2
		\, \bigg| \, \II_k^c \bigg\}
		\\
		& 
		=
		O_P( r_{\pi,\NC}^2)  + O_P(r_{g,\NC}^2 )  \ .
	\end{align*}
	Third, the rate of \eqref{proof-gap3} is $O_P(r_{g,\NC}^2)$.
	\begin{align*}
		&
		\EXP \bigg\{
		\Big\| \widehat{\bI}_i^{(-k),\intercal} \widehat{B}_i^{\ddagger,(-k)} + \bI_i^{*,\intercal} B_i'^{,(-k)} \Big\|_2^2
		\Big\| \widehat{\bm{R}}_i^{(-k)} - \bm{R}_i^* \Big\|_2^2
		\, \bigg| \, \II_k^c \bigg\}
		\leq
		4C_\pi^2 C_B^2
		\EXP \bigg\{
		\Big\| \widehat{\bm{R}}_i^{(-k)} - \bm{R}_i^* \Big\|_2^2
		\, \bigg| \, \II_k^c \bigg\} = 
		O_P(r_{g,\NC}^2)  \ .
	\end{align*}
	Lastly, \eqref{proof-gap4} is $O_P(r_{g,\NC}^2)$ from the established result. As a consequence, by plugging in the rate in \eqref{proof-AN-007}, we have $\EXP ( [B]^2 \cond \II_k^c \big) =O_P( r_{\pi,\NC}^2)  + O_P(r_{g,\NC}^2 ) $. Moreover, this implies $[B] = O_P( r_{\pi,\NC}^2)  + O_P(r_{g,\NC}^2 ) $ from Lemma 6.1 of \citet{Victor2018}.
	
	\newpage
	
	\item  \textbf{(Rate of $[C]$)}
		
		$[C]$ is represented as 
		\begin{align*}
		[C]
		&
		=
		\EXP \Big\{ \widehat{\phi}_k(\bO_i \con \beta^{\ddagger,(-k)} ) - \phi^*(\bO_i \con \beta'^{,(-k)}) \, \Big| \, \II_k^c \Big\}
		\\
		&
		=
		\EXP \Bigg[ w(\bC_i) \bigg[
		\widehat{\bI}_i^{(-k),\intercal} B_i^{\ddagger,(-k)} \widehat{\bm{R}}_i^{(-k)}
	+
	\frac{\bm{1}\T}{\NI_i} \Big\{ \widehat{\bg}_i^{(-k)}(\bm{1}) - \widehat{\bg}_i^{(-k)}(\bm{0}) \Big\} 		
	-
	\frac{\bm{1}\T}{\NI_i} \Big\{ \bg_i^*(\bm{1}) - \bg_i^*(\bm{0}) \Big\} 		
	\bigg]
	\, \Bigg| \, \II_k^c \Bigg] \ .
		\end{align*}	
	The first term in the second line is decomposed into 
	\begin{align}								\label{proof-AN-008}
	\EXP \Big\{
	w(\bC_i) \widehat{\bI}_i^{(-k),\intercal} B_i^{\ddagger,(-k)} \widehat{\bm{R}}_i^{(-k)} \, \Big| \, \II_k^c \Big\}
	& = 
	\EXP \Big[ w(\bC_i) \bI_i^{*,\intercal} B_i^{\ddagger,(-k)} \big\{ \bg_i^*(\bA_i) - \widehat{\bg}_i^{(-k)}(\bA_i) \big\} \, \Big| \, \II_k^c \Big]
	\\
	& \hspace*{1cm}
	+
	\EXP \Big[ w(\bC_i) \big\{ \widehat{\bI}_i^{(-k)} -\bI_i^* \big\} \T B_i^{\ddagger,(-k)} \big\{ \bg_i^*(\bA_i) - \widehat{\bg}_i^{(-k)}(\bA_i) \big\} \, \Big| \, \II_k^c \Big] \ .
	\nonumber
	\end{align}
	Taking $g' = g^* - \widehat{g}^{(-k)}$ in Section \ref{proof-thm:2-1}, the first term of \eqref{proof-AN-008} is equivalent as
	\begin{align*}
		&
		\EXP \Big[ w(\bC_i) \bI_i^{*,\intercal} B_i^{\ddagger,(-k)} \big\{ \bg_i^*(\bA_i) - \widehat{\bg}_i^{(-k)}(\bA_i) \big\} \, \Big| \, \II_k^c \Big]
		=
		\EXP \Bigg[
		\frac{\bm{1}\T}{\NI_i} \Big\{ \bg_i^*(\bm{1}) - \bg_i^*(\bm{0}) \Big\} 		
		-
	\frac{\bm{1}\T}{\NI_i} \Big\{ \widehat{\bg}_i^{(-k)}(\bm{1}) - \widehat{\bg}_i^{(-k)}(\bm{0}) \Big\} 		
	\, \Bigg| \, \II_k^c \Bigg] \ .
	\end{align*}
	The second term of \eqref{proof-AN-008} is 
	\begin{align*}
		&
		\bigg| 
		\EXP \Big[ \big\{ \widehat{\bI}_i^{(-k)} -\bI_i^* \big\}\T B_i^{\ddagger,(-k)} \big\{ \bg_i^*(\bA_i) - \widehat{\bg}_i^{(-k)}(\bA_i) \big\} \, \Big| \, \II_k^c \Big]
		\bigg|
		\\
		&
		\leq 
		\EXP \Big[ \big\| \widehat{\bI}_i^{(-k)} -\bI_i^* \big\|_2 \big\|B_i^{\ddagger,(-k)}\big\|_2 \big\| \bg_i^*(\bA_i) - \widehat{\bg}_i^{(-k)}(\bA_i) \big\|_2 \, \Big| \, \II_k^c \Big]
		\\
		& 
		\leq		
		C'  \big\| \widehat{\bI}_i^{(-k)} -\bI_i^* \big\|_{P,2}  \big\| \bg_i^*(\bA_i) - \widehat{\bg}_i^{(-k)}(\bA_i) \big\|_{P,2}
		\\
		& = O_P(r_{\pi,\NC}r_{g,\NC}) \ .
	\end{align*}
	
	\end{itemize}

We use the established rates of $[B]$ and $[C]$ in \eqref{proof-AN-006} which leads to the following result.
\begin{align*}
	\frac{1}{\sqrt{\NC/2}} \sum_{i \in \II_k} \Big\{ \widehat{\phi}_k(\bO_i \con \beta^{\ddagger,(-k)} )  - \phi^*(\bO_i \con \beta'^{,(-k)}) \Big\}
	&
	=
	[B]
	+
	\sqrt{\frac{\NC}{2}} [C]
	\\
	&
	=
	O_P( r_{\pi,\NC} ) + O_P( r_{g,\NC} )
	+
	\sqrt{\NC} O_P(r_{\pi,\NC} r_{g,\NC} ) 
	\\
	&
	=
	o_P(1)  \ .
\end{align*}
From \eqref{proof-AN-004} and \eqref{proof-AN-005}, we find
\begin{align}									\label{proof-AN-009}
	\sqrt{ \frac{\NC}{2} } \big( \widehat{\tau}_{\PARASUB k} - \tau_{\PARASUB}^* \big) 
	&=
		\frac{1}{\sqrt{\NC/2}} \sum_{i \in \II_k} \Big\{ \widehat{\phi}_k(\bO_i \con \widehat{\beta}^{(k)} )  -\tau_{\PARASUB}^* \Big\}
		\nonumber
		\\
		&
		=
		\frac{1}{\sqrt{\NC/2}} \sum_{i \in \II_k} \Big\{ \phi^*(\bO_i \con \beta'^{,(-k)}) - \tau_{\PARASUB}^* \Big\}
		+
		o_P(1) 
		\nonumber
		\\
		&
		\stackrel{D}{\rightarrow}
		N \Big( 0 , \VAR\big\{  \phi^*(\bO_i \con \beta'^{,(-k)}) \big\} \Big) 
\end{align}
and this implies
\begin{align*}
	\sqrt{\NC}\big( \widehat{\tau}_{\PARASUB} - \tau_{\PARASUB}^* \big)
		&
		=
		\frac{1}{\sqrt{2}} \bigg\{
			\sqrt{\frac{\NC}{2}} \big( \widehat{\tau}_{\PARASUB 1} - \tau_{\PARASUB}^* \big) 
			+
			\sqrt{\frac{\NC}{2}} \big( \widehat{\tau}_{\PARASUB 2} - \tau_{\PARASUB}^* \big) 
		\bigg\}
		\\
		&
		=
		\frac{1}{\sqrt{2}} \sum_{k=1}^2 
		\frac{1}{\sqrt{\NC/2}} \sum_{i \in \II_k} \Big\{ \phi^*(\bO_i \con \beta'^{,(-k)}) - \tau_{\PARASUB}^* \Big\}
		+
		o_P(1)  
		\\
		&
		\stackrel{D}{\rightarrow}
		N \Big( 0 , \sigma^2(\beta') \Big) 
		\quad, \quad
		\sigma^2 (\beta') =
		\frac{1}{2} \Big[
		\VAR \big\{ \phi(\bO_i \con \beta'^{,(-1)}) \big\} + \VAR \big\{ \phi(\bO_i \con \beta'^{,(-2)}) \big\}
		\Big] \ .
\end{align*}
From Lemma \ref{lemma:consistency1}, we have $\beta'^{,(-1)} = \beta'^{,(-2)} = \beta^*$ under the assumption in the main paper, and this concludes the asymptotic normality.

	\item \textbf{[Step 3]}: Lastly, we show that the variance estimator is consistent. Let $\sigma_k^2$ be $ \VAR\big\{  \phi^*(\bO_i \con \beta'^{,(-k)}) \big\} $ and $\widehat{\sigma}_k^2 = (\NC/2)^{-1} \sum_{i \in \II_k} \{
			\widehat{\phi}_k(\bO_i \con \widehat{\beta}^{(k)} ) - \widehat{\tau}_{\PARASUB k} \}^2 $. We decompose $\widehat{\sigma}_k^2 - \sigma_k^2 $ as $\widehat{\sigma}_k^2 - S_k^2 + S_k^2 - \sigma_k^2$ where
	\begin{align*}
		S_k^2
		=
		\frac{1}{\NC/2} \sum_{i \in \II_k} \Big\{ \phi^* (\bO_i \con \beta'^{,(-k)} )  - \tau_{\PARASUB}^* \Big\}^2
		 \ .
	\end{align*}
	From the law of large numbers, we have $S_k^2 - \sigma_k^2 = o_P(1)$ so it is sufficient to show $\widehat{\sigma}_k^2 - S_k^2 = o_P(1)$ which is represented as follows.
	\begin{align*}
		\big|
		\widehat{\sigma}_k^2 - S_k^2
		\big|
		&
		=
		\bigg|
		\frac{1}{\NC/2} \sum_{i \in \II_k} 
		\Big\{
			\widehat{\phi}_k(\bO_i \con \widehat{\beta}^{(k)} ) - \widehat{\tau}_{\PARASUB k}
		\Big\}^2
		-
		\frac{1}{\NC/2} \sum_{i \in \II_k} \Big\{ \phi^* (\bO_i \con \beta'^{,(-k)} )  - \tau_{\PARASUB}^* \Big\}^2
		\bigg|
		\\
		& =
		\bigg|
		\underbrace{ 
		\frac{1}{\NC/2} \sum_{i \in \II_k} 
		\Big\{
			\widehat{\phi}_k(\bO_i \con \widehat{\beta}^{(k)} ) - \widehat{\tau}_{\PARASUB k}
			-
			 \phi^* (\bO_i \con \beta'^{,(-k)} )  + \tau_{\PARASUB}^* 
		\Big\}^2
		}_{V_\NC}
		\\
		& \hspace*{1cm}
		+
		\frac{2}{\NC/2} \sum_{i \in \II_k} 
		\Big\{
			\widehat{\phi}_k(\bO_i \con \widehat{\beta}^{(k)} ) - \widehat{\tau}_{\PARASUB k}
			-
			 \phi^* (\bO_i \con \beta'^{,(-k)} )  + \tau_{\PARASUB}^* 
		\Big\} \Big\{ \phi^* (\bO_i \con \beta'^{,(-k)} )  - \tau_{\PARASUB}^* \Big\}
		\bigg|
		\\
		& 
		\leq V_\NC + 2 \sqrt{V_\NC} S_k^2 \ .
	\end{align*}
	The inequality holds from the H\"older's inequality. Since $S_k^2 = \sigma_k^2 + o_P(1) = O_P(1)$, it suffices to show that $V_\NC = o_P(1)$.
	
	We observe that $V_\NC$ is upper bounded by
	\begin{align}								\label{proof-AN-101}
		V_\NC
		\leq 
		\frac{2}{\NC/2} \sum_{i \in \II_k} \Big\{ \widehat{\phi}_k(\bO_i \con \widehat{\beta}^{(k)} ) - \phi^* (\bO_i \con \beta'^{,(-k)} ) \Big\}^2
		+
		\Big(  \widehat{\tau}_{\PARASUB k} - \tau_{\PARASUB}^*  \Big)^2 \ .
	\end{align}
	To bound the empirical mean in \eqref{proof-AN-101}, we again empirical process methods. Let $\mathcal{S}$ be the collection of functions $\big\{ \widehat{\phi}_k(\bO_i \con \beta ) - \phi^* (\bO_i \con \beta'^{,(-k)} )  \big\}^2$ for $\beta \in \mathcal{B}$, i.e.
	\begin{align*}
		\mathcal{S} : = \Big\{ \big\{ \widehat{\phi}_k(\bO_i \con \beta ) - \phi^* (\bO_i \con \beta'^{,(-k)} )  \big\}^2 \, \Big| \, \beta \in \mathcal{B} \Big\} \ .
	\end{align*}
	We find that $\big\{ \widehat{\phi}_k(\bO_i \con \beta ) - \phi^* (\bO_i \con \beta'^{,(-k)} )  \big\}^2$ has a form
	\begin{align*}
		&
		\big\{ \widehat{\phi}_k(\bO_i \con \beta ) - \phi^* (\bO_i \con \beta'^{,(-k)} )  \big\}^2
		\\
		&
		=
		\underbrace{
		w(\bC_i)^2
		\widehat{\bI}_i^{(-k),\intercal} (I - \bm{1}\bm{1}\T) \big\{
			\bY_i - \widehat{\bg}_i^{(-k)} (\bA_i) 
		\big\}^{\otimes 2} (I - \bm{1}\bm{1}\T) \widehat{\bI}_i^{(-k)} }_{\widehat{\psi}_{3,k}(\bO_i)} \beta(\bC_i)^2
		\\
		& 
		\hspace*{1cm}
		-
		2
		\underbrace{
		w(\bC_i)^2
		\widehat{\bI}_i^{(-k),\intercal} (I - \bm{1}\bm{1}\T) \big\{
			\bY_i - \widehat{\bg}_i^{(-k)} (\bA_i) 
		\big\} \big\{ \bY_i - \bg_i^*(\bA_i)  \big\} (I - \bm{1}\bm{1}\T) \bI_i^*  \beta'^{,(-k)}(\bC_i)
		}_{\widehat{\psi}_{4,k}(\bO_i)} \beta(\bC_i)
		\\
		&
		\hspace*{1cm}
		+
		\underbrace{
		w(\bC_i)^2
		\bI_i^{*,\intercal}  (I - \bm{1}\bm{1}\T)  \big\{ \bY_i - \bg_i^*(\bA_i)  \big\}^{\otimes 2} (I - \bm{1}\bm{1}\T) \bI_i^* \beta'^{,(-k)}(\bC_i)^2 }_{\psi_{5,k}(\bO_i)}
		\\
		& 
		=
		\lambda( \widehat{\psi}_{3,k}, \widehat{\psi}_{4,k}, \psi_{5,k} , \beta ) 
	\end{align*}
	where $\lambda: \R^4 \rightarrow \R$ is a continuous map with $\lambda(x_1,x_2,x_3,x_4) = x_1 x_4^2 - 2 x_2x_4 + x_3$. Using $\lambda$, we find $\mathcal{S} = \lambda \big( \big\{ \widehat{\psi}_{3,k} \big\}, \big\{ \widehat{\psi}_{4,k} \big\}, \big\{ \psi_{5,k} \big\} , \mathcal{B} \big)$. We find that $\big\{ \widehat{\psi}_{3,k} \big\}$, $\big\{ \widehat{\psi}_{4,k} \big\}$, and $\big\{ \psi_{5,k} \big\}$ are Glivenko-Cantelli because they are singleton sets and integrable \citep[page 270]{Vaart1998}. Moreover, $\mathcal{B}$ is Glivenko-Cantelli because it is Donsker \citep[page 82]{VW1996}. Lastly, for any $\beta \in \mathcal{B}$, we find $\big| \beta(\bC_i) \big| \leq B_0$ and, as a consequence, we have $ \big| \big\{ \widehat{\phi}_k(\bO_i \con \beta ) - \phi^* (\bO_i \con \beta'^{,(-k)} )  \big\}^2 \big| \leq S(\bO_i) := B_0^2 \widehat{\psi}_{3,k}(\bO_i) + 2B_0 \widehat{\psi}_{4,k}(\bO_i) + \psi_{5,k}(\bO_i)$, i.e. $S(\bO_i)$ is the envelope function of $\mathcal{S}$ which is integrable as follows
	\begin{align*}
		\EXP \Big\{ S(\bO_i) \, \Big| \, \II_k^c \Big\}
		& =
		B_0^2 \EXP \Big\{ \widehat{\psi}_{3,k} (\bO_i) \, \Big| \, \II_k^c \Big\}
		+
		2
		B_0 \EXP \Big\{ \widehat{\psi}_{4,k} (\bO_i)  \, \Big| \, \II_k^c \Big\}
		+
		\EXP \Big\{ \widehat{\psi}_{5,k} (\bO_i) \, \Big| \, \II_k^c \Big\}
		\\
		&
		\leq
		B_0^2 C_w^2 \NT^2 \EXP \Big\{ \bm{1}\T \Sigma(\bA_i, \bX_i) \bm{1} + \big\| \bg_i^*(\bA_i) - \widehat{\bg}_i^{(-k)}(\bA_i) \big\|_2^2  \, \Big| \, \II_k^c \Big\}
		\\
		& \hspace*{2cm}
		+
		2
		B_0^2 C_w^2 \NT^2 \EXP \Big\{  \bm{1}\T \Sigma(\bA_i, \bX_i) \bm{1} \, \Big| \, \II_k^c \Big\}
		+
		B_0^2 C_w^2 \NT^2
		\EXP \Big\{  \bm{1}\T \Sigma(\bA_i, \bX_i) \bm{1} \, \Big| \, \II_k^c \Big\}
		\\
		&
		\leq
		4 B_0^2 C_w^2 \NT^4 C_2 + B_0^2 C_w^2 \NT^2 C_g^2
		 \ .
	\end{align*}
	Therefore, Theorem 3 of \citet{VW2000} can be applied so $\mathcal{S}$ is Glivenko-Cantelli. As a consequence, we have the following result for the empirical mean in \eqref{proof-AN-101}.
	\begin{align*}
		\frac{2}{\NC/2} \sum_{i \in \II_k} \Big\{ \widehat{\phi}_k(\bO_i \con \widehat{\beta}^{(k)} ) - \phi^* (\bO_i \con \beta'^{,(-k)} ) \Big\}^2
		=
		2 
	\int 
	\Big\{ \widehat{\phi}_k(\bo_i \con \widehat{\beta}^{(k)} ) - \phi^*(\bo_i \con \beta'^{,(-k)}) \Big\}^2 
	\, dP(\bo_i)
	+
	o_P(1) \ . 
	\end{align*}
	The integral in the right hand side has the following rate.
\begin{align}									\label{proof-AN-102}
	&
	\int 
	\Big\{ \widehat{\phi}_k(\bo_i \con \widehat{\beta}^{(k)} ) - \phi^*(\bo_i \con \beta'^{,(-k)}) \Big\}^2 
	\, dP(\bo_i)
	\nonumber
	\\
	&
	\leq 
	2
	\int 
	\Big\{ \widehat{\phi}_k(\bo_i \con \widehat{\beta}^{(k)} ) - \widehat{\phi}_k(\bo_i \con \beta^{\ddagger,(-k)}) \Big\}^2 
	\, dP(\bo_i)
	+
	2
	\int 
	\Big\{\widehat{\phi}_k(\bo_i \con \beta^{\ddagger,(-k)}) - \phi^*(\bo_i \con \beta'^{,(-k)}) \Big\}^2 
	\, dP(\bo_i)
	\nonumber
	\\
	& =
	O_P(r_{\beta,\NC}^2) + O_P(r_{\pi,\NC}^2) + O_P(r_{g,\NC}^2) \ .
\end{align}
To establish the rate we observe that the first term of \eqref{proof-AN-102} is $O_P(r_{\beta,\NC}^2)$ from \eqref{proof-AN-010} and that the second term \eqref{proof-AN-102} is equal to $2\EXP \big[  \big\{ \widehat{\phi}_k(\bO_i \con \beta^{\ddagger,(-k)}) - \phi^*(\bO_i \con \beta'^{,(-k)} ) \big\}^2  \cond \II_k^c \big] = 2 \EXP \big\{ [B]^2 \cond \II_k^c \big\} = O_P(r_{\pi,\NC}^2) + O_P(r_{g,\NC}^2)$ from \eqref{proof-AN-011}. Therefore, the first term of \eqref{proof-AN-101} is $o_P(1)$. The second term of \eqref{proof-AN-101} is also $o_P(1)$  from \eqref{proof-AN-009}. This implies $V_\NC = o_P(1)$. This concludes the proof.

\end{itemize}

\subsubsection{Proof of (1) Double Robustness}						\label{proof-thm:3-(1)}

Similar to the proof of (2) Asymptotic Normality, we consider the following decomposition.
\begin{align*}
	& \frac{2}{\NC} \sum_{i \in \II_k} 
	\big\{ 
	\phi(\bO_i, \widehat{\pi}^{(-k)}, \widehat{g}^{(-k)}, \widehat{\beta}^{(k)})
	- 
	\tau^* 
	\big\}
	\\
	& = 
	\underbrace{
	\frac{2}{\NC} \sum_{i \in \II_k} 
	\Bigg[
	\bigg\{
	\begin{array}{l}
	\phi(\bO_i, \widehat{\pi}^{(-k)}, \widehat{g}^{(-k)}, \widehat{\beta}^{(k)})
	\\
	-
	\phi(\bO_i, \widehat{\pi}^{(-k)}, \widehat{g}^{(-k)}, \beta^{\ddagger,(-k)})
	\end{array}	
	\bigg\}
	-
	\EXP \bigg\{
	\begin{array}{l}
	\phi(\bO_i, \widehat{\pi}^{(-k)}, \widehat{g}^{(-k)}, \widehat{\beta}^{(k)})
	\\
	-
	\phi(\bO_i, \widehat{\pi}^{(-k)}, \widehat{g}^{(-k)}, \beta^{\ddagger,(-k)})
	\end{array}	
	\, \bigg| \, \II_k^c \bigg\}
	\Bigg]
	}_{[A] =o_P(1)}
	\\
	&
	+
	\underbrace{
	\EXP \bigg\{
	\begin{array}{l}
	\phi(\bO_i, \widehat{\pi}^{(-k)}, \widehat{g}^{(-k)}, \widehat{\beta}^{(k)})
	\\
	-
	\phi(\bO_i, \widehat{\pi}^{(-k)}, \widehat{g}^{(-k)}, \beta^{\ddagger,(-k)})
	\end{array}	
	\, \bigg| \, \II_k^c \bigg\}
	}_{[B] = O_P( r_{g,\NC} r_{\beta,\NC})}
	\\
	&
	+
	\underbrace{
	\frac{2}{\NC} \sum_{i \in \II_k} 
	\Bigg[
	\bigg\{ 
	\begin{array}{l}
	\phi(\bO_i, \widehat{\pi}^{(-k)}, \widehat{g}^{(-k)}, \beta^{\ddagger,(-k)})
	\\
	\hspace*{1cm}
	-	\phi(\bO_i, \pi', g',  \beta^{\ddagger,(-k)})
	\end{array}
	\bigg\}
	-
	\EXP
	\bigg\{ 
	\begin{array}{l}
	\phi(\bO_i, \widehat{\pi}^{(-k)}, \widehat{g}^{(-k)}, \beta^{\ddagger,(-k)})
	\\
	\hspace*{1cm}
	-	\phi(\bO_i, \pi', g',  \beta^{\ddagger,(-k)})
	\end{array}
	\, \bigg| \, \II_k^c
	\bigg\}
	\Bigg]
	}_{[C] =o_P(1)}
	\\
	&
	+ 
	\underbrace{
	 \EXP
	\bigg\{ 
	\begin{array}{l}
	\phi(\bO_i, \widehat{\pi}^{(-k)}, \widehat{g}^{(-k)}, \beta^{\ddagger,(-k)})
	\\
	\hspace*{1cm}
	-	\phi(\bO_i, \pi', g',  \beta^{\ddagger,(-k)})
	\end{array}
	\, \bigg| \, \II_k^c
	\bigg\} 
	}_{[D] = O_P( r_{\pi,\NC} r_{g,\NC})}
	\\
	&
	+
	\underbrace{
	\frac{2}{\NC} \sum_{i \in \II_k} 
	\big\{ \phi(\bO_i, \pi', g',  \beta^{\ddagger,(-k)})
	-
	\tau^*
	\big\}
	}_{[E] =o_P(1) } \ .
\end{align*}

In \textbf{[Case PS]} and \textbf{[Case OR]}, we assume \hyperlink{Main-(EN2.PS)}{(EN2.PS)} (i.e., $\pi' = \pi^*$) and \hyperlink{Main-(EN2.OR)}{(EN2.OR)} (i.e., $g'=g^*$), respectively. The proof is similar to the proof of Theorem  \ref{Main-thm:3}-(b) in Section \ref{proof-thm:3-(2)} except (i) we consider the empirical mean instead of $\sqrt{\NC}$-scaled empirical mean and (ii) we use $ \phi'(\bO_i \con \beta^{\ddagger, (-k)} ) $ instead of $\phi^*(\bO_i \con \beta'^{,(-k)})$ in equation \eqref{proof-AN-005}.

\begin{itemize}[leftmargin=0.5cm]
	\item \textbf{[Case PS]}: We assume $\pi' = \pi^*$. 
By following \textbf{[Step 1]} and \textbf{[Step 2]} in the proof of Theorem \ref{Main-thm:3}, we find that $[A]/\sqrt{\NC/2}$ in \eqref{proof-AN-003} is upper bounded by
\begin{align*}
		\bigg|
	\frac{[A]}{\sqrt{\NC/2}}
	\bigg|
	&
	=
		\Big|
		m \big( \widehat{\beta}^{(k)} \big)
		-
		m \big( \beta^{\ddagger, (-k)} \big)
		\Big|
		 \\
		 &
		 =
		 \Bigg|
		 \int \widehat{\phi}_{2,k} (\bo_i) \Big\{ \widehat{\beta}^{(k)}(\bc_i) - \beta^{\ddagger, (-k)} (\bc_i) \Big\} \, dP(\bo_i)
		 \Bigg|
		 \\
		 & 
		 = 
		 \Bigg|
		 \int  \bI (\ba_i, \bx_i \con \widehat{\pi}^{(-k)})\T (I - \bm{1}\bm{1}\T ) \Big\{ \bg^*(\ba_i, \bx_i) - \widehat{\bg}^{(-k)} (\ba_i, \bx_i) \Big\}
		 \Big\{ \widehat{\beta}^{(k)}(\bc_i) - \beta^{\ddagger, (-k)}(\bc_i) \Big\} \, dP(\bo_i)
		 \Bigg|
		  \\
		 & 
		 =
			C_\pi \NT \,
			\underbrace{
		 \Big\| \bg^*(\ba_i, \bx_i) - \widehat{\bg}^{(-k)} (\ba_i, \bx_i) \Big\|_{P,2}
		 }_{\leq C_g}
		 \bigg[
 		 \int 
		 \Big\{ \widehat{\beta}^{(k)}(\bc_i) - \beta^{\ddagger, (-k)}(\bc_i) \Big\}^2 \, dP(\bo_i)
		 \bigg]^{1/2}
		   \\
		 & 
		= O_P (r_{\beta,\NC})  \ .
\end{align*}
	Therefore, \eqref{proof-AN-003} divided by $\sqrt{\NC/2}$ is expressed as
	\begin{align*}
		\frac{1}{\NC/2} \sum_{i \in \II_k} \Big\{ \widehat{\phi}_k(\bO_i \con \widehat{\beta}^{(k)} )  - \tau_{\PARASUB}^* \Big\}
		&
		=
		\frac{1}{\NC/2} \sum_{i \in \II_k} \Big\{ \widehat{\phi}_k (\bO_i \con \beta^{\ddagger, (-k)} ) - \tau_{\PARASUB}^* \Big\}
		+
		\frac{[A]}{\sqrt{\NC/2}} 
		+
		 o_P(\NC^{-1/2}) 
		 \\
		 &
		 =
		 \frac{1}{\NC/2} \sum_{i \in \II_k} \Big\{ \widehat{\phi}_k (\bO_i \con \beta^{\ddagger, (-k)} ) - \tau_{\PARASUB}^* \Big\}
		 +
		 o_P(1) \ .
	\end{align*}

	In \textbf{[Step 2]} in the proof of Theorem \ref{Main-thm:3}, we use $\phi'(\bO_i \con \beta^{\ddagger, (-k)} )$ instead of $\phi^*(\bO_i \con \beta'^{, (-k)})$ where
	\begin{align*}
		\phi' (\bO_i \con \beta^{\ddagger, (-k)} )
		=
		w(\bC_i)
	\bigg[
	\bI (\bA_i, \bX_i \con \pi^* )\T B (\bC_i \con \beta^{\ddagger, (-k)} ) \Big\{ \bY_i - \bg' (\bA_i, \bX_i) \Big\}
	+
	\frac{\bm{1}\T}{\NI_i} \Big\{ \bg' (\bm{1}, \bX_i) - \bg' (\bm{0}, \bX_i) \Big\}
	\bigg]  \ .
	\end{align*}
	Specifically, \eqref{proof-AN-005} divided by $\sqrt{\NC/2}$ becomes
	\begin{align*}
		&
		\frac{1}{\NC/2} \sum_{i \in \II_k} \Big\{ \widehat{\phi}_k (\bO_i \con \beta^{\ddagger, (-k)} ) - \tau_{\PARASUB}^* \Big\}
		\\
		&
		=
		\frac{1}{\NC/2} \sum_{i \in \II_k} \Big\{ \widehat{\phi}_k(\bO_i \con \beta^{\ddagger, (-k)} )  - \phi'(\bO_i \con \beta^{\ddagger, (-k)} ) \Big\}
		+
		\frac{1}{\NC/2}  \sum_{i \in \II_k} \Big\{ \phi'(\bO_i \con \beta^{\ddagger, (-k)} )  - \tau_{\PARASUB}^* \Big\} \ .
	\end{align*}
	Since $\EXP \big\{  \phi'(\bO_i \con \beta^{\ddagger, (-k)} ) \cond \II_k^c \big\} = \tau_{\PARASUB}^*$, the second term in the right hand side is $o_P(1)$ from the law of large numbers. Consequently, it suffices to show that the first term in the right hand side is $o_P(1)$, which is expressed as follows from the law of large numbers.
		\begin{align*}
		&
		\frac{1}{\NC/2} \sum_{i \in \II_k} \Big\{ \widehat{\phi}_k(\bO_i \con \beta^{\ddagger, (-k)} )  - \phi'(\bO_i \con \beta^{\ddagger, (-k)} ) \Big\}
		\\
		&
		=
		\EXP \Big\{ \widehat{\phi}_k(\bO_i \con \beta^{\ddagger, (-k)} ) - \phi'(\bO_i \con \beta^{\ddagger, (-k)} )\, \Big| \, \II_k^c \Big\} + o_P(1)
		\\
		&
		=
		\EXP \Bigg[ w(\bC_i) \bigg[
		\widehat{\bI}_i^{(-k),\intercal} B_i^{\ddagger, (-k)} \widehat{\bm{R}}_i^{(-k)}
	+
	\frac{\bm{1}\T}{\NI_i} \Big\{ \widehat{\bg}_i^{(-k)}(\bm{1}) - \widehat{\bg}_i^{(-k)}(\bm{0}) \Big\} 		
	-
	\frac{\bm{1}\T}{\NI_i} \Big\{ \bg_i^*(\bm{1}) - \bg_i^*(\bm{0}) \Big\} 		
	\bigg]
	\, \Bigg| \, \II_k^c \Bigg] 
	+ o_P(1)
	\ .
		\end{align*}	
		The last inequality holds because $\EXP \big\{  \phi'(\bO_i \con \beta^{\ddagger, (-k)} ) \cond \II_k^c  \big\} = \tau_{\PARASUB}^* = \EXP \big[ w(\bC_i) \bm{1}\T \big\{ \bg_i^*(\bm{1}) - \bg_i^*(\bm{0}) \big\} / \NI_i \big]$. The first term in the last line is equivalent to
	\begin{align*}
	\EXP \Big\{
	w(\bC_i) \widehat{\bI}_i^{(-k),\intercal} B_i^{\ddagger, (-k)} \widehat{\bm{R}}_i^{(-k)} \, \Big| \, \II_k^c \Big\}
	& = 
	\EXP \Big[ w(\bC_i) \bI_i^{*,\intercal} B_i^{\ddagger, (-k)} \big\{ \bg_i^*(\bA_i) - \widehat{\bg}_i^{(-k)}(\bA_i) \big\} \, \Big| \, \II_k^c \Big]
	\\
	& \hspace*{1cm}
	+
	\EXP \Big[ w(\bC_i) \big\{ \widehat{\bI}_i^{(-k)} -\bI_i^* \big\} \T B_i^{\ddagger, (-k)} \big\{ \bg_i^*(\bA_i) - \widehat{\bg}_i^{(-k)}(\bA_i) \big\} \, \Big| \, \II_k^c \Big] 
	\\
	& =
		\EXP \Bigg[
		\frac{\bm{1}\T}{\NI_i} \Big\{ \bg_i^*(\bm{1}) - \bg_i^*(\bm{0}) \Big\} 		
		-
	\frac{\bm{1}\T}{\NI_i} \Big\{ \widehat{\bg}_i^{(-k)}(\bm{1}) - \widehat{\bg}_i^{(-k)}(\bm{0}) \Big\} 		
	\, \Bigg| \, \II_k^c \Bigg]
	\\
	&
	\hspace*{1cm}
	+
	\EXP \Big[ w(\bC_i) \big\{ \widehat{\bI}_i^{(-k)} -\bI_i^* \big\} \T B_i^{\ddagger, (-k)} \big\{ \bg_i^*(\bA_i) - \widehat{\bg}_i^{(-k)}(\bA_i) \big\} \, \Big| \, \II_k^c \Big]  
	\end{align*}
	where the second inequality holds from Theorem \ref{Main-thm:2}. The last term has a following rate.
	\begin{align*}
		&
		\bigg|
		\EXP \Big[ \big\{ \widehat{\bI}_i^{(-k)} -\bI_i^* \big\}\T B_i^{\ddagger, (-k)} \big\{ \bg_i^*(\bA_i) - \widehat{\bg}_i^{(-k)}(\bA_i) \big\} \, \Big| \, \II_k^c \Big]
		\bigg|
		\\
		&
		\leq 
		\EXP \Big[ \big\| \widehat{\bI}_i^{(-k)} -\bI_i^* \big\|_2 \big\|B_i^{\ddagger, (-k)} \big\|_2 \big\| \bg_i^*(\bA_i) - \widehat{\bg}_i^{(-k)}(\bA_i) \big\|_2 \, \Big| \, \II_k^c \Big]
		\\
		& 
		\leq		
		C'  \big\| \widehat{\bI}_i^{(-k)} -\bI_i^* \big\|_{P,2}  
		\\
		& = O_P(r_{\pi,\NC})  \ .
	\end{align*}
	The second inequality holds from Assumption in the theorem. Therefore, we find
	\begin{align*}
		\frac{1}{\NC/2} \sum_{i \in \II_k} \Big\{ \widehat{\phi}_k(\bO_i \con \beta^{\ddagger, (-k)} )  - \phi'(\bO_i \con \beta^{\ddagger, (-k)} ) \Big\}
		=
		O_P(r_{\pi,\NC}) + o_P(1)
		=
		o_P(1) \ .
	\end{align*}
	This concludes the proof. 
	
	\item 	\textbf{[Case OR]}: We assume $g' = g^*$. 
By following \textbf{[Step 1]} and \textbf{[Step 2]} in the proof of Theorem \ref{Main-thm:3}, we find that $[A]$ in \eqref{proof-AN-003} has the same rate $[A] = o_P(1)$. Therefore, \eqref{proof-AN-003} divided by $\sqrt{\NC/2}$ is expressed as
	\begin{align*}
		\frac{1}{\NC/2} \sum_{i \in \II_k} \Big\{ \widehat{\phi}_k(\bO_i \con \widehat{\beta}^{(k)} )  - \tau_{\PARASUB}^* \Big\}
		&
		=
		\frac{1}{\NC/2} \sum_{i \in \II_k} \Big\{ \widehat{\phi}_k (\bO_i \con \beta^{\ddagger, (-k)} ) - \tau_{\PARASUB}^* \Big\}
		+
		\frac{[A]}{\sqrt{\NC/2}} 
		+
		 o_P(\NC^{-1/2}) 
		 \\
		 &
		 =
		 \frac{1}{\NC/2} \sum_{i \in \II_k} \Big\{ \widehat{\phi}_k (\bO_i \con \beta^{\ddagger, (-k)} ) - \tau_{\PARASUB}^* \Big\}
		 +
		 o_P(1) \ .
	\end{align*}

	In \textbf{[Step 2]} in the proof of Theorem \ref{Main-thm:3}, we use $\phi'(\bO_i \con \beta^{\ddagger, (-k)} )$ instead of $\phi^*(\bO_i \con \beta'^{, (-k)})$ where
	\begin{align*}
		\phi' (\bO_i \con \beta^{\ddagger, (-k)} )
		=
		w(\bC_i)
	\bigg[
	\bI (\bA_i, \bX_i \con \pi' )\T B (\bC_i \con \beta^{\ddagger, (-k)}  ) \big\{ \bY_i - \bg^* (\bA_i, \bX_i) \big\}
	+
	\frac{\bm{1}\T}{\NI_i} \Big\{ \bg^* (\bm{1}, \bX_i) - \bg^* (\bm{0}, \bX_i) \Big\}
	\bigg]  \ .
	\end{align*}
	Specifically, \eqref{proof-AN-005} divided by $\sqrt{\NC/2}$ becomes
	\begin{align*}
		&
		\frac{1}{\NC/2} \sum_{i \in \II_k} \Big\{ \widehat{\phi}_k (\bO_i \con \beta^{\ddagger, (-k)}  ) - \tau_{\PARASUB}^* \Big\}
		\\
		&
		=
		\frac{1}{\NC/2} \sum_{i \in \II_k} \Big\{ \widehat{\phi}_k(\bO_i \con \beta^{\ddagger, (-k)}  )  - \phi'(\bO_i \con \beta^{\ddagger, (-k)}  ) \Big\}
		+
		\frac{1}{\NC/2}  \sum_{i \in \II_k} \Big\{ \phi'(\bO_i \con \beta^{\ddagger, (-k)}  )  - \tau_{\PARASUB}^* \Big\} \ .
	\end{align*}
	Since $\EXP \big\{  \phi'(\bO_i \con \beta^{\ddagger, (-k)} ) \cond \II_k^c  \big\} = \tau_{\PARASUB}^*$, the second term in the right hand side is $o_P(1)$ from the law of large numbers. Consequently, it suffices to show that the first term in the right hand side is $o_P(1)$, which is expressed as follows from the law of large numbers.
		\begin{align*}
		&
		\frac{1}{\NC/2} \sum_{i \in \II_k} \Big\{ \widehat{\phi}_k(\bO_i \con \beta^{\ddagger, (-k)}  )  - \phi'(\bO_i \con \beta^{\ddagger, (-k)}  ) \Big\}
		\\
		&
		=
		\EXP \Big\{ \widehat{\phi}_k(\bO_i \con \beta^{\ddagger, (-k)}  ) - \phi'(\bO_i \con \beta^{\ddagger, (-k)}  )\, \Big| \, \II_k^c \Big\} + o_P(1)
		\\
		&
		=
		\EXP \Bigg[ w(\bC_i) \bigg[
		\widehat{\bI}_i^{(-k),\intercal} B_i^{\ddagger, (-k)}  \widehat{\bm{R}}_i^{(-k)}
	+
	\frac{\bm{1}\T}{\NI_i} \Big\{ \widehat{\bg}_i^{(-k)}(\bm{1}) - \widehat{\bg}_i^{(-k)}(\bm{0}) \Big\} 		
	-
	\frac{\bm{1}\T}{\NI_i} \Big\{ \bg_i^*(\bm{1}) - \bg_i^*(\bm{0}) \Big\} 		
	\bigg]
	\, \Bigg| \, \II_k^c \Bigg] 
	+ o_P(1)
	\ .
		\end{align*}	
		The last inequality holds because $\EXP \big\{  \phi'(\bO_i \con \beta^{\ddagger, (-k)} ) \cond \II_k^c  \big\} = \tau_{\PARASUB}^* = \EXP \big[ w(\bC_i) \bm{1}\T \big\{ \bg_i^*(\bm{1}) - \bg_i^*(\bm{0}) \big\} / \NI_i \big]$. The first term in the last line is equivalent to
	\begin{align*}
	\EXP \Big\{
	w(\bC_i) \widehat{\bI}_i^{(-k),\intercal} B_i^{\ddagger, (-k)}  \widehat{\bm{R}}_i^{(-k)} \, \Big| \, \II_k^c \Big\}
	& = 
	\EXP \Big[ w(\bC_i) \bI_i^{*,\intercal} B_i^{\ddagger, (-k)}  \big\{ \bg_i^*(\bA_i) - \widehat{\bg}_i^{(-k)}(\bA_i) \big\} \, \Big| \, \II_k^c \Big]
	\\
	& \hspace*{1cm}
	+
	\EXP \Big[ w(\bC_i) \big\{ \widehat{\bI}_i^{(-k)} -\bI_i^* \big\} \T B_i^{\ddagger, (-k)}  \big\{ \bg_i^*(\bA_i) - \widehat{\bg}_i^{(-k)}(\bA_i) \big\} \, \Big| \, \II_k^c \Big] 
	\\
	& =
		\EXP \Bigg[
		\frac{\bm{1}\T}{\NI_i} \Big\{ \bg_i^*(\bm{1}) - \bg_i^*(\bm{0}) \Big\} 		
		-
	\frac{\bm{1}\T}{\NI_i} \Big\{ \widehat{\bg}_i^{(-k)}(\bm{1}) - \widehat{\bg}_i^{(-k)}(\bm{0}) \Big\} 		
	\, \Bigg| \, \II_k^c \Bigg]
	\\
	&
	\hspace*{1cm}
	+
	\EXP \Big[ w(\bC_i) \big\{ \widehat{\bI}_i^{(-k)} -\bI_i^* \big\} \T B_i^{\ddagger, (-k)}  \big\{ \bg_i^*(\bA_i) - \widehat{\bg}_i^{(-k)}(\bA_i) \big\} \, \Big| \, \II_k^c \Big]  
	\end{align*}
	where the second inequality holds from Lemma \ref{Main-thm:2}. The last term has a following rate.
	\begin{align*}
		&
		\bigg|
		\EXP \Big[ \big\{ \widehat{\bI}_i^{(-k)} -\bI_i^* \big\}\T B_i^{\ddagger, (-k)} \big\{ \bg_i^*(\bA_i) - \widehat{\bg}_i^{(-k)}(\bA_i) \big\} \, \Big| \, \II_k^c \Big]
		\bigg|
		\\
		&
		\leq 
		\EXP \Big[ \big\| \widehat{\bI}_i^{(-k)} -\bI_i^* \big\|_2 \big\|B_i^{\ddagger, (-k)} \big\|_2 \big\| \bg_i^*(\bA_i) - \widehat{\bg}_i^{(-k)}(\bA_i) \big\|_2 \, \Big| \, \II_k^c \Big]
		\\
		& 
		\leq		
		C'  \big\|  \bg_i^*(\bA_i) - \widehat{\bg}_i^{(-k)}(\bA_i)  \big\|_{P,2}  
		\\
		& = O_P(r_{g,\NC})  \ .
	\end{align*}
	The second inequality holds from Assumption in the theorem. Therefore, we find
	\begin{align*}
		\frac{1}{\NC/2} \sum_{i \in \II_k} \Big\{ \widehat{\phi}_k(\bO_i \con \beta^{\ddagger, (-k)} )  - \phi'(\bO_i \con \beta^{\ddagger, (-k)} ) \Big\}
		=
		O_P(r_{g,\NC}) + o_P(1)
		=
		o_P(1)  \ .
	\end{align*}
	This concludes the proof. 
\end{itemize}

\subsubsection{Proof of (3) Efficiency Gain Under Known Treatment Assignment Mechanism}						\label{proof-thm:3-(4)}

Let $\bm{\epsilon}_i = \bY_i - \bg(\bA_i, \bX_i)$ be the true residual and $\bm{\epsilon}_i$ satisfies $\EXP( \bm{\epsilon}_i \cond \bA_i, \bX_i  ) = \bm{0}$. Note that $\beta^*$ is
	\begin{align*}
	\beta^*
	\in
	\argmin_{\beta \in \mathcal{B} }
	\EXP
	\Big\{
	w(\bC_i)^2 \bI (\bA_i, \bX_i \con \pi^*) \T  B(\bC_i \con \beta) 
	\bm{\epsilon}_i^{\otimes 2} 
	B(\bC_i \con \beta)  \bI (\bA_i, \bX_i \con \pi^*)
	\Big\}
	 \ .
		\end{align*}
We first show that $\beta^*$ is also the minimizer of $\VAR \big\{ \widehat{\tau}(\pi^*,g^*,\beta ) \big\}$ over $\beta \in \mathcal{B}$. Note that $\VAR  \big\{ \widehat{\tau}(\pi^*,g^*,\beta ) \big\} $ is
\begin{align*}
	&
\VAR  \big\{
\widehat{\tau}(\pi^*,g^*,\beta )
\big\}
\\
	& =
\VAR \Bigg[
	w(\bC_i) \bigg[
		\bI(\bA_i, \bX_i \con \pi^*) B(\bC_i \con \beta) \bm{\epsilon}_i 
	+
	\frac{\bm{1}\T}{\NI_i} \big\{ \bg (\bm{1}, \bX_i) - \bg (\bm{0}, \bX_i) \big\} 
	\bigg]
	\Bigg]
	\\
	& =  
	\VAR \bigg\{
	w(\bC_i) 
		\bI(\bA_i, \bX_i \con \pi^*) B(\bC_i \con \beta) \bm{\epsilon}_i 
	\bigg\}
	+
	\VAR 
	\bigg[
	\frac{w(\bC_i) \bm{1}\T}{\NI_i} \big\{ \bg (\bm{1}, \bX_i) - \bg (\bm{0}, \bX_i) \big\} 
	\bigg]
	\\
	&
	\hspace*{1cm}
	+
	2
	\text{Cov}
	\bigg[
	w(\bC_i) 
		\bI(\bA_i, \bX_i \con \pi^*) B(\bC_i \con \beta) \bm{\epsilon}_i 
		\ , \
		\frac{w(\bC_i) \bm{1}\T}{\NI_i} \big\{ \bg (\bm{1}, \bX_i) - \bg (\bm{0}, \bX_i) \big\} 
		\bigg] \ .
\end{align*}
Wince $\EXP( \bm{\epsilon}_i \cond \bA_i, \bX_i  ) = \bm{0}$, we find the last covariance is zero. As a consequence,
\begin{align*}
	\argmin_{\beta \in \mathcal{B}} \VAR  \big\{
\widehat{\tau}(\pi^*,g^*,\beta )
\big\}
&=
\argmin_{\beta \in \mathcal{B}}
\VAR \bigg\{
	w(\bC_i) 
		\bI(\bA_i, \bX_i \con \pi^*) B(\bC_i \con \beta) \bm{\epsilon}_i 
		\bigg\}
	\\
	&=
	\argmin_{\beta \in \mathcal{B} }
	\EXP
	\Big\{
	w(\bC_i)^2 \bI (\bA_i, \bX_i \con \pi^*) \T  B(\bC_i \con \beta) 
	\bm{\epsilon}_i^{\otimes 2} 
	B(\bC_i \con \beta)  \bI (\bA_i, \bX_i \con \pi^*)
	\Big\}
	\\
	&
	\ni \beta^* \ .
\end{align*}
As a result, we use Lemma \ref{Main-thm:2} with $g=g^*$, i.e.,
\begin{align*}
	{\rm ARE} (\widehat{\tau}_{\PARASUB} , \overline{\tau}_{\PARASUB} )
	=
	\frac{a\VAR(\overline{\tau}_{\PARASUB})}{a\VAR(\widehat{\tau}_{\PARASUB})}
	=
	\frac{\VAR\big\{\varphi (\bO_i, e^*, g^*) \big\}}{\VAR \big\{ \phi(\bO_i, \pi^*, g^*, \beta^*) \}} \geq 1
	\ .
\end{align*}
This concludes the proof. 

\newpage

\section{Proof of Lemmas in the Supplementary Material}						\label{sec:prooflemma}

\subsection{Proof of Lemma \ref{lemma:taubar}}

	Since condition (a) is trivially satisfied, we only need to show that conditions (b), (c), and (d) hold. First, we show condition \hyperlink{Main-(M2)}{(M2)} of the main paper implies condition (b) of Lemma \ref{lemma:taubar}. It suffices to show $ \big\| \Sigma(\bA_i, \bX_i) \big\|_2 \leq C_2$, which is given as
	\begin{align*}
		 \big\| \Sigma(\bA_i, \bX_i) \big\|_2
		 &=
		 \big\| 
		\EXP \big[ \big\{  \bY_i - \bg^*(\bA_i, \bX_i)  \big\}^{\otimes 2} \cond \bA_i, \bX_i \big]
		\big\|_2
		\\
		&
		\leq
		\EXP \big\{ \big\| \bY_i - \bg^*(\bA_i, \bX_i) \big\|_2^2 \cond \bA_i, \bX_i \big\}
		\\
		&
		\leq
		\Big[
		\EXP \big\{ \big\| \bY_i - \bg^*(\bA_i, \bX_i) \big\|_2^4 \cond \bA_i, \bX_i \big\}
		\Big]^{1/2} 
		\\
		&
		\leq \sqrt{C_4} \ .
	\end{align*}
	The first inequality holds from the Jensen's inequality and the second inequality holds from the H\"older's inequality.
	
	Second, we show condition \hyperlink{Main-(E1)}{(E1)} of the main paper implies condition (c) of Lemma \ref{lemma:taubar}. First, we find $e^*(A_{ij} \cond \bX_{ij})$ obeys the positivity assumption with some constant $c_e$.
	\begin{align*}
		e^*(A_{ij} \cond \bX_{ij})
		=
		\int P(A_{ij} \cond \bX_{i}) \, dP(\bX_\eij)
		=
		\int \sum_{\bA_\eij} P(\bA_i \cond \bX_{i}) \, dP(\bX_\eij)
		\geq 
		\int \sum_{\bA_\eij} \delta \, dP(\bX_\eij)
		\geq 
		c_e \ .
	\end{align*}
	This implies $ e^* (A_{ij} \cond \bX_{ij})  \in [\delta_e, 1-\delta_e]$. Therefore, we find the result from the following algebra
	\begin{align*}
			&
			\big\| \widehat{\bg}^{(-k)} (\bA_i, \bX_i)  - \bg^*(\bA_i, \bX_i) \big\|_2^2
			=
			\sum_{j=1}^{\NI_i} \Big\{ 
				\widehat{g}^{(-k)}(A_{ij}, \bX_{ij})
				-
				g^*(A_{ij}, \bX_{ij})
			\Big\}^2
			\leq
			\NT c_g^2  \equiv C_g^2 \\
			&
			\big\|
				\bI(\bA_i, \bX_i \con e^*)
			\big\|_2^2
			=
			\sum_{j=1}^{\NI_i}
			\frac{1}{e^*(A_{ij} \cond \bX_{ij} )^2}
			\leq
			\frac{\NT}{c_e^2} \equiv C_e^2
			\\
			&
			\big\|
				\bI(\bA_i, \bX_i \con \widehat{e}^{(-k)})
			\big\|_2^2
			=
			\sum_{j=1}^{\NI_i}
			\frac{1}{\widehat{e}^{(-k)}(A_{ij} \cond \bX_i)^2}
			\leq
			\frac{\NT }{c_e^2} \equiv C_e^2 \ .
		\end{align*}

	Lastly, we only show condition \hyperlink{Main-(E2.PS)}{(E2.PS)} of the main paper implies condition (d.PS) of Lemma \ref{lemma:taubar} but other cases can be shown in a similar manner. Let $\text{supp}(\NI_i) \subset \{1,\ldots,\NT\}$ be the support of $\NI_i$. We find
	{\small
		\begin{align*}
			& 
			 \Big\| \widehat{e}^{(-k)} \big( 1 \cond \bX_{ij} \big) - e^* \big( 1 \cond \bX_{ij} \big) \Big\|_{P,2}^2
			 \\
			 &
			 =
			 \int \Big\{  \widehat{e}^{(-k)} \big( 1 \cond \bX_{ij} \big) - e^* \big( 1 \cond \bX_{ij} \big) \Big\}^2 \, dP(\bX_{ij})
			 \\
			 & =
			 \sum_{m \in \text{supp}(\NI_i)}  P(\NI_i=m)  \int\Big\{  \widehat{e}^{(-k)} \big( 1 \cond \bX_{ij} , \NI_i=m \big) - e^* \big( 1 \cond \bX_{ij} , \NI_i=m \big) \Big\}^2 \, dP(\bX_{ij} \cond \NI_i=m)
			 \\
			 &
			 \geq
			 P(\NI_i = m) \underbrace{ \int\Big\{  \widehat{e}^{(-k)} \big( 1 \cond \bX_{ij} , \NI_i=m \big) - e^* \big( 1 \cond \bX_{ij} , \NI_i=m \big) \Big\}^2 \, dP(\bX_{ij} \cond \NI_i=m) }_{ \equiv R_m(ij) } \ .
		\end{align*}	}
	Similarly, by choosing $g' = g$, we find
	\begin{align*}	
		&
		\Big\| \widehat{g}^{(-k)}(A_{ij}, \bX_{ij}) - g'(A_{ij},\bX_{ij}) \Big\|_{P,2}^2
		\\
		& =
		\int \Big\{
				\widehat{g}^{(-k)}(A_{ij}, \bX_{ij}) - g'(A_{ij}, \bX_{ij})
			\Big\}^2 \, dP(A_{ij}, \bX_{ij})
			\\
			& =
			\sum_{m \in \text{supp}(\NI_i)} P(\NI_i = m)
			\int \Big\{
				\widehat{g}^{(-k)}(A_{ij}, \bX_{ij}, \NI_i=m) - g'(A_{ij}, \bX_{ij}, \NI_i=m)
			\Big\}^2 \, dP(A_{ij}, \bX_{ij} \cond \NI_i=m)
			\\
			&
			\geq
			 P(\NI_i = m)
			 \underbrace{
			\int \Big\{
				\widehat{g}^{(-k)}(A_{ij}, \bX_{ij}, \NI_i=m) - g'(A_{ij}, \bX_{ij}, \NI_i=m)
			\Big\}^2 \, dP(A_{ij}, \bX_{ij} \cond \NI_i=m) }_{\equiv T_m(ij) }
	\end{align*}
	Since $P(\NI_i = m)$ is a positive number, $R_m(ij) = O_P(r_{e,\NC}^2)$ and $T_m(ij) = O_P(r_{g,\NC}^2)$ for any $j=1,\ldots,m$ and $m \in \text{supp}(\NI_i)$. As a consequence, $ \sum_{j=1}^m R_m(ij) = O_P(r_{e,\NC}^2)$ and $\sum_{j=1}^m T_m(ij) = O_P(r_{g,\NC}^2)$ for $m \in \text{supp}(\NI_i)$. Therefore,
		\begin{align*}
			&
			 \Big\| \bI(\bA_i, \bX_i \con \widehat{e}^{(-k)}) - \bI(\bA_i, \bX_i \con e^*) \Big\|_ {P,2}^2
			 \\
			 & =
			 \int \sum_{j=1}^{\NI_i} \frac{ \big\{ \widehat{e}^{(-k)}(A_{ij} \cond \bX_{ij}) -  e^*(A_{ij} \cond \bX_{ij}) \big\}^2 }{ \big\{ e^*(A_{ij} \cond \bX_{ij}) \widehat{e}^{(-k)}(A_{ij} \cond \bX_{ij}) \big\}^2 }
			 \, dP(\bA_i, \bX_i)
			 \\
			 & \leq \frac{1}{c_e^4}
			\int \sum_{j=1}^{\NI_i} \Big\{ \widehat{e}^{(-k)}(A_{ij} \cond \bX_{ij}) -  e^*(A_{ij} \cond \bX_{ij})\Big\}^2 \, dP(\bA_i, \bX_i)
			 \\
			 & = \frac{1}{c_e^4}
			 \sum_{m \in \text{supp}(\NI_i)}
			 P(\NI_i=m)
			 \sum_{j=1}^m
			R_m(ij)
			\\
			& = O_P(r_{e,\NC}^2) \ .
		\end{align*}
		Similarly, we have
		\begin{align*}	
			\Big\| \widehat{\bg}^{(-k)}(\bA_i, \bX_i) - \bg'(\bA_i, \bX_i) \Big\|_{P,2}^2
			& =
			\int
			\sum_{j=1}^{\NI_i}
			\Big\{
				\widehat{g}^{(-k)}(A_{ij}, \bX_{ij}) - g'(A_{ij}, \bX_{ij})
			\Big\}^2 \, dP(\bA_i, \bX_i)
			\\
			& =
			\sum_{m \in \text{supp}(\NI_i)}  P(\NI_i = m) \sum_{j=1}^m T_m(ij)
			\\
			& = 
			O_P(r_{g,\NC}^2) \ . 
		\end{align*}		
		This concludes the proof.

\subsection{Proof of Lemma \ref{lemma:tauhat}}

	The proof of condition (a) and (b) are the same as Lemma \ref{lemma:taubar}. 
	
	We show condition \hyperlink{Main-(EN1)}{(EN1)} of the main paper implies condition (c) of Lemma \ref{lemma:tauhat}. From the following algebra, we find the result.
	\begin{align*}
			&
			\big\| \widehat{\bg}^{(-k)} (\bA_i, \bX_i)  - \bg^*(\bA_i, \bX_i) \big\|_2^2
			=
			\sum_{j=1}^{\NI_i} \Big\{ 
				\widehat{g}^{(-k)}(A_{ij}, \bX_{ij})
				-
				g^*(A_{ij}, \bX_{ij})
			\Big\}^2
			\leq
			\NT c_g^2 \equiv C_g^2 \\
			&
			\big\|
				\bI(\bA_i, \bX_i \con \pi^*)
			\big\|_2^2
			=
			\sum_{j=1}^{\NI_i}
			\frac{1}{\pi^*(A_{ij} \cond \bA_\eij , \bX_{ij} , \bX_\eij)^2}
			=
			\sum_{j=1}^{\NI_i}
			\frac{P(\bA_\eij \cond \bX_i)^2}{P(\bA_i \cond \bX_i)^2}
			\leq
			\frac{\NT (1-\delta)^2}{\delta^2} 
			\\
			&
			\big\|
				\bI(\bA_i, \bX_i \con \widehat{\pi}^{(-k)})
			\big\|_2^2
			=
			\sum_{j=1}^{\NI_i}
			\frac{1}{\widehat{\pi}^{(-k)}(A_{ij} \cond \bA_\eij , \bX_{ij}, \bX_\eij)^2}
			\leq
			\frac{\NT }{c_\pi^2} \ .
		\end{align*}
		We take $C_\pi^2 = \max \big\{ \NT(1-\delta)^2/\delta^2 , \NT/c_\pi^2 \}$. 
		
	
		Next, we only show condition \hyperlink{Main-(EN2.PS)}{(EN2.PS)} of the main paper implies condition (d.PS) of Lemma \ref{lemma:taubar} but other cases can be shown in a similar manner. Let $\text{supp}(\NI_i) \subset \{1,\ldots,\NT\}$ be the support of $\NI_i$. We find
		\begin{align*}
			& 
			 \Big\| \widehat{\pi}^{(-k)} \big( 1 \cond \bA_\eij, \bX_{ij}, \bX_\eij \big) - \pi^* \big( 1 \cond \bA_\eij, \bX_{ij}, \bX_\eij \big) \Big\|_{P,2}^2
			 \\
			 &
			 =
			 \int \left\{ \begin{matrix}
			  \widehat{\pi}^{(-k)} \big( 1 \cond \bA_\eij, \bX_{ij}, \bX_\eij \big)
			  \quad \quad
			  \\
			  \quad \quad
			  - \pi^* \big( 1 \cond \bA_\eij, \bX_{ij}, \bX_\eij \big)
			 \end{matrix}  \right\}^2 \, dP(\bA_\eij, \bX_i)
			 \\
			 & =
			 \sum_{m \in \text{supp}(\NI_i)} P(\NI_i=m)  \int 
			 \left\{ \begin{matrix}
			  \widehat{\pi}^{(-k)} \big( 1 \cond \bA_\eij, \bX_{ij}, \bX_\eij , \NI_i = m \big)
			  \quad \quad
			  \\
			  \quad \quad
			  - \pi^* \big( 1 \cond \bA_\eij, \bX_{ij}, \bX_\eij , \NI_i = m \big)
			 \end{matrix}  \right\}^2
			 \, dP(\bA_\eij, \bX_i \cond \NI_i=m)
			 \\
			 &
			 \geq
			 P(\NI_i = m) \underbrace{ \int  \left\{ \begin{matrix}
			  \widehat{\pi}^{(-k)} \big( 1 \cond \bA_\eij, \bX_{ij}, \bX_\eij , \NI_i = m \big)
			  \quad \quad
			  \\
			  \quad \quad
			  - \pi^* \big( 1 \cond \bA_\eij, \bX_{ij}, \bX_\eij , \NI_i = m \big)
			 \end{matrix}  \right\}^2 \, dP(\bA_\eij, \bX_i \cond \NI_i=m) }_{ \equiv R_m(ij) } \ .
		\end{align*}	
	Similarly, by choosing $g' = g$, we find
	\begin{align*}	
		&
		\Big\| \widehat{g}^{(-k)}(A_{ij}, \bX_{ij}) - g'(A_{ij},\bX_{ij}) \Big\|_{P,2}^2
		\\
		& =
		\int \Big\{
				\widehat{g}^{(-k)}(A_{ij}, \bX_{ij}) - g'(A_{ij}, \bX_{ij})
			\Big\}^2 \, dP(A_{ij}, \bX_{ij})
			\\
			& =
			\sum_{m \in \text{supp}(\NI_i)} P(\NI_i = m)
			\int \Big\{
				\widehat{g}^{(-k)}(A_{ij}, \bX_{ij}, \NI_i=m) - g'(A_{ij}, \bX_{ij}, \NI_i=m)
			\Big\}^2 \, dP(A_{ij}, \bX_{ij} \cond \NI_i=m)
			\\
			&
			\geq
			 P(\NI_i = m)
			 \underbrace{
			\int \Big\{
				\widehat{g}^{(-k)}(A_{ij}, \bX_{ij}, \NI_i=m) - g'(A_{ij}, \bX_{ij}, \NI_i=m)
			\Big\}^2 \, dP(A_{ij}, \bX_{ij} \cond \NI_i=m) }_{\equiv T_m(ij) } \ .
	\end{align*}
	Since $P(\NI_i = m)$ is a positive number, $R_m(ij) = O_P(r_{\pi,\NC}^2)$ and $T_m(ij) = O_P(r_{g,\NC}^2)$ for any $j=1,\ldots,m$ and $m \in \text{supp}(\NI_i)$. As a consequence, $ \sum_{j=1}^m R_m(ij) = O_P(r_{\pi,\NC}^2)$ and $\sum_{j=1}^m T_m(ij) = O_P(r_{g,\NC}^2)$ for $m \in \text{supp}(\NI_i)$. Therefore,
		\begin{align*}
			&
			 \Big\| \bI(\bA_i, \bX_i \con \widehat{\pi}^{(-k)}) - \bI(\bA_i, \bX_i \con \pi^*) \Big\|_ {P,2}^2
			 \\
			 & =
			 \int \sum_{j=1}^{\NI_i} \frac{ \big\{ \widehat{\pi}^{(-k)}(A_{ij} \cond \bA_\eij, \bX_{ij}, \bX_\eij) -  \pi^*(A_{ij} \cond \bA_\eij, \bX_{ij}, \bX_\eij) \big\}^2 }{ \big\{ \pi^*(A_{ij} \cond \bA_\eij, \bX_{ij}, \bX_\eij) \widehat{\pi}^{(-k)}(A_{ij} \cond \bA_\eij, \bX_{ij}, \bX_\eij) \big\}^2 }
			 \, dP(\bA_i, \bX_i)
			 \\
			 & \leq \frac{(1-\delta)^2}{\delta^2 c_\pi^2}
			\int \sum_{j=1}^{\NI_i} \Big\{  \widehat{\pi}^{(-k)} \big( 1 \cond \bA_\eij, \bX_{ij}, \bX_\eij \big) - \pi^* \big( 1 \cond \bA_\eij, \bX_{ij}, \bX_\eij \big) \Big\}^2 \, dP(\bA_i, \bX_i)
			 \\
			 & = \frac{(1-\delta)^2}{\delta^2 c_\pi^2}
			 \sum_{m \in \text{supp}(\NI_i)}
			 P(\NI_i=m)
			 \sum_{j=1}^m
			R_m(ij)
			\\
			& = O_P(r_{\pi,\NC}^2) \ .
		\end{align*}
		Similarly, we have
		\begin{align*}	
			\Big\| \widehat{\bg}^{(-k)}(\bA_i, \bX_i) - \bg'(\bA_i, \bX_i) \Big\|_{P,2}^2
			& =
			\int
			\sum_{j=1}^{\NI_i}
			\Big\{
				\widehat{g}^{(-k)}(A_{ij}, \bX_{ij}) - g'(A_{ij}, \bX_{ij})
			\Big\}^2 \, dP(\bA_i, \bX_i)
			\\
			& =
			\sum_{m \in \text{supp}(\NI_i)}  P(\NI_i = m) \sum_{j=1}^m T_m(ij)
			\\
			& = 
			O_P(r_{g,\NC}^2) \ . 
		\end{align*}		

%

Lastly, we establish conditions (e) and (f). From  Lemma \ref{lemma:consistency1}, we establish condition (e) with $\beta^{\ddagger,(-k)} = \beta^{\dagger,(-k)}$ and $\beta'^{,(-k)} = \beta^*$, respectively. Condition (f) is identical to \hyperlink{Main-(EN4)}{(EN4)}. This concludes the proof.

\subsection{Proof of Lemma \ref{lemma:consistency1}}

To construct the estimator of $\beta^{\dagger,(-k)}$, we find an alternative representation of $\beta^{\dagger,(-k)}$ from the law of total expectation and the first order condition of \eqref{eq-defbeta0} and \eqref{eq-303}.
		  \begin{align}									\label{eq-defbeta}
		  0 & = 2 \EXP \Big\{  w(\bC_i)^2 \widehat{\bI}_i^{(-k),\intercal}
	( I - \bm{1}\bm{1}\T ) \widehat{\bm{S}}_i^{(-k)}
		 ( I - \bm{1}\bm{1}\T )
		\widehat{\bI}_i^{(-k)} \, \Big| \, \sigma(\mathcal{B}) , \II_k^c \Big\} \beta^{\dagger,(-k)} (\bC_i)
		\\
		&
		\hspace*{2cm}
		-
		\EXP \Big[  w(\bC_i)^2 \widehat{\bI}_i^{(-k),\intercal}
	\Big\{ 2 \widehat{\bm{S}}_i^{(-k)} - \bm{1}\bm{1}\T \widehat{\bm{S}}_i^{(-k)} - \widehat{\bm{S}}_i^{(-k)} \bm{1}\bm{1}\T \Big\}
		\widehat{\bI}_i^{(-k)} \, \Big| \, \sigma(\mathcal{B}) , \II_k^c \Big]
		\nonumber
		\\
				  0 & = 2 \EXP \Big\{  w(\bC_i)^2 {\bI}_i^{*,\intercal}
	( I - \bm{1}\bm{1}\T ) {\bm{S}}_i
		 ( I - \bm{1}\bm{1}\T )
		{\bI}_i^{*} \, \Big| \, \sigma(\mathcal{B}) \Big\} \beta^{*} (\bC_i)
		\nonumber
		\\
		&
		\hspace*{2cm}
		-
		\EXP \Big[  w(\bC_i)^2 {\bI}_i^{*,\intercal}
	\Big\{ 2 {\bm{S}}_i - \bm{1}\bm{1}\T {\bm{S}}_i - {\bm{S}}_i \bm{1}\bm{1}\T \Big\}
		{\bI}_i^{*} \, \Big| \, \sigma(\mathcal{B}) \Big] \ .
		\nonumber
		 \end{align}

We define $\Psi(\bO_i \con \beta)$ and $\widehat{\Psi}^{(-k)}(\bO_i \con \beta)$ as follows.
\begin{align}										\label{eq-popEE}
& 	\Psi(\bO_i \con \beta)
	=
		w(\bC_i)^2 
		\Big\{
		2	\beta(\bC_i)
		\Big( \bI_i^{*,\intercal}
		\mathbb{S}_i^*
		\bI_i^*
		\Big)
		-
		 \bI_i^{*,\intercal} \mathbb{T}_i^*  \bI_i^*
		 \Big\} \ , \
		 \\
		 \nonumber
& 		 \widehat{\Psi}^{(-k)} (\bO_i \con \beta)
	=
		w(\bC_i)^2 \Big\{
		2	\beta(\bC_i)
		\Big( \widehat{\bI}_i^{(-k),\intercal}
	\widehat{\mathbb{S}}_i^{(-k)}
		\widehat{\bI}_i^{(-k)}
		\Big)
		-
		\widehat{\bI}_i^{(-k),\intercal} \widehat{\mathbb{T}}_i^{(-k)}  \widehat{\bI}_i^{(-k)}
		\Big\} \ .
\end{align}
Here $\mathbb{S}_i^* = ( I - \bm{1}\bm{1}\T ) \bm{S}_i
		 ( I - \bm{1}\bm{1}\T )$, $\mathbb{T}_i^* = 2 \bm{S}_i - \bm{1}\bm{1}\T\bm{S}_i - \bm{S}_i \bm{1}\bm{1}\T$, and $\bm{S}_i = \bm{\epsilon}_i\bm{\epsilon}_i\T =  \big\{ \bY_i - \bg^* (\bA_i, \bX_i) \big\}^{\otimes 2}$. Similarly, $\widehat{\mathbb{S}}_i^{(-k)} = ( I - \bm{1}\bm{1}\T ) \widehat{\bm{S}}_i^{(-k)}
		 ( I - \bm{1}\bm{1}\T )$, $\widehat{\mathbb{T}}_i^{(-k)} = 2 \widehat{\bm{S}}_i^{(-k)} - \bm{1}\bm{1}\T\widehat{\bm{S}}_i^{(-k)} - \widehat{\bm{S}}_i^{(-k)} \bm{1}\bm{1}\T$, and $\widehat{\bm{S}}_i^{(-k)} = \big\{ \bY_i - \widehat{\bg}^{(-k)} (\bA_i, \bX_i) \big\}^{\otimes 2}$. Then, we find $\beta^*$ defined in \eqref{eq-303} and $\beta^{\dagger,(-k)}$ defined in \eqref{eq-defbeta0} solve the following estimating equations, respectively.
\begin{align}											\label{eq-popEE1}
	&
	\EXP \big\{ \Psi (\bO_i \con \beta^*) \cond \sigma(\mathcal{B}) \big\}
	=
	0
	\ , \
	\EXP \big\{ \widehat{\Psi}^{(-k)} (\bO_i \con \beta^{\dagger,(-k)}) \cond \sigma(\mathcal{B}) , \II_k^c \big\}
	= 0	 \ .
\end{align}
As a consequence, if $w(\bC_i)>0$ with positive probability given $\sigma(\mathcal{B})$, we have
	  \begin{align}									\label{eq-solbeta}
		  &
		 	\beta^*(\bC_i)
		 	=
		 	\frac{1}{2} \frac{  \EXP \big\{  w(\bC_i)^2 \bI_i^{*,\intercal}
		\mathbb{T}_i^*
		\bI_i^* \cond \sigma(\mathcal{B}) \big\}  }{ \EXP \big\{  w(\bC_i)^2 \bI_i^{*,\intercal}
		\mathbb{S}_i^*
		\bI_i^* \cond \sigma(\mathcal{B}) \big\} }
		\ , \
		\nonumber
		\\
		&
				 	\beta^{\dagger,(-k)}(\bC_i)
		 	=
		 	\frac{1}{2} \frac{ \EXP \big\{  w(\bC_i)^2 \widehat{\bI}_i^{(-k),\intercal}
	\widehat{\mathbb{T}}_i^{(-k)}
		\widehat{\bI}_i^{(-k)} \cond \sigma(\mathcal{B}) , \II_k^c \big\}  }{ \EXP \big\{  w(\bC_i)^2 \widehat{\bI}_i^{(-k),\intercal}
	\widehat{\mathbb{S}}_i^{(-k)}
		\widehat{\bI}_i^{(-k)} \cond \sigma(\mathcal{B}) , \II_k^c \big\} }
		\ .
		 \end{align}
	If $w(\bC_i)=0$ for any $\bC_i$ given $\sigma(\mathcal{B})$, there is nothing to prove because $\beta^*(\bC_i) = \beta^{\dagger,(-k)}(\bC_i) = 0$ from the definition of $\beta^*$.

	From the definition of $\beta^*$ and $\beta^{\dagger,(-k)}$ in \eqref{eq-solbeta}, we find
	\begin{align}						\label{proof-consist-001}
		&
		\Big|
			\beta^{\dagger,(-k)}(\bC_i) -
			\beta^*(\bC_i) 
		\Big|
		\nonumber
		\\
		&
		 	=
		 		 	\frac{1}{2} \Bigg|
		 		 	 \frac{  \EXP \big\{ w(\bC_i)^2 \widehat{\bI}_i^{(-k),\intercal}
	\widehat{\mathbb{T}}_i^{(-k)}
		\widehat{\bI}_i^{(-k)} \cond \sigma(\mathcal{B}) , \II_k^c \big\}  }{ \EXP \big\{w(\bC_i)^2 \widehat{\bI}_i^{(-k),\intercal}
	\widehat{\mathbb{S}}_i^{(-k)}
		\widehat{\bI}_i^{(-k)} \cond \sigma(\mathcal{B}) , \II_k^c \big\} }
		-
		\frac{  \EXP \big\{ w(\bC_i)^2 \bI_i^{*,\intercal}
		\mathbb{T}_i^*
		\bI_i^* \cond \sigma(\mathcal{B}) \big\}  }{ \EXP \big\{ w(\bC_i)^2 \bI_i^{*,\intercal}
		\mathbb{S}_i^*
		\bI_i^* \cond \sigma(\mathcal{B}) \big\} }
		\Bigg|
		\nonumber
		\\
		&
		=
		\frac{ 
		\displaystyle{
		\left| 
		\begin{matrix}
				 \EXP \big\{ w(\bC_i)^2 \bI_i^{*,\intercal}
		\mathbb{S}_i^*
		\bI_i^* \cond \sigma(\mathcal{B}) \big\}
		 \EXP \big\{ w(\bC_i)^2 \widehat{\bI}_i^{(-k),\intercal}
	\widehat{\mathbb{T}}_i^{(-k)}
		\widehat{\bI}_i^{(-k)} \cond \sigma(\mathcal{B}), \II_k^c \big\}
		\hspace*{1cm}
		\\
		\hspace*{1cm}
		-
		 \EXP \big\{w(\bC_i)^2 \bI_i^{*,\intercal}
		\mathbb{T}_i^*
		\bI_i^* \cond \sigma(\mathcal{B}) \big\}
		 \EXP \big\{w(\bC_i)^2 \widehat{\bI}_i^{(-k),\intercal}
	\widehat{\mathbb{S}}_i^{(-k)}
		\widehat{\bI}_i^{(-k)} \cond \sigma(\mathcal{B}) , \II_k^c \big\}		
		\end{matrix}
		\right|
		}
		}{2 \EXP \big\{ w(\bC_i)^2 \bI_i^{*,\intercal} \mathbb{S}_i ^* \bI_i^* \cond \bC_i \big\}  
		 \EXP \big\{ w(\bC_i)^2 \widehat{\bI}_i^{(-k),\intercal}
		\widehat{\mathbb{S}}_i^{(-k)}
		\widehat{\bI}_i^{(-k)} \cond \bC_i , \II_k^c \big\} }
		\ .
	\end{align} 
	
	We find the first term of the denominator of \eqref{proof-consist-001} is bounded below by
	\begin{align}							\label{proof-consist-002}
		&
		\EXP \big\{ w(\bC_i)^2 \bI_i^{*,\intercal}
		\mathbb{S}_i^*
		\bI_i^* \cond \sigma(\mathcal{B}) \big\} 
		\nonumber
		\\
		&
		=
		\EXP \big\{ w(\bC_i)^2 \bI_i^{*,\intercal}
		(I - \bm{1}\bm{1}\T)
		\Sigma(\bA_i, \bX_i)
		(I - \bm{1}\bm{1}\T)
		\bI_i^* \cond \sigma(\mathcal{B}) \big\} 
		\nonumber
		\\
		&
		=
		\EXP \bigg[ w(\bC_i)^2
		\EXP \bigg\{ 
		\sum_{\bA_i}		
		P(\bA_i \cond \bX_i)
		\bI_i^{*,\intercal}
		(I - \bm{1}\bm{1}\T)
		\Sigma(\bA_i, \bX_i)
		(I - \bm{1}\bm{1}\T)
		\bI_i^* \, \bigg| \, \bA_i, \bX_i, \sigma(\mathcal{B}) \bigg\} 
		\, \bigg| \, \sigma(\mathcal{B})
		\bigg]
		\nonumber
		\\
		&
		\geq
		\frac{\delta}{C_2}		
		\EXP \Big[ w(\bC_i)^2 \EXP \big\{ \big\| \bI_i^{*,\intercal}
		(I - \bm{1}\bm{1}\T) \big\|_2^2 \cond \bA_i = \ba_i, \bX_i, \sigma(\mathcal{B}) \big\}
		\, \Big| \, \sigma(\mathcal{B}) \Big]
		\ .
	\end{align}
	The inequality holds for some $\ba_i$ from condition (b) of Lemma \ref{lemma:tauhat} and the positivity assumption, i.e. $\delta \leq P(\bA_i = \bm{1} \cond \bX_i ) \leq 1- \delta$. Moreover, the positivity assumption implies that 
	\begin{align*}
		P(A_{ij} = a_{ij} \cond \bA_\eij = \ba_{ij}, \bX_i)^{-1}
		=
		\frac{P(\bA_\eij = \ba_{ij} \cond \bX_i)}{P(\bA_i = \ba_i \cond \bX_i) }
		\geq \frac{\delta}{1-\delta} \ .
	\end{align*}
	Therefore, at $\bA_i = \ba_i$, we find
	\begin{align*}
		\big\| 
		\bI(\bA_i = \ba_i, \bX_i \con \pi^*)\T
		(I - \bm{1}\bm{1}\T) 
		\big\|_2
		=	
		\frac{1}{\NI_i}
		\left\|
		\begin{bmatrix}
			\sum_{j \neq 1} \big\{ P(A_{ij} = a_{ij} \cond \bA_\eij = \ba_\eij, \bX_i) \big\}^{-1}
			\\ \vdots \\
			\sum_{j \neq \NI_i} \big\{ P(A_{ij} = a_{ij} \cond \bA_\eij = \ba_\eij, \bX_i) \big\}^{-1}
		\end{bmatrix}
		\right\|_2
		\geq \frac{\delta}{\NT(1-\delta)} \ .
	\end{align*}
	Therefore, equation \eqref{proof-consist-002} is bounded below by 
		\begin{align*}
		\EXP \big\{ w(\bC_i)^2 \bI_i^{*,\intercal}
		\mathbb{S}_i^*
		\bI_i^* \cond \sigma(\mathcal{B}) \big\} 
		\geq
		\frac{\delta^2}{C_2\NT(1-\delta)}
		\EXP \big\{ w(\bC_i)^2 \cond \sigma(\mathcal{B} ) \big\}
		>
		0
		\ ,
		\end{align*}
		and  we find $\EXP \big\{ \bI_i^{*,\intercal}
		\mathbb{S}_i^*
		\bI_i^* \cond \sigma(\mathcal{B}) \big\}^{-1} \leq C_1'$ for some constant $C_1'>0$.

	Similarly, the second term of the denominator of \eqref{proof-consist-001} is bounded below by
			\begin{align}										\label{proof-consist-003}
		&
		\EXP \big\{ w(\bC_i)^2 \widehat{\bI}_i^{(-k),\intercal}
	\widehat{\mathbb{S}}_i^{(-k)}
		\widehat{\bI}_i^{(-k)}  \cond \sigma(\mathcal{B}) , \II_k^c \big\} 
		\nonumber
		\\
		&
		=
		\EXP \Big[ w(\bC_i)^2  \widehat{\bI}_i^{(-k),\intercal}
		(I - \bm{1}\bm{1}\T)
		\big[
		\big\{
			\bg_i^*(\bA_i) - \widehat{\bg}_i^{(-k)}(\bA_i)
		\big\}^{\otimes 2} + 
		\Sigma(\bA_i, \bX_i)
		\big]
		(I - \bm{1}\bm{1}\T)
		\widehat{\bI}_i^{(-k)}  \, \Big| \, \sigma(\mathcal{B}), \II_k^c \Big]
		\nonumber
		\\
		&
		=
		\EXP \bigg[
		w(\bC_i)^2
		\EXP \bigg\{ 
		\sum_{\bA_i}		
		P(\bA_i \cond \bX_i)
		\widehat{\bI}_i^{(-k),\intercal}
		(I - \bm{1}\bm{1}\T)
		\Sigma(\bA_i, \bX_i)
		(I - \bm{1}\bm{1}\T)
		\widehat{\bI}_i^{(-k)} \, \bigg| \, \bA_i, \bX_i, \sigma(\mathcal{B}) \bigg\} 
		\, \bigg| \, \sigma(\mathcal{B})
		\bigg]
		\nonumber
		\\
		&
		\geq
		\frac{\delta}{C_2}		
		\EXP \Big[ w(\bC_i)^2
		\EXP \big\{ \big\| \widehat{\bI}_i^{(-k),\intercal}
		(I - \bm{1}\bm{1}\T) \big\|_2^2 \cond \bA_i = \ba_i, \bX_i, \sigma(\mathcal{B}) \big\}
		\, \Big| \, \sigma(\mathcal{B}) \Big]
		\ .
	\end{align}	
	The inequality holds for some $\ba_i$ from condition (b) of Lemma \ref{lemma:tauhat} and the positivity assumption, i.e. $\delta \leq P(\bA_i = \bm{1} \cond \bX_i ) \leq 1- \delta$. We find that $\big\| \bI (\bA_i, \bX_i \con \widehat{\pi}^{(-k)} ) \big\|_2 \leq C_\pi$  implies $C_\pi^{-1} \leq \widehat{\pi}^{(-k)} (A_{ij} \cond \bA_\eij, \bX_i)$ and, as a consequence, $C_\pi^{-1} \leq \widehat{\pi}^{(-k)}(A_{ij} \cond \bA_\eij, \bX_i) \leq 1- C_\pi^{-1}$ for all $(\bA_i, \bX_i)$. 	Moreover, at $\bA_i = \ba_i$, we find 
	\begin{align*}
		\big\| 
		\bI(\bA_i = \ba_i, \bX_i \con \widehat{\pi}^{(-k)} )\T
		(I - \bm{1}\bm{1}\T) 
		\big\|_2
		& =	
		\frac{1}{\NI_i}
		\left\|
		\begin{bmatrix}
			\sum_{j \neq 1} \big\{ \widehat{\pi}^{(-k)} (A_{ij} = a_{ij} \cond \bA_\eij = \ba_\eij, \bX_i) \big\}^{-1}
			\\ \vdots \\
			\sum_{j \neq \NI_i} \big\{ \widehat{\pi}^{(-k)} (A_{ij} = a_{ij} \cond \bA_\eij = \ba_\eij, \bX_i) \big\}^{-1}
		\end{bmatrix}
		\right\|_2
		\\
		&
		\geq 
		\frac{1}{\NT(C_\pi-1)}
		\ .
	\end{align*}
		Therefore, equation \eqref{proof-consist-003} is bounded below by 
		\begin{align}										\label{proof-consist-103}
		\EXP \big\{ w(\bC_i)^2 \widehat{\bI}_i^{(-k),\intercal}
	\widehat{\mathbb{S}}_i^{(-k)}
		\widehat{\bI}_i^{(-k)}  \cond \sigma(\mathcal{B}) , \II_k^c \big\} 
		\geq
		\frac{\delta}{C_2\NT(C_\pi-1)} 
		\EXP \big\{ w(\bC_i)^2 \cond \sigma(\mathcal{B} ) \big\}
		>
		0
		\end{align}
		and we find $\EXP \big\{ w(\bC_i)^2 \widehat{\bI}_i^{(-k),\intercal}
	\widehat{\mathbb{S}}_i^{(-k)}
		\widehat{\bI}_i^{(-k)}  \cond \sigma(\mathcal{B}) , \II_k^c \big\}^{-1}  \leq C_2'$ for some constant $C_2'>0$. 	Therefore, \eqref{proof-consist-001} is upper bounded by 
		\begin{align}
			&
		\Big|
			\beta^{\dagger,(-k)}(\bC_i) -
			\beta^*(\bC_i) 
		\Big|
		\nonumber
		\\
		&
		\leq
		\frac{C_1'C_2'}{2}
		\Big|  \beta(\bC_i)
				 \EXP \big\{ w(\bC_i)^2 \bI_i^{*,\intercal}
		\mathbb{S}_i^*
		\bI_i^* \cond \sigma(\mathcal{B}) \big\}
		 \EXP \big\{ w(\bC_i)^2 \widehat{\bI}_i^{(-k),\intercal}
	\widehat{\mathbb{T}}_i^{(-k)}
		\widehat{\bI}_i^{(-k)} \cond \sigma(\mathcal{B}), \II_k^c \big\}
		\nonumber
		\\
		& \hspace*{2cm}
		-
		 \EXP \big\{w(\bC_i)^2 \bI_i^{*,\intercal}
		\mathbb{T}_i^*
		\bI_i^* \cond \sigma(\mathcal{B}) \big\}
		 \EXP \big\{w(\bC_i)^2 \widehat{\bI}_i^{(-k),\intercal}
	\widehat{\mathbb{S}}_i^{(-k)}
		\widehat{\bI}_i^{(-k)} \cond \sigma(\mathcal{B}) , \II_k^c \big\}		
		\Big|
		\nonumber
		\\
		& 
		=
		\frac{C_1'C_2' C_w^4}{2}
		\Big| 
		 \EXP \big\{ \bI_i^{*,\intercal}
		\mathbb{S}_i^*
		\bI_i^* \cond \sigma(\mathcal{B}) \big\}
		 \EXP \big\{ \widehat{\bI}_i^{(-k),\intercal}
	\widehat{\mathbb{T}}_i^{(-k)}
		\widehat{\bI}_i^{(-k)} - \bI_i^{*,\intercal}
		\mathbb{T}_i^*
		\bI_i^* \cond \sigma(\mathcal{B}) , \II_k^c \big\}
		\nonumber
		\\
		& \hspace*{2cm} 
		-
		 \EXP \big\{ \bI_i^{*,\intercal}
		\mathbb{T}_i^*
		\bI_i^* \cond \sigma(\mathcal{B}) \big\}
		\EXP \big\{ \widehat{\bI}_i^{(-k),\intercal}
	\widehat{\mathbb{S}}_i^{(-k)}
		\widehat{\bI}_i^{(-k)} -  \bI_i^{*,\intercal}
		\mathbb{S}_i^*
		\bI_i^*  \cond  \sigma(\mathcal{B}) , \II_k^c \big\}
		\Big| 
		\nonumber
		\\
		&
		\leq
		\frac{C_1'C_2'C_w^4}{2}
		\Big[
		\Big| 
		 \EXP \big\{ \bI_i^{*,\intercal}
		\mathbb{S}_i^*
		\bI_i^* \cond  \sigma(\mathcal{B}) \big\}
		\Big| 
		\EXP \Big\{ \big| \widehat{\bI}_i^{(-k),\intercal}
	\widehat{\mathbb{T}}_i^{(-k)}
		\widehat{\bI}_i^{(-k)} - \bI_i^{*,\intercal}
		\mathbb{T}_i^*
		\bI_i^* \big|  \, \Big| \,  \sigma(\mathcal{B}) , \II_k^c \Big\}
		\nonumber
		\\
		& \hspace*{2cm}
		+
		\Big|
		 \EXP \big\{ \bI_i^{*,\intercal}
		\mathbb{T}_i^*
		\bI_i^* \cond  \sigma(\mathcal{B}) \big\}
		\Big|
		\EXP \Big\{ \big| \widehat{\bI}_i^{(-k),\intercal}
	\widehat{\mathbb{S}}_i^{(-k)}
		\widehat{\bI}_i^{(-k)} - \bI_i^{*,\intercal}
		\mathbb{S}_i^*
		\bI_i^* \big|  \, \Big| \,  \sigma(\mathcal{B}) , \II_k^c \Big\}
		\Big]
		\nonumber
		\\
		&
		\leq
		\frac{C_1'C_2'C_w^4}{2} \Big[
		C_\pi^2 \NT^2 C_2	
		\EXP \Big\{ \big| \widehat{\bI}_i^{(-k),\intercal}
	\widehat{\mathbb{T}}_i^{(-k)}
		\widehat{\bI}_i^{(-k)} - \bI_i^{*,\intercal}
		\mathbb{T}_i^*
		\bI_i^* \big|  \, \Big| \,  \sigma(\mathcal{B}), \II_k^c \Big\}
		\nonumber
		\\
		& \hspace*{2cm}
		+
		2 C_\pi^2 \NT C_2	
		\EXP \Big\{ \big| \widehat{\bI}_i^{(-k),\intercal}
	\widehat{\mathbb{S}}_i^{(-k)}
		\widehat{\bI}_i^{(-k)} - \bI_i^{*,\intercal}
		\mathbb{S}_i^*
		\bI_i^* \big|  \, \Big| \,  \sigma(\mathcal{B}) , \II_k^c \Big\}
		\Big]			\ .							\label{proof-consist-004}
		\end{align}
		In the last inequality of \eqref{proof-consist-004}, we used
		\begin{align*}		
		\EXP \big\{ \bI_i^{*,\intercal}
		\mathbb{S}_i^*
		\bI_i^* \cond \sigma(\mathcal{B})\big\} 
		&
		=
		\EXP \big\{ \bI_i^{*,\intercal}
		(I - \bm{1}\bm{1}\T)
		\Sigma(\bA_i, \bX_i)
		(I - \bm{1}\bm{1}\T)
		\bI_i^* \cond \sigma(\mathcal{B}) \big\} 
		\nonumber
		\\
		&
		\leq
		\EXP \Big\{ \big\| \bI_i^* \big\|_2^2
		\big\| I - \bm{1}\bm{1}\T \big\|_2^2
		\big\| \Sigma(\bA_i, \bX_i) \big\|_2
		\, \Big| \, \sigma(\mathcal{B}) \Big\}
		\nonumber
		\\
		&
		\leq
		C_\pi^2 \NT^2 C_2	\ ,
		\\
		\EXP \big\{ \bI_i^{*,\intercal}
		\mathbb{T}_i^*
		\bI_i^* \cond \sigma(\mathcal{B}) \big\} 
		&
		=
		\EXP \big\{ \bI_i^{*,\intercal}
		\big\{
		2 \Sigma(\bA_i, \bX_i) - \bm{1}\bm{1}\T \Sigma(\bA_i, \bX_i) - \Sigma(\bA_i, \bX_i) \bm{1}\bm{1}\T
		\big\}
		\bI_i^* \cond \sigma(\mathcal{B}) \big\} 
		\nonumber
		\\
		&
		\leq
		2 
		\EXP \Big\{ \big\| \bI_i^* \big\|_2^2
		\big\| I - \bm{1}\bm{1}\T \big\|_2
		\big\| \Sigma(\bA_i, \bX_i) \big\|_2
		\, \Big| \,  \sigma(\mathcal{B}) \Big\}
		\nonumber
		\\
		&
		\leq
		2 C_\pi^2 \NT C_2	\ .
	\end{align*}
	By taking square, we have the following result for some constants $c_1$ and $c_2$.
	\begin{align}
		&
		\Big\{
			\beta^{\dagger,(-k)}(\bC_i) -
			\beta^*(\bC_i) 
		\Big\}^2
		\label{proof-consist-005}
		\\
		&
		\leq c_1 \EXP \Big\{ \big| \widehat{\bI}_i^{(-k),\intercal}
	\widehat{\mathbb{S}}_i^{(-k)}
		\widehat{\bI}_i^{(-k)} - \bI_i^{*,\intercal}
		\mathbb{S}_i^*
		\bI_i^* \big|  \, \Big| \, \sigma(\mathcal{B}) , \II_k^c \Big\}^2 + 
		c_2 
		\EXP \Big\{ \big| \widehat{\bI}_i^{(-k),\intercal}
	\widehat{\mathbb{T}}_i^{(-k)}
		\widehat{\bI}_i^{(-k)} - \bI_i^{*,\intercal}
		\mathbb{T}_i^*
		\bI_i^* \big|  \, \Big| \, \sigma(\mathcal{B}) , \II_k^c \Big\}^2
		\nonumber
		\\
		&
		\leq c_1 \EXP \Big\{ \big| \widehat{\bI}_i^{(-k),\intercal}
	\widehat{\mathbb{S}}_i^{(-k)}
		\widehat{\bI}_i^{(-k)} - \bI_i^{*,\intercal}
		\mathbb{S}_i^*
		\bI_i^* \big|^2  \, \Big| \, \sigma(\mathcal{B}) , \II_k^c \Big\} + 
		c_2 
		\EXP \Big\{ \big| \widehat{\bI}_i^{(-k),\intercal}
	\widehat{\mathbb{T}}_i^{(-k)}
		\widehat{\bI}_i^{(-k)} - \bI_i^{*,\intercal}
		\mathbb{T}_i^*
		\bI_i^* \big|^2  \, \Big| \, \sigma(\mathcal{B}) , \II_k^c \Big\}
		 \ .
		\nonumber		
	\end{align}

	We study upper bounds of two terms in \eqref{proof-consist-005}. Since the two terms in the conditional expectations have the same form, so we only present the upper bound of the second term which is represented as
\begin{align}									\label{proof-consist-006}
		&
		\EXP \Big\{
		\big|
		 \widehat{\bI}_i^{(-k),\intercal} \widehat{\mathbb{S}}_i^{(-k)} \widehat{\bI}_i^{(-k)}
		 -
		 \bI_i^{*,\intercal} \mathbb{S}_i^* \bI_i^*
		 \big|^2
		 \, \Big| \, \sigma(\mathcal{B}), \II_k^c
		\Big\}
		\\
		& = 
		\EXP \Big\{
		\big|
			\big( \bI_i^* + R_\bI) \T \big( \mathbb{S}_i^* + R_\mathbb{S} \big) \big( \bI_i^* + R_\bI)
			-
			 \bI_i^{*,\intercal} \mathbb{S}_i^* \bI_i^*
			 \big|^2
		 \, \Big| \, \sigma(\mathcal{B}), \II_k^c
		\Big\}
		\nonumber
		\\
		& = 
		\EXP \Big\{
		\big|		
			\bI_i^{*,\intercal} R_\mathbb{S}^* \bI_i^*
			 + 2 R_\bI\T \mathbb{S}_i^* \bI_i ^*
			+ 2 R_\bI\T R_\mathbb{S} \bI_i^*
			+
			R_\bI\T \mathbb{S}_i^* R_\bI
			+
			R_\bI\T R_\mathbb{S} R_\bI
			\big|^2
		 \, \Big| \, \sigma(\mathcal{B}),  \II_k^c
		\Big\}
		\nonumber
		\\
		& 
		\leq
		16
		\EXP \Big\{ \big\|
			\bI_i^{*,\intercal} R_\mathbb{S} \bI_i^* \big\|_2^2
		+ 4 \big\|
			R_\bI \T \mathbb{S}_i^* \bI_i^* \big\|_2^2
		+ 4 \big\|
			R_\bI \T R_\mathbb{S} \bI_i^* \big\|_2^2
		+ \big\|
			R_\bI \T \mathbb{S}_i^* R_\bI \big\|_2^2
		+ \big\|
			R_\bI \T R_\mathbb{S} R_\bI \big\|_2^2
		 \, \Big| \, \sigma(\mathcal{B}), \II_k^c
		\Big\} \ .
		\nonumber
\end{align}
Therefore, it is sufficient to obtain the convergence rate of $R_\mathbb{S}$ and $R_\mathbb{T}$; note that $\| R_\bI \|_{P,2} = O_P(r_{\pi,\NC})$ from condition (d) of Lemma \ref{lemma:tauhat}. 

We first derive the convergence rate of $\widehat{\bm{S}}_i^{(-k)} - \bm{\epsilon}_i \bm{\epsilon}_i\T = \widehat{\bm{S}}_i^{(-k)} - \bm{S}_i$. We have the following upper bound with some constant $c_1$ and $c_2$.
	\begin{align*}
		\big\| \widehat{\bm{S}}_i^{(-k)} - \bm{S}_i \big\|_2^2
		&
		\leq
		c_1 \big\{ \bg_i^*(\bA_i) - \widehat{\bg}_i^{(-k)}(\bA_i) \big\}\T \bm{\epsilon}_i^{\otimes 2} \big\{ \bg_i^*(\bA_i) - \widehat{\bg}_i^{(-k)}(\bA_i) \big\} 
		+ c_2 \big\| \bg_i^*(\bA_i) - \widehat{\bg}_i^{(-k)}(\bA_i) \big\|_2^4
		\\
		&
		\leq
		c_1 \big\{ \bg_i^* (\bA_i)- \widehat{\bg}_i^{(-k)}(\bA_i) \big\}\T \bm{\epsilon}_i^{\otimes 2} \big\{ \bg_i^*(\bA_i) - \widehat{\bg}_i^{(-k)}(\bA_i) \big\} 
		+ c_2 C_g^2 \big\| \bg_i^*(\bA_i) - \widehat{\bg}_i^{(-k)}(\bA_i) \big\|_2^2 \ .
	\end{align*}
	The second inequality is from condition (c) of Lemma \ref{lemma:tauhat}. Given $\II_k^c$, we find
	\begin{align*}
		&
		\EXP \Big\{	\big\| \widehat{\bm{S}}_i^{(-k)} - \bm{S}_i \big\|_2^2
		\, \Big| \, \II_k^c \Big\}
		\\
		&
		\leq
		c_1
		\EXP \Big[  \big\{ \bg_i^*(\bA_i) - \widehat{\bg}_i^{(-k)}(\bA_i) \big\}\T \Sigma(\bA_i, \bX_i)  \big\{ \bg_i^*(\bA_i) - \widehat{\bg}_i^{(-k)}(\bA_i) \big\} \, \Big| \, \II_k^c \Big]
		+
		c_2 C_g^2
		\EXP \Big\{  \big\| \bg_i^*(\bA_i) - \widehat{\bg}_i^{(-k)}(\bA_i) \big\|_2^2 \, \Big| \, \II_k^c \Big\}
		\\
		&
		\leq
		(c_1 C_2 + c_2 C_g^2)
		\EXP \Big\{  \big\| \bg_i^*(\bA_i) - \widehat{\bg}_i^{(-k)}(\bA_i) \big\|_2^2 \, \Big| \, \II_k^c \Big\}
		\\
		&
		= O_P(r_{g,\NC}^2) \ .
	\end{align*}
	The first inequality is from $\Sigma(\bA_i, \bX_i) = \EXP(\bm{\epsilon}_i^{\otimes 2} \cond \bA_i , \bX_i)$ and the second inequality is from condition (c) of Lemma \ref{lemma:tauhat}. The last equality holds from condition (d) of Lemma \ref{lemma:tauhat}. Moreover,
\begin{align*}	
		\big \| \mathbb{S}_i^* - \widehat{\mathbb{S}}_i^{(-k)} \big\|_2^2
		 \leq
		 \big\| I - \bm{1}\bm{1}\T \big\|_2^4 \big\| \bm{S}_i -  \widehat{\bm{S}}_i^{(-k)} \big\|_2^2
		 \leq
		 \big\| I - \bm{1}\bm{1}\T \big\|_F^4 \big\| \bm{S}_i -  \widehat{\bm{S}}_i^{(-k)} \big\|_2^2
		 \leq \NT^4 \big\| \bm{S}_i -  \widehat{\bm{S}}_i^{(-k)} \big\|_2^2
\end{align*}
and
\begin{align*}
	\big\| \mathbb{T}_i^* - \widehat{\mathbb{T}}_i^{(-k)} \big\|_2^2
	&
	\leq
	8 \big\| \bm{S}_i -  \widehat{\bm{S}}_i^{(-k)} \big\|_2^2
	+ 8 \big\| \bm{1}\bm{1}\T \big\|_2^2 \big\| \bm{S}_i -  \widehat{\bm{S}}_i^{(-k)} \big\|_2^2
	\leq
	(8 + 8\NT^2)  \big\| \bm{S}_i -  \widehat{\bm{S}}_i^{(-k)} \big\|_2^2
	\ .
\end{align*}
Therefore, 
\begin{align}								\label{proof-consist-007}
	&
	\big \|  \mathbb{S}_i^* - \widehat{\mathbb{S}}_i^{(-k)}  \big\|_{P,2}^2
		 =
		 O_P(r_{g,\NC}^2) \quad , \quad
	\big \|  \mathbb{T}_i^* - \widehat{\mathbb{T}}_i^{(-k)}  \big\|_{P,2}^2
	 =
	 O_P(r_{g,\NC}^2)  \ .
\end{align}

Moreover, for any $\bm{a}_i = \bm{a}(\bA_i, \bX_i)$ and $\bm{b}_i = \bm{b}(\bA_i, \bX_i)$ be $\NI_i$-dimensional random and bounded vectors, there exists a constant $C'$ so that
\begin{align}								\label{proof-consist-008}
	\EXP \big( \bm{a}_i \T \mathbb{S}_i^* \bm{b}_i \big)
	&
	\leq
	\EXP \big\{ \bm{a}_i \T (I-\bm{1}\bm{1}\T) \Sigma(\bA_i, \bX_i) (I-\bm{1}\bm{1}\T) \bm{b}_i \big\}
	\leq C' 
	\EXP \big\{ \| \bm{a}_i \|_2 \| \bm{b}_i \|_2 \big\}
	\leq  C'  \| \bm{a}_i \|_{P,2} \| \bm{b}_i \|_{P,2} 
\end{align}
and $\EXP \big( \bm{a}_i \T \mathbb{T}_i ^* \bm{b}_i \big) \leq C'  \| \bm{a}_i \|_{P,2} \| \bm{b}_i \|_{P,2}$ from similar manner.

The five terms in the last line of \eqref{proof-consist-006} have the following convergence rate by using conditions (c) and (d) of Lemma \ref{lemma:tauhat}, \eqref{proof-consist-007}, \eqref{proof-consist-008}, and the properties of matrix norms.
\begin{align*}
	&
		\EXP \big\{ \big\|
			\bI_i^* \big\|_2^4 \big\| R_\mathbb{S} \big\|_2^2
		 \, \big| \, \II_k^c
		\big\}
		\stackrel{(b)}{\leq}
		C_\pi^4 \big\| R_\mathbb{S} \big\|_{P,2} ^2
		\stackrel{\eqref{proof-consist-007}}{=} O_P(r_{g,\NC}^2)
		\\
		&
		\EXP \big\{ \big\|
			R_\bI \T \mathbb{S}_i \bI_i^* \big\|_2^2
		 \, \big| \, \II_k^c
		\big\}
		\stackrel{ \eqref{proof-consist-008} }{\leq}
		C'^2  \big\| R_\bI \big\|_{P,2}^2 \big\| \bI_i^* \big\|_{P,2}^2
		\stackrel{(a),(b)}{=}
		O_P(r_{\pi,\NC}^2)
		\\
		&
		\EXP \big\{ \big\|
			R_\bI \T R_\mathbb{S} \bI_i^* \big\|_2^2
		 \, \big| \, \II_k^c
		\big\}
		\leq
		\EXP \big\{ \big\|
			R_\bI \big\|_2^2 \big\| R_\mathbb{S} \big\|_2^2 \big\| \bI_i^* \big\|_2^2
		 \, \big| \, \II_k^c
		\big\}
		\stackrel{(b)}{\leq}
		C_\pi^2
		\big\| R_\bI \big\|_{P,2}^2 \big\| R_\mathbb{S} \big\|_{P,2}^2
		\stackrel{\eqref{proof-consist-007}}{=} O_P( r_{\pi,\NC}^2 r_{g,\NC}^2 )
		\\
		&
		\EXP \big\{ \big\|
			R_\bI \T \mathbb{S}_i R_\bI \big\|_2 ^2
			\, \big| \, \II_k^c
		\big\}
		\stackrel{ \eqref{proof-consist-008} }{\leq}
		C'  \big\| R_\bI \big\|_{P,2}^4
		\stackrel{(a)}{=} O_P( r_{\pi,\NC}^4 )
		\\
		&
		\EXP \big\{ \big\|
			R_\bI \T R_\mathbb{S} R_\bI \big\|_2 ^2
			\, \big| \, \II_k^c
		\big\}
		\leq
		\EXP \big\{ \big\|
			R_\bI \big\|_2^4 \big\| R_\mathbb{S} \big\|_2 ^2
			\, \big| \, \II_k^c
		\big\}
		\leq
		\big\|
			R_\bI \big\|_{P,2}^4 \big\| R_\mathbb{S} \big\|_{P,2} ^2
			\stackrel{(a),\eqref{proof-consist-007}}{=} O_P( r_{\pi,\NC}^4 r_{g,\NC}^2 ) \ .
\end{align*}
Therefore, the expectation of \eqref{proof-consist-006} with respect to $\beta(\bC_i)$ is $O_P(r_{\pi,\NC}) + O_P(r_{g,\NC})$.
\begin{align*}
	&
	\EXP \Big\{
		\big|
		 \widehat{\bI}_i^{(-k),\intercal} \widehat{\mathbb{S}}_i^{(-k)} \widehat{\bI}_i^{(-k)}
		 -
		 \bI_i^{*,\intercal} \mathbb{S}_i^* \bI_i^*
		 \big|^2
		 \, \Big| \, \II_k^c
		\Big\}
		\\
		&
		\leq
		16
		\EXP \bigg[
		\EXP \Big\{ \big\|
			\bI_i^{*,\intercal} R_\mathbb{S} \bI_i^* \big\|_2^2
		+ 4 \big\|
			R_\bI \T \mathbb{S}_i^* \bI_i^* \big\|_2^2
		+ 4 \big\|
			R_\bI \T R_\mathbb{S} \bI_i^* \big\|_2^2
		+ \big\|
			R_\bI \T \mathbb{S}_i^* R_\bI \big\|_2^2
		+ \big\|
			R_\bI \T R_\mathbb{S} R_\bI \big\|_2^2
		 \, \Big| \, \beta( \bC_i ), \II_k^c
		\Big\}
		\, \bigg| \, \II_k^c \bigg]
		\\
		&
		=
		16
		\EXP \Big\{ \big\|
			\bI_i^{*,\intercal} R_\mathbb{S} \bI_i^* \big\|_2^2
		+ 4 \big\|
			R_\bI \T \mathbb{S}_i^* \bI_i^* \big\|_2^2
		+ 4 \big\|
			R_\bI \T R_\mathbb{S} \bI_i^* \big\|_2^2
		+ \big\|
			R_\bI \T \mathbb{S}_i^* R_\bI \big\|_2^2
		+ \big\|
			R_\bI \T R_\mathbb{S} R_\bI \big\|_2^2
		 \, \Big| \, \II_k^c
		\Big\}
		\\
		& 
		= 
		O_P(r_{\pi,\NC}^2) + O_P(r_{g,\NC}^2) + O_P(r_{\pi,\NC} ^2r_{g,\NC}^2) + O_P(r_{\pi,\NC}^4) + O_P(r_{\pi,\NC}^4 r_{g,\NC}^2)
		\\
		&
		= 
		O_P(r_{\pi,\NC}^2) + O_P(r_{g,\NC}^2) \ . 
\end{align*}
Similarly, we find $\EXP \big\{
		\big|
		 \widehat{\bI}_i^{(-k),\intercal} \widehat{\mathbb{T}}_i^{(-k)} \widehat{\bI}_i^{(-k)}
		 -
		 \bI_i^{*,\intercal} \mathbb{T}_i^* \bI_i^*
		 \big|^2
		 \, \big| \, \sigma(\mathcal{B}) , \II_k^c
		\big\} = O_P(r_{\pi,\NC}^2) + O_P(r_{g,\NC})^2$. Plugging the result in \eqref{proof-consist-005} after taking the expectation with respect to $\beta(\bC_i)$, we find the desired result.
		\begin{align*}
		\Big\|
		\beta^{\dagger,(-k)}-
			\beta^*
		\Big\|_{P,2}^2
		=
		\EXP \bigg[
		\Big\{
			\beta^{\dagger,(-k)}(\bC_i) -
			\beta^*(\bC_i) 
		\Big\}^2
		\, \bigg| \, \II_k^c
		\bigg]
		= 
		O_P(r_{\pi,\NC}^2) + O_P(r_{g,\NC}^2) \ .
		\end{align*}

\subsection{Proof of Lemma \ref{lemma:consistency2}}

	The outline of the proof is given as follows. In \textbf{[Step 1]}, we show that $\bm{\gamma}^{\dagger,(-k)}$ is identifiable. In \textbf{[Step 2]}, we show that the estimating equation in \eqref{eq-EEbeta} is uniformly consistent.  In \textbf{[Step 3]}, we show that $\widehat{\bm{\gamma}}^{(k)}$ is consistent for $\bm{\gamma}^{\dagger,(-k)}$. In \textbf{[Step 4]}, we find the rate of $\big\|
		\widehat{\beta}^{(k)} (\bC_i) - \beta^{\dagger,(-k)} (\bC_i)
	\big\|_{P,2}$.
	
	\begin{itemize}[leftmargin=0.5cm]
		\item \textbf{[Step 1]}: Identifiability of $\bm{\gamma}^{\dagger,(-k)}$. 

If $\EXP \big\{ w(\bC_i)^2 \cond L(\bC_i) =\ell \big\}=0$ for some $\ell$, the average treatment effect does not use the clusters in $\ell$th stratum so $\beta(\bC_i)$ is not defined in $\ell$th stratum. Therefore, $\EXP \big\{ w(\bC_i)^2 \cond L(\bC_i) =\ell \big\}>0$ for all $\ell = 1,\ldots,\MT$. We re-write \eqref{eq-popEE1} in terms of $\bm{\gamma}$ as follows.
\begin{align*}
	&
	\EXP \big\{ \widehat{\Psi}^{(-k)}(\bO_i \con \bm{\gamma}^{\dagger,(-k)}) \cond \sigma(\paraB) , \II_k^c \big\} 
	=
	\begin{bmatrix}
		\EXP \big\{	 \widehat{\Psi}_1^{(-k)}(\bO_i \con \gamma_1^{\dagger,(-k)}) \cond \sigma(\paraB) , \II_k^c \big\}
		\\ \vdots \\
		\EXP \big\{	\widehat{\Psi}_\MT^{(-k)} (\bO_i \con \gamma_\MT^{\dagger,(-k)})  \cond \sigma(\paraB) , \II_k^c \big\}
	\end{bmatrix}
	=
	0
	\ , \\
	&
	\widehat{\Psi}_\ell^{(-k)} (\bO_i \con \gamma_\ell)
	=
\ind \big\{ L(\bC_i) = \ell \big\}
	w(\bC_i)^2	
	\Big[
		2 \gamma_\ell
		\Big\{ \widehat{\bI}_i^{(-k),\intercal}
		 \widehat{\mathbb{S}}_i^{(-k)} \widehat{\bI}_i^{(-k)}
		\Big\}
		-
		\widehat{\bI}_i^{(-k),\intercal}
		\widehat{\mathbb{T}}_i^{(-k)}
		\widehat{\bI}_i^{(-k)}
	\Big] \ .
\end{align*}
Accordingly, $\widehat{\bm{\gamma}}^{(k)}$ is the solution to the following estimating equation.
\begin{align}																		\label{eq-EEbeta}
		\frac{1}{\NC/2} \sum_{i\in \II_k} \widehat{\Psi}^{(-k)} (\bO_i \con \widehat{\bm{\gamma}}^{(k)} )
		=
		0
		\ .
\end{align}
And we find that the $\ell$th component of $\bm{\gamma}^{\dagger,(-k)}$ is 
\begin{align*}
	\gamma_\ell^{\dagger,(-k)}
		 =
		 \frac{ \ind \big\{ L(\bC_i) =\ell \big\} }{2}  
		 \frac{  \EXP \big\{ w(\bC_i)^2 \widehat{\bI}_i^{(-k),\intercal}
	\widehat{\mathbb{T}}_i^{(-k)}
		\widehat{\bI}_i^{(-k)} \cond L(\bC_i) =\ell , \II_k^c \big\}  }{ \EXP \big\{ w(\bC_i)^2 \widehat{\bI}_i^{(-k),\intercal}
	\widehat{\mathbb{S}}_i^{(-k)}
		\widehat{\bI}_i^{(-k)} \cond L(\bC_i) =\ell , \II_k^c \big\} }
		\ .
\end{align*}
From similar steps in \eqref{proof-consist-003}, we can show that $\EXP \big\{ w(\bC_i)^2 \widehat{\bI}_i^{(-k),\intercal}
	\widehat{\mathbb{S}}_i^{(-k)}
		\widehat{\bI}_i^{(-k)} \cond L(\bC_i) =\ell , \II_k^c \big\}$ is strictly bounded away from zero. Therefore,  $\bm{\gamma}^{\dagger,(-k)}$ is uniquely defined. 
		
		\item \noindent \textbf{[Step 2]}: Uniform consistency.

To bound the first term, we use Lemma 2.4 of \citet{NMcF1994} which is restated as follows in our context. Suppose  that  (i) the data is i.i.d., (ii) $\Omega \equiv [-B_0,B_0]^{\otimes \MT}$ is compact, (iii) $\widehat{\Psi}^{(-k)} (\bO_i \con \bm{\gamma})$ is continuous at each $\bm{\gamma} \in \Omega$ with probability 1, and (iv) there exists $F(\bO_i)$ with $\big\| \widehat{\Psi}^{(-k)} (\bO_i \con \bm{\gamma}) \big\|_2 \leq F(\bO_i)$ for all $\bm{\gamma} \in \Omega$ and $\EXP \big\{ F(\bO_i) \big\} < \infty$. Then, 
		\begin{align}							\label{proof-UC-decompose1}
		\sup_{\bm{\gamma} \in \Omega}
		\Bigg\|
				\frac{1}{\NC/2} \sum_{i \in \II_k} \widehat{\Psi}^{(-k)} (\bO_i \con \bm{\gamma})
				-
				\EXP \Big\{ \widehat{\Psi}^{(-k)} (\bO_i \con \bm{\gamma}) \, \Big| \, \II_k^c \Big\}
				\Bigg\|_2 = o_P(1) \ .
		\end{align}
	Conditions (i), (ii), and (iii) are trivially satisfied, so it suffices to show condition (iv). From some algebra, we find
	\begin{align}								\label{proof-UC-001}
		\widehat{\bm{S}}_i^{(-k)} 
		&
		=
		\bm{\epsilon}_i^{\otimes 2}
		+
		\big\{ \bm{g}^*(\bA_i, \bX_i) - \widehat{\bm{g}}^{(-k)}(\bA_i, \bX_i) \big\} \bm{\epsilon}_i\T
		\\
		& \hspace*{3cm}
		+
		\bm{\epsilon}_i \big\{ \bm{g}^*(\bA_i, \bX_i) - \widehat{\bm{g}}^{(-k)}(\bA_i, \bX_i) \big\} \T
		+
		\big\{ \bm{g}^*(\bA_i, \bX_i) - \widehat{\bm{g}}^{(-k)}(\bA_i, \bX_i) \big\}^{\otimes 2}
		\nonumber
	\end{align}
	which leads to
	\begin{align}										\label{proof-UC-101}
		\Big\| \widehat{\mathbb{S}}_i^{(-k)} \Big\|_2
		&
		\leq
		\big\| I - \bm{1}\bm{1}\T \big\|_2^2 \Big\| \widehat{\bm{S}}_i^{(-k)}  \Big\|_2
		\nonumber
		\\
		&
		\leq
		\big\| I - \bm{1}\bm{1}\T \big\|_F^2\Big\{
			\big\| \bm{\epsilon}_i \big\|_2^2
			+
			2
			\big\| \bm{\epsilon}_i \big\|_2 \big\| \bm{g}^*(\bA_i, \bX_i) - \widehat{\bm{g}}^{(-k)}(\bA_i, \bX_i) \big\|_2 
			+
			\big\| \bm{g}^*(\bA_i, \bX_i) - \widehat{\bm{g}}^{(-k)}(\bA_i, \bX_i) \big\|_2^2
		\Big\}
		\nonumber
		\\
		&
		\leq
		\NT^2 \Big\{ \big\| \bm{\epsilon}_i \big\|_2^2 + 2C_g \big\| \bm{\epsilon}_i \big\|_2 + C_g \Big\} \ .
	\end{align}
	The first and second inequalities hold from the properties of matrix norms and the last inequality holds from condition (c) of Lemma \ref{lemma:tauhat}. Similarly,
		\begin{align*}
		\Big\| \widehat{\mathbb{T}}_i^{(-k)} \Big\|_2
		&
		\leq
		2 \big\| I - \bm{1}\bm{1}\T \big\|_2 \Big\| \widehat{\bm{S}}_i^{(-k)}  \Big\|_2
		\\
		&
		\leq
		2 \big\| I - \bm{1}\bm{1}\T \big\|_F\Big\{
			\big\| \bm{\epsilon}_i \big\|_2^2
			+
			2
			\big\| \bm{\epsilon}_i \big\|_2 \big\| \bm{g}^*(\bA_i, \bX_i) - \widehat{\bm{g}}(\bA_i, \bX_i) \big\|_2 
			+
			\big\| \bm{g}^*(\bA_i, \bX_i) - \widehat{\bm{g}}(\bA_i, \bX_i) \big\|_2^2
		\Big\}
		\\
		&
		\leq
		2 \NT \Big\{ \big\| \bm{\epsilon}_i \big\|_2^2 + 2C_g \big\| \bm{\epsilon}_i \big\|_2 + C_g \Big\} \ .
	\end{align*}	
	As a consequence, we find the following inequality holds for any $\bm{\beta}$.
	\begin{align}									\label{proof-UC-002}
		\big\| \widehat{\Psi}^{(-k)} (\bO_i \con \bm{\beta}) \big\|_2
		&		
		\leq
		C_w^2 \Big\{ 
		2 \big\| \bm{\beta} \big\|_2 \Big\| \widehat{\bI}_i^{(-k)} \Big\|_2^2 
		\Big\| \widehat{\mathbb{S}}_i^{(-k)} \Big\|_2
		+
		\Big\| \widehat{\bI}_i^{(-k)} \Big\|_2^2 
		\Big\| \widehat{\mathbb{T}}_i^{(-k)} \Big\|_2
		\Big\}
		\nonumber
		\\
		&
		\leq
		C_w^2 C_\pi^2 \Big[
		2 \text{diam}(\Omega)
		\Big\| \widehat{\mathbb{S}}_i^{(-k)} \Big\|_2
		\NT^2 \Big\{ \big\| \bm{\epsilon}_i \big\|_2^2 + 2C_g \big\| \bm{\epsilon}_i \big\|_2 + C_g \Big\}
		+
		2 \NT \Big\{ \big\| \bm{\epsilon}_i \big\|_2^2 + 2C_g \big\| \bm{\epsilon}_i \big\|_2 + C_g \Big\}
		\Big]
		\nonumber
		\\
		&
		\leq
		c_2 \big\| \bm{\epsilon}_i \big\|_2^2 + c_1 \big\| \bm{\epsilon}_i \big\|_2 + c_0 
	\end{align}
	where $c_0$, $c_1$, and $c_2$ are generic constants. Since $\EXP \big\{\big\|  \bm{\epsilon}_i \big\|_2^2 \cond \II_k^c \big\}$ and $\EXP \big\{ \big\| \bm{\epsilon}_i \big\|_2 \cond \II_k^c \big\}$ are integrable, $F(\bO_i) = c_2 \big\| \bm{\epsilon}_i \big\|_2^2 + c_1 \big\| \bm{\epsilon}_i \big\|_2 + c_0 $ satisfies condition (iv) of Lemma 2.4 of \citet{NMcF1994}. As a consequence, \eqref{proof-UC-decompose1} holds.
	
	\item \textbf{[Step 3]}: Consistency of $\widehat{\bm{\gamma}}^{(k)}$ 
	
	Conditioning on $\II_k^c$, $\widehat{\pi}^{(-k)}$ and $\widehat{g}^{(-k)}$ can be understood as fixed quantities. Therefore, we find $\widehat{\bm{\gamma}}^{(k)} = \bm{\gamma}^{\dagger,(-k)} + o_P(1)$  from Theorem 5.9 of \citet{Vaart1998}. \\

\noindent 	\textbf{[Step 4]}: Rate of $\big\| \widehat{\beta}^{(k)} (\bC_i) - \beta^{\dagger,(-k)} (\bC_i) \big\|_{P,2}^2 = \int \big\| \widehat{\beta}^{(k)}(\bc_i) - \beta^{\dagger,(-k)}(\bc_i) \big\|_2^2 \, dP(\bc_i)$ \\
	
	We check conditions for Theorem 5.21 of \citet{Vaart1998}. First, for any $\bm{\gamma}_1$ and $\bm{\gamma}_2$ in $\Omega \equiv [-B_0,B_0]^{\otimes \MT}$, we find
	\begin{align*}
		\widehat{\Psi}^{(-k)} (\bO_i \con \bm{\gamma}_1)
		-
		\widehat{\Psi}^{(-k)} (\bO_i \con \bm{\gamma}_2)
		=
		2 w(\bC_i)^2 
		\Big\{ \widehat{\bI}_i^{(-k),\intercal}	 \widehat{\mathbb{S}}_i^{(-k)} \widehat{\bI}_i^{(-k)}
		\Big\}
		\begin{bmatrix}
			\ind \{ L(\bC_i) = 1 \} \big( \gamma{1,1} - \gamma_{1,2} \big)
			\\ \vdots \\ 
			\ind \{ L(\bC_i) = \MT \} \big( \gamma{\MT,1} - \gamma_{\MT,2} \big)
		\end{bmatrix} \ ,
	\end{align*}
	and this implies
	\begin{align*}
		\Big\| 
		\widehat{\Psi}^{(-k)} (\bO_i \con \bm{\gamma}_1)
		-
		\widehat{\Psi}^{(-k)} (\bO_i \con \bm{\gamma}_2)
		\Big\|_2
		 & \leq
		2 C_w^2 C_\pi^2 \Big\|   \widehat{\mathbb{S}}_i^{(-k)}  \Big\|_2
		\Big\| \bm{\gamma}_1 - \bm{\gamma}_2 \Big\|_2
		\\
		& \leq
		\Big\{ c_2\big\| \bm{\epsilon}_i \big\|_2^2 + c_1 \big\| \bm{\epsilon}_i \big\|_2 + c_0 \Big\}
		\Big\| \bm{\gamma}_1 - \bm{\gamma}_2 \Big\|_2
		\ .
	\end{align*}
	The second inequality holds from \eqref{proof-UC-101} with some constants $c_0$, $c_1$, and $c_2$. Therefore, we find the function on the right hand side is squared integrable as follows.
	\begin{align*}
		&
		\EXP \bigg[ \Big\{ c_2\big\| \bm{\epsilon}_i \big\|_2^2 + c_1 \big\| \bm{\epsilon}_i \big\|_2 + c_0 \Big\}^2 \bigg]
		=
		\EXP \Big\{ c_4'\big\| \bm{\epsilon}_i \big\|_2^4 +  c_3'\big\| \bm{\epsilon}_i \big\|_2^3 +  c_2'\big\| \bm{\epsilon}_i \big\|_2^2 + c_1' \big\| \bm{\epsilon}_i \big\|_2 + c_0' \Big\}
		< \infty
	\end{align*}
	where $c_0', \ldots, c_4'$ are some constants.
	
	Second, $\EXP \big\{ \big\| \widehat{\Psi}^{(-k)} (\bO_i \con \bm{\gamma}^*) \big\|_2 \cond \II_k^c \big\} < \infty$ because \eqref{proof-UC-002} holds with $\bm{\gamma} = \bm{\gamma}^*$.
	
	Third, $\EXP \big\{ \widehat{\Psi}^{(-k)} (\bO_i \con \bm{\gamma})  \cond \II_k^c \big\}$ is differentiable for any $\bm{\gamma}$ with the derivative
	\begin{align*}
		\frac{\partial}{\partial \bm{\gamma}}
		\EXP \big\{ \widehat{\Psi}^{(-k)} (\bO_i \con \bm{\gamma})  \cond \II_k^c \big\}
		=
		2 w(\bC_i)^2 
		\Big\{ \widehat{\bI}_i^{(-k),\intercal}	 \widehat{\mathbb{S}}_i^{(-k)} \widehat{\bI}_i^{(-k)}
		\Big\}
		\text{diag} \Big[ \ind \big\{ L(\bC_i) = 1 \big\} ,\ldots, \ind \big\{ L(\bC_i) = \MT \big\}  \Big] \ .
	\end{align*}
	The expectation of the $\ell$th diagonal of the above term is lower bounded by a positive constant $C_L$ as follows.
	\begin{align}											\label{proof-UC-003}
	\nonumber
		&
		\EXP \Big[ 2 w(\bC_i)^2  \ind \big\{ L(\bC_i) = \ell \big\} \Big\{ \widehat{\bI}_i^{(-k),\intercal}	 \widehat{\mathbb{S}}_i^{(-k)} \widehat{\bI}_i^{(-k)}
		\Big\} \, \Big| \, \II_k^c \Big]
		\\
		&
		=
		2 P\big\{ L(\bC_i) = \ell \big\} \EXP \Big[ w(\bC_i)^2 \Big\{ \widehat{\bI}_i^{(-k),\intercal}	 \widehat{\mathbb{S}}_i^{(-k)} \widehat{\bI}_i^{(-k)}
		\Big\} \, \Big| \, L(\bC_i) = \ell , \II_k^c \Big]
		\nonumber
		\\
		&
		\geq 
		C_L \equiv
		\frac{ 2 P\big\{ L(\bC_i) = \ell \big\} \delta}{C_2\NT(C_\pi-1)} 
		\EXP \big\{ w(\bC_i)^2 \cond L(\bC_i) = \ell , \II_k^c \big\} > 0 \ .
	\end{align}		
	The inequality holds from \eqref{proof-consist-103} and $\EXP \big\{ w(\bC_i)^2 \cond L(\bC_i) = \ell , \II_k^c \big\}>0$ (if this quantity is zero, the investigator should not define a parameter of statum $\ell$). This implies the derivative of $\EXP \big\{ \widehat{\Psi}^{(-k)} (\bO_i \con \bm{\gamma})  \cond \II_k^c \big\}$ is nonsingular at $\bm{\gamma}=\bm{\gamma}^*$. 
	
	Lastly, we have $\sum_{i \in \II_k}  \widehat{\Psi}^{(-k)} (\bO_i \con \widehat{\bm{\gamma}}^{(k)})=0$ from the definition of $\widehat{\bm{\gamma}}^{(k)}$ and $\widehat{\bm{\gamma}}^{(k)} = \bm{\gamma}^{\dagger,(-k)} + o_P(1)$ from \textbf{[Step 3]}. Therefore, all conditions of Theorem 5.21 of \citet{Vaart1998} are satisfied. Thus, we have the following asymptotic normality for some variance matrix $V_0$.
	\begin{align*}
		\sqrt{ \frac{\NC}{2} }
		\Big\{ \widehat{\bm{\gamma}}^{(k)} - \bm{\gamma}^{\dagger,(-k)}  \Big\}
		\stackrel{D}{\rightarrow} N(0, V_0) \ .
	\end{align*}
	This implies $\widehat{\bm{\gamma}}^{(k)} - \bm{\gamma}^{\dagger,(-k)} = O_P(\NC^{-1/2})$. 
	
	\item \textbf{[Step 4]}: Rate of $\big\|
		\widehat{\beta}^{(k)} (\bC_i) - \beta^{\dagger,(-k)} (\bC_i)
	\big\|_{P,2}$
	
	Since $\beta(\bC_i) = \sum_{\ell=1}^\MT \ind \big\{ L(\bC_i) = \ell \big\} \gamma_\ell$ , we find
	\begin{align*}
		\int \big\| \widehat{\beta}^{(k)}(\bc_i) - \beta^{\dagger,(-k)}(\bc_i) \big\|_2^2 \, dP(\bc_i)
		&
		=
		\int \sum_{\ell=1}^\MT \ind \big\{ L(\bC_i) = \ell \big\} \Big\{ \widehat{\gamma}_\ell^{(k)} - \gamma_\ell^{\dagger,(-k)} \Big\}^2 \, dP(\bc_i)
		\\
		&
		\leq
		\MT
		\sum_{\ell=1}^{\MT} \Big\{ \widehat{\gamma}_\ell^{(k)} - \gamma_\ell^{\dagger,(-k)} \Big\}^2 
		\leq
		\NT 
		\Big\| \widehat{\bm{\gamma}}^{(k)} - \bm{\gamma}^{\dagger,(-k)} \Big\|_2^2
		=
		O_P(\NC^{-1}) \ .
	\end{align*}
	As a consequence, $\big\| \widehat{\beta}^{(k)} (\bC_i) - \beta^{\dagger,(-k)} (\bC_i) \big\|_{P,2} = O_P(\NC^{-1/2})$. This concludes the proof.

	\end{itemize}

\newpage
\bibliographystyle{apa}
\bibliography{ClusterEff.bib}

\end{document}